\documentstyle[12pt]{article}
\newcommand{\pbar}{\overline{p}}
\newcommand{\pbarp}{\overline{p}p}
\newcommand{\nbar}{\overline{n}}
\newcommand{\nnbar}{n\overline{n}}
\newcommand{\pbarn}{\overline{p}n}
\newcommand{\pbard}{\overline{p}d}
\newcommand{\NNbar}{\overline{N}N}
\newcommand{\KKbar}{K\overline{K}}
\newcommand{\qqbar}{q\overline{q}}
\newcommand{\uubar}{u\overline{u}}
\newcommand{\ddbar}{d\overline{d}}
\newcommand{\ssbar}{s\overline{s}}
\newcommand{\nbarn}{\overline{n}n}
\newcommand{\twoqqbars}{\overline{q}^2 q^2 }
\newcommand{\threeqqbars}{\overline{q}^3 q^3 }
\newcommand{\ro}{\rho -\omega}
\newcommand{\oeg}{\omega\rightarrow\eta\gamma}
\newcommand{\reg}{\rho^0\rightarrow\eta\gamma}
\newcommand{\opg}{\omega\rightarrow\pi^0\gamma}
\newcommand{\po}{\pi^0\omega}
\newcommand{\eo}{\eta\omega}
\newcommand{\otg}{\omega\rightarrow 3\gamma}

\newcommand{\etap}{\eta'}
\begin{document}

\title{Proton-Antiproton Annihilation \\
and Meson Spectroscopy\\
with the Crystal Barrel\footnote{Submitted to Reviews of Modern Physics}}
 
\author{Claude Amsler\\
Physik-Institut der Universit\"at Z\"urich, \\
Winterthurerstrasse 190, CH-8057 Z\"urich, Switzerland}
\maketitle

\abstract{This report reviews the achievements of the Crystal Barrel experiment at 
the Low Energy Antiproton Ring (LEAR) at CERN. During seven years of operation Crystal 
Barrel has collected very large statistical samples in $\pbarp$ annihilation, especially at 
rest and with emphasis on final states with high neutral multiplicity. The measured rates 
for annihilation into various two-body channels and for electromagnetic processes have 
been used to test simple models for the annihilation mechanism based on the quark 
internal structure of hadrons. From three-body annihilations three scalar 
mesons, $a_0(1450)$, $f_0(1370)$ and $f_0(1500)$ have been established in various decay 
modes. One of them, $f_0(1500)$, may be identified with the expected ground state 
scalar glueball.}

\tableofcontents

\section{Introduction}
Low energy antiproton-proton annihilation at rest is a valuable tool 
to investigate phenomena in the low energy regime of Quantum 
Chromodynamics (QCD). Due to the absence of Pauli blocking, the 
antiproton and proton overlap and one expects the interactions 
between constituent quarks and antiquarks (annihilation, pair 
creation or rearrangement) to play an important role in the 
annihilation process. From bubble chamber experiments 
performed in the sixties (Armenteros and French, 1969) one knows 
that annihilation proceeds through $\qqbar$ intermediate meson 
resonances. The $\omega(782) $, $f_1(1285)$,  $E/\eta(1440)$ and 
$K_1(1270)$ mesons were discovered  and numerous properties of 
other mesons ($a_0(980)$, $K^*(892)$, $\phi(1020)$, $a_2(1320)$) 
were studied in low energy $\pbarp$ annihilation. With the advent 
of QCD one now also predicts states made exclusively of gluons 
(glueballs), of a mixture of quarks and gluons (hybrids) and 
multiquark states, all of which can be produced in $\pbarp$ 
annihilation.

With the invention of stochastic cooling and the operation of 
the Low Energy Antiproton Ring (LEAR) from 1983 to 1996, intense 
and pure accelerator beams of low momentum antiprotons between 60 
and 1940 MeV/s were available at CERN. It is impressive to 
compare the high flux of today's antiproton beams ($>$ 10$^6$ 
$\pbar$/s) with the rate of about 1 $\pbar$ every 15 minutes in 
the early work when the antiproton was discovered, back in 1955 
(Chamberlain, 1955). 

This survey covers the results obtained with the Crystal Barrel, designed to study low 
energy $\pbarp$ annihilation with very high statistics, in particular 
annihilation into $n$ charged particles ($n$-prong) and $m$ neutrals ($\pi^0, \eta, \etap$ 
or $\omega$) with m $\geq$ 2,  leading to final states with several photons. These 
annihilation channels occur with a probability of  
about  50\% and have not been investigated previously. They are often   
simpler to analyze due to $C$-parity conservation which limits 
the range of possible quantum numbers for the intermediate 
resonances and the $\pbarp$ initial states.

The experiment started data taking in late 1989 and was completed 
in autumn 1996 with the closure of LEAR. Most of the data 
analyzed so far  
were taken with stopping antiprotons in liquid hydrogen on which I 
shall therefore concentrate. This article is organized as follows: 
After a brief reminder of the physical processes involved when 
antiprotons are stopped in liquid hydrogen (section 2), I shall describe in 
section 3 the Crystal Barrel apparatus and its 
performances. The review then covers results relevant to the 
annihilation mechanism and the roles of quarks in the annihilation 
process (section \ref{seq:anni}). Electromagnetic processes are covered in 
section \ref{sec:em}. The observation of a strangeness enhancement 
may possibly be related to the presence of strange quarks in the 
nucleon (section \ref{seq:phi}). After 
describing the mathematical tools for extracting masses and spins of 
intermediate resonances (section \ref{sec:spectro}) I shall review in sections  
8 to 10 what is considered to 
be the main achievement, the discovery of several new 
mesons,  in particular a scalar ($J^P=0^+$) state around 
1500 MeV, which is generally interpreted as the ground state 
glueball. Section 11 finally describes the status of pseudoscalars 
in the 1400 MeV region.

In this review I shall concentrate on results published by 
the Crystal Barrel Collaboration 
or submitted for publication before summer 1997\footnote{All Crystal Barrel publications 
are listed with their full titles in the reference section.}. Alternative 
analyses have been performed by other groups 
using more flexible parametrizations and also data from previous experiments (e.g. Bugg 
(1994, 1996), Abele (1996a)). I shall only refer to them without describing them in detail 
since they basically lead to the same results. Results on $\pbard$ annihilation will not be 
reviewed here. They include the observation of the channels $\pbard\rightarrow$ 
$\pi^0 n, \eta n, \omega n$ (Amsler, 1995a),  $\pbard\rightarrow\Delta(1232)\pi^0$ 
(Amsler, 1995b) which involve both nucleons in the reaction process.

\section{Proton-Antiproton Annihilation at Rest}
Earlier investigations of low energy $\pbarp$ annihilation have dealt  
mainly with final states involving charged mesons ($\pi^{\pm}, 
K^{\pm}$) or $K_S\rightarrow \pi^+\pi^-$, with at most one missing 
(undetected) $\pi^0$, due to the lack of a good $\gamma$ detection 
facility (for reviews, see Armenteros and French (1969), Sedl\'ak 
and \v{S}im\'ak (1988) and  
Amsler and Myhrer (1991)). 

The average 
charged pion multiplicity is 3.0 $\pm$ 0.2 for annihilation at rest and the 
average $\pi^0$ multiplicity is 2.0 $\pm$ 0.2. The fraction of purely neutral annihilations 
(mainly from channels like $3\pi^0$,  $5\pi^0$, $2\pi^0\eta$ and $4\pi^0\eta$ decaying to 
photons only) is (3.9 $\pm$ 0.3) \% (Amsler, 1993a). This number is in good agreement 
with an earlier estimate from bubble chambers, (4.1 $^{+ 0.2}_{- 0.6}$) \% (Ghesqui\`ere, 
1974). In addition to pions, $\eta$ mesons are produced with a rate of about 7 \%  (Chiba, 
1987) and kaons with a rate of about 6\% of all annihilations (Sedl\'ak 
and \v{S}im\'ak, 1988). 

In fireball models the pion multiplicity $N = N_+ + N_- + N_0$ follows a Gaussian 
distribution (Orfanidis and Rittenberg, 1973). The pion multiplicity distribution at rest in 
liquid hydrogen is shown in Fig. \ref{mult}. Following the model of  Pais (1960) one 
expects on statistical grounds the branching ratios to be distributed according to 1/($N_+! 
N_-! N_0!$) for a given multiplicity $N$. The open squares show the predictions normalized 
to the measured branching ratios from channels with charged pions and $N_0\leq 1$ 
(Armenteros and French, 1969). The full circles show the data from bubble chambers, 
together with Crystal Barrel results for $N_0>1$. The fit to the data (curve) leads to $\sigma 
= 1$ for a Gauss distribution assuming $\langle N\rangle = 5$. The open circles 
show an estimate from bubble chamber experiments which appears to overestimate the 
contribution from $N$ = 5 (Ghesqui\`ere, 1974). 

\subsection{S- and P-wave annihilation at rest}
Stopping antiprotons in hydrogen are captured  to form  
antiprotonic hydrogen atoms (protonium). The probability of forming a $\pbarp$ 
atom is highest for states with principal quantum number $n \sim 30$ 
corresponding to the binding energy (13.6 eV) of the K-shell 
electron ejected during the capture process. Two competing 
de-excitation mechanisms occur: (i) the cascade to 
lower levels by X-ray or external Auger emission of electrons from 
neighbouring H$_2$-molecules and (ii) Stark mixing between the 
various angular momentum states due to collisions with 
neighbouring H$_2$ molecules. Details on the cascade process can be found in Batty (1989). 
In liquid hydrogen, Stark mixing 
dominates (Day, 1960) and the $\pbarp$ system annihilates with 
the angular momentum $\ell = 0$ from high S levels (S-wave 
annihilation) due to the absence of angular momentum barrier.
The initial states are the spin singlet ($s=0$) $^1S_0$ and the spin 
triplet ($s=1$) $^3S_1$ levels with parity $P=(-1)^{\ell +1}$ and 
$C$-parity $C=(-1)^{\ell +s}$, hence with quantum numbers 
\begin{equation} 
J^{PC}(^1S_0)  =  0^{-+} \ {\rm and} \ J^{PC}(^3S_1) =  1^{--}.
\end{equation}

The cascade, important in low density hydrogen (e.g. in gas),   
populates mainly the $n=2$ level (2P) from which the $\pbarp$ 
atom annihilates due its small size: The K$_{\alpha}$ transition from 
2P to 1S has been observed at LEAR in gaseous hydrogen (Ahmad, 1985; Baker, 1988). 
Compared to annihilation, it is suppressed with a probability of (98 $\pm$ 1) \% at 
atmospheric pressure. Annihilation with relative angular 
momentum $\ell=1$ (P-wave annihilation) can therefore be 
selected by detecting the L X-rays to the 2P levels, in coincidence 
with the annihilation products. This procedure permits the 
spectroscopy of intermediate meson resonances produced from the 
P-states
\begin{eqnarray}
J^{PC}(^1P_1) & = & 1^{+-}, \ \ J^{PC}(^3P_0) = 0^{++}, \nonumber\\
J^{PC}(^3P_1) & = & 1^{++} \ {\rm and} \ J^{PC}(^3P_2) = 2^{++}.
\end{eqnarray}
Annihilation from P-states has led to the discovery of the 
$f_2(1565)$ meson (May, 1989).

The much reduced Stark mixing in low density hydrogen also 
allows annihilation from higher P levels. At 1 bar, 
S- and P-waves each contribute about 50\% to annihilation (Doser, 1988). 
Annihilation from D-waves is negligible due to the very small 
overlap of the $p$ and $\pbar$ wave functions.

The assumption of S-wave dominance in liquid hydrogen is often a 
crucial ingredient to the amplitude analyses when determining the 
spin and parity of an intermediate resonance in the annihilation 
process, since the quantum numbers of the initial state must be 
known. The precise fraction of P-wave annihilation in liquid has 
been the subject of a longstanding controversy. The reaction 
$\pbarp\rightarrow\pi^0\pi^0$ can only proceed through the P-states 
$0^{++}$ or $2^{++}$ (see section \ref{sec:ann}) while 
$\pi^+\pi^-$   also 
proceeds from S-states ($1^{--}$). The annihilation rate $B(\pi^0\pi^0)$ 
for this channel in liquid has been measured earlier by several 
groups but with inconsistent results (Devons, 1971; Adiels, 1987; 
Chiba, 1988). 

Crystal Barrel has determined the branching ratio 
for $\pbarp\rightarrow\pi^0\pi^0$ in liquid 
by measuring the angles and energies of the four 
decay photons (Amsler, 1992a). The main 
difficulties in selecting this channel are annihilation into $3\pi^0$ 
which occurs with a much higher rate and, most importantly,  
P-wave annihilation in flight. The former background source can be 
reduced with the good $\gamma$ detection efficiency and large solid angle
of Crystal 
Barrel, while the latter can be eliminated thanks to the very narrow 
stop distribution  from cooled low-energy antiprotons from LEAR 
(0.5 mm at 200 MeV/c). The small contamination from 
annihilation in flight can easily be subtracted from the 
stopping distribution by measuring the annihilation vertex. The latter
was determined by performing a 5 
constraints (5C) fit to $\pbarp\rightarrow\pi^0\pi^0$, assuming 
energy conservation, two invariant 2$\gamma$-masses consistent 
with $2\pi^0$ and momentum conservation perpendicular to the 
beam axis. The branching ratio for $\pi^0\pi^0$ is
\begin{equation}
B(\pi^0\pi^0) = (6.93 \pm 0.43) \times 10^{-4},
\label{30}
\end{equation}
in  agreement with Devons (1971) but much larger than 
Adiels (1987) and Chiba (1988). From the annihilation rate 
$B(\pi^+\pi^-)_{2P}$ into $\pi^+\pi^-$ from atomic 2P-states (Doser, 
1988) one can, in principle, extract the fraction $f_p$ of $P$-wave 
annihilation in liquid\footnote{In Doser (1988) $f_p$ was 
found to be (8.6 $\pm$ 1.1) \%, 
when using the most precise measurement for $\pi^0\pi^0$ 
available at that time (Adiels, 1987).}:
\begin{equation}
f_p = 2 \frac{B(\pi^0\pi^0)}{B(\pi^+\pi^-)_{2P}} = (28.8 \pm 3.5) 
\%.
\label{1}
\end{equation}
This is a surprisingly large contribution. However, Eq. (\ref{1}) assumes 
that the population of the fine and hyperfine structure states  is the 
same for the 2P as for the higher P levels which is in general  
not true. In liquid, strong Stark mixing constantly repopulates the 
levels. A P-state with large hadronic width, for instance $^3P_0$ 
(Carbonell, 1989), will therefore contribute more to annihilation 
than expected from a pure statistical population. On the other hand, 
in low pressure gas or for states with low principal quantum 
numbers the levels are populated according to their statistical 
weights. The branching ratio for annihilation into a given final state 
is given in terms of the branching ratios $B_i^S$ and $B_i^P$  from 
the two S-, respectively the four P-states (Batty, 1996): 
\begin{eqnarray}
B = [1-f_P(\rho)]\sum_{i=1}^2w_i^S E_i^S(\rho) B_i^S\nonumber\\
 + f_P(\rho)\sum_{i=1}^4w_i^P E_i^P(\rho) B_i^P,
\end{eqnarray}
where $\rho$ is the target density. The purely statistical weights are 
\begin{equation}
w_i^S = \frac{2J_i+1}{4}, \ w_i^P = \frac{2J_i+1}{12}.
\end{equation}
The enhancement factors $E_i$ describe the 
departure from pure statistical population ($E_i$=1). For 
$\pi^0\pi^0$ in liquid one obtains 
\begin{eqnarray}
B(\pi^0\pi^0) = f_P({\rm liq}) 
[\frac{1}{12}E_{^3P_0}({\rm liq})B_{^3P_0}(\pi^0\pi^0)\nonumber\\
 + \frac{5}{12}E_{^3P_2}({\rm liq})B_{^3P_2}(\pi^0\pi^0)],
\end{eqnarray}
and for $\pi^+\pi^-$ from 2P states
\begin{equation}
B(\pi^+\pi^-)_{2P} = 2  [\frac{1}{12}B_{^3P_0}(\pi^0\pi^0) + 
\frac{5}{12}B_{^3P_2}(\pi^0\pi^0)].
\end{equation}
It is obviously not possible to determine $f_P({\rm liq})$ unless the 
enhancement factors are unity (and hence Eq. (\ref{1}) follows). The enhancement 
factors 
have been calculated with an X-ray cascade calculation (Batty, 
1996) using the observed yields of $K$ and $L$ X-rays in 
antiprotonic atoms and the predicted hadronic widths from optical 
potential models of the $\pbarp$ interaction (Carbonell, 1989). For 
example, Batty (1996) finds typically $E_{^3P_0}({\rm liq})\sim 2.3$ and 
$E_{^3P_2}({\rm liq})\sim 1.0$.  The branching ratios $B_i$ and 
$f_P(\rho)$ were then fitted to the measured two-body branching 
ratios for $\pbarp\rightarrow\pi^0\pi^0, \pi^+\pi^-$, $K^+K^-$, 
$K_SK_S$ and $K_SK_L$ at various target densities, with and without 
L X-ray coincidence. The fraction of P-wave annihilation is shown in 
Fig. \ref{fp} as a function of density. In liquid hydrogen one obtains
\begin{equation}
f_P({\rm liq}) = (13 \pm 4) \%,
\end{equation} 
a more realistic value when compared to Eq. (\ref{1}). 

\section{The Crystal Barrel Experiment}
\subsection{Detector}
Figure \ref{Barrel} shows a sketch of the Crystal Barrel detector (Aker, 1992). 
The incoming antiprotons entered a 1.5 T solenoidal magnet 
along its axis and interacted in a  liquid hydrogen target, 44 mm 
long and 17 mm in diameter. A segmented silicon counter in front 
of the target defined the incoming beam. The final state charge multiplicity was 
determined online with two cylindrical proportional wire chambers (PWC). The 
charged particle momentum was measured by a jet drift chamber 
(JDC) which also provided $\pi$/K separation below 500 MeV/c by 
ionization sampling.

Photons were detected in a barrel-shaped assembly of 1,380 CsI(Tl) 
crystals, 16.1 radiation lengths long (30 cm), with photodiode 
readout. The crystals were oriented towards 
the interaction point and covered a solid angle of 0.97 $\times 4\pi$. 
Each crystal, wrapped in teflon and aluminized mylar, was  
enclosed in a 100 $\mu$m thick titanium container. The light 
(peaking at 550 nm) was collected at the rear end by a wavelength 
shifter and the re-emitted light was detected by a photodiode glued on 
the edge of the wavelength shifter. With the low electronic noise of 
typically 220 keV the energy resolution was
\begin{equation} 
\frac{\sigma}{E} = \frac{0.025}{E [{\rm GeV}]^{\frac{1}{4}}}
\end{equation}
and photons could be detected efficiently down to 4 MeV. The 
angular resolution was typically $\sigma$ = 20 mrad for both polar 
and azimuthal angles. The mass resolution was $\sigma$ = 10 MeV 
for $\pi^0$ and 17 MeV for $\eta\rightarrow 2\gamma$.

A rough calibration of the electromagnetic calorimeter was first obtained 
with traversing minimum ionizing pions which deposit 170 MeV in 
the crystals. The final calibration was achieved with 0-prong events 
using $2\gamma$ invariant masses from $\pi^0$ decays. An 
energy dependent correction was applied to take shower leakage at 
the rear end of the crystals into account. The stability of the 
calibration was monitored with a light pulser system.

The JDC had 30 sectors, each with 23 sense wires at radial 
distances between 63 mm and 239 mm) read out on both ends by 
100 MHz flash ADC's. The position resolution in the plane transverse 
to the beam axis ($r\phi$ coordinates) was $\sigma$ = 125 
$\mu$m using slow gas, a 90:10 \% CO$_{2}$/isobutane mixture. The 
coordinate $z$ along the wire was determined by charge division 
with a resolution of $\sigma$ = 
8 mm. This led to a momentum resolution for pions of 
$\sigma/p \simeq $2 \% at 200 MeV/c, rising to $\simeq$ 7\% at 1 
GeV/c for those tracks that traversed all JDC layers. 

The $z$ coordinates were calibrated by fitting straight 
tracks from 4-prong events without magnetic field. The momentum calibration was 
performed with monoenergetic pions and 
kaons from the two-body final states $\pi^+\pi^-$ and $K^+K^-$.
Pressure and temperature dependent drift time tables were 
generated and fitted to the measured momentum distribution.

In 1994 the JDC was replaced by a new jet drift chamber with  
only 15 sectors for the 6 innermost layers. In 1995 the PWC's were 
also replaced by a microstrip vertex detector 
(SVX) consisting of 15 single-sided silicon detectors 
arranged in a windmill configuration at a radial distance of 13 mm 
around the target (Fig. \ref{SVX}). Each detector had 128 strips 
with a pitch of 50 $\mu$m running parallel to the beam axis. The 
increase of charge multiplicity between the  
SVX and the inner layers of the JDC permitted to trigger on
$K_S\rightarrow\pi^+\pi^-$. The SVX  also provided an improved 
vertex resolution in $r\phi$ and a better momentum resolution.  

For annihilation at rest in liquid hydrogen the $\pbar$ incident 
momentum was 200 MeV/c  with typically $10^4$ incident 
$\pbar$/s to minimize pile-up in the crystals. For annihilation in 
gaseous hydrogen the liquid target was replaced by a hydrogen 
flask at 13 bar. The incident momentum was 105 MeV/c. Since the 
annihilation rate was higher than the maximum 
possible data acquisition speed, a multilevel trigger could be used.
The two PWC's and the inner layers (2 - 5)  of the JDC determined 
the charged multiplicity of the final state.  Events with long tracks 
could be selected for optimum momentum resolution by counting 
the charged multiplicity in the outer layers (20 and 21) of the JDC. 
A hardwired processor  determined the cluster multiplicity in the 
barrel. A  processor then fetched the digitized energy deposits in the barrel, 
computed all two-photon invariant masses thus providing a trigger 
on the $\pi^{0}$ or $\eta$ multiplicity (Urner, 1995).

\subsection{Photon reconstruction}
We now briefly describe the photon reconstruction which is 
particularly relevant to the results reviewed in this article.
Photon induced electromagnetic showers spread
out over several crystals. The size of a cluster depends on the 
photon energy and varies from 1 to about 20 crystals.
The reconstruction of photons is done by searching for clusters of
neighbouring crystals with energy deposits of at least 1 MeV. The 
threshold for cluster identification (typically between 4 and 20 
MeV) depends on the annihilation channel being studied. Local 
maxima with a predefined threshold (typically between 10 and 20 
MeV) are then searched for within clusters. When only one local 
maximum is found, the photon energy is defined as the cluster 
energy and the direction is given by the center of gravity of the 
crystals, weighted by their energies. When $n$ local maxima are 
found within a cluster,  the latter is assumed to contain showers 
from $n$ photons. In this case the cluster energy $E_C$
is shared between the $n$ subclusters of nine crystals with energies $E_i$ 
around the local maxima. Hence the photon energies are 
given by  
\begin{equation}
E_{\gamma,i}=\frac{E_i}{\sum_{j=1}^{n} E_j} E_C.
\end{equation}

Additional clusters mocking photons are due to shower fluctuations 
which may develop small but well separated satellites in the 
vicinity of the main shower. These ``split-offs" can be removed by 
requiring a minimum separation between the showers. However, 
this cut may reduce the detection efficiency for high energy $\pi^0$'s since 
photons from $\pi^0$ decay cluster around the minimum opening 
angle. The opening angle between two photons with energies $E_1 
\leq E_2$ from $\pi^0$ decay is given by
\begin{equation}
\cos\phi = 1 - \frac{(1+R)^2}{2\gamma^2 R} \ {\rm with} \ R = 
\frac{E_1}{E_2},
\label{2}
\end{equation}
where $\gamma$ = $E_{\pi^0}/m_{\pi^0}$. Hence for all pairs of 
neighbouring clusters one calculates $R$ and removes the low energy clusters whenever 
cos$\phi$ is larger than given by Eq. (\ref{2}), assuming the maximum possible 
value of $\gamma$ in the annihilation channel under consideration 
(Pietra, 1996).

Clusters generated by ionizing particles can be removed by 
matching the impact points extrapolated from the reconstructed 
tracks in the JDC. However, split-offs from charged particles are 
more cumbersome to eliminate. They are initiated, for example,  
from neutrons which travel long distances before being absorbed.  
These split-offs can   
be suppressed by requiring momentum and energy conservation in 
the annihilation process (kinematic fits). 

\subsection{Available data}
The bulk of the Crystal Barrel data consists in $\pbarp$ annihilation 
at rest and in flight in liquid hydrogen. As discussed above, 
annihilation from initial P-states is enhanced when using a gaseous 
target. Annihilation in deuterium at rest allows the formation of 
$\NNbar$ bound states below 2$m_N$, the  
spectator neutron (or proton) removing the excess energy (for a review on baryonium 
states, see Amsler (1987)). With a spectator 
proton one gains access to $\pbarn$ annihilation, a pure isospin $I=1$ initial state.  

The data collected by Crystal Barrel are shown in Table 
\ref{data}. Data were taken in liquid hydrogen, gaseous hydrogen 
(13 bar) and in liquid deuterium  with a minimum bias trigger 
(requiring only an incident antiproton) or with the multiplicity 
trigger requiring 0-prong or $n$-prong with long tracks in the JDC. In addition, data were 
collected with specialized triggers enhancing specific final states. 
As a comparison, the largest earlier sample 
of annihilations at rest in liquid was obtained by the CERN-Coll\`ege 
de France collaboration with about 100,000 pionic events and 80,000 events containing 
at least one $K_S\rightarrow\pi^+\pi^-$ (Armenteros and French, 1969). The Asterix 
collaboration  
collected some $10^7$ pionic events in 
gaseous hydrogen at 1 bar (for a review and references see 
Amsler and Myhrer (1991)). The total number of annihilations at rest 
in liquid hydrogen collected by Crystal Barrel is $10^8$. The triggered 0-prong sample 
alone corresponds to 6.3$\times 10^8$ annihilations.  

\section{Annihilation into Two Mesons}
\label{seq:anni}
Consider a pair $M\overline{M}$ of charge conjugated mesons in the eigenstate of isospin 
$I$. The $P$-, $C$- and $G$-parities are: 
\begin{eqnarray}
P(M\overline{M}) & = & (-1)^L,\\
C(M\overline{M}) & = & (-1)^{L+S}, 
\label{6}\\
G(M\overline{M})  & = & (-1)^{L+S+I},
\label{8}
\end{eqnarray}
where $L$ is the relative angular momentum and $S$ the total spin.  For the $\pbarp$ 
system with angular momentum $\ell$ and  spin $s$ one has
\begin{eqnarray}
P(\pbarp) & = &  (-1)^{\ell + 1}, 
\label{5}\\
C(\pbarp) & = & (-1)^{\ell + s},
\label{80}\\
G(\pbarp) & = & (-1)^{\ell + s + I}.
\label{79}
\end{eqnarray}
For annihilation into two mesons the two sets of equations relate the quantum numbers of 
the initial state to those of the final state since $P$,  $C$, $G$ and $I$ are conserved. In 
addition, $L$, $S$, $\ell$ and $s$ must be chosen so that the total angular momentum $J$
is conserved: 
\begin{equation}
|L-S| \leq J \leq L+S, \ \ |\ell-s| \leq J \leq \ell + s.
\label{7}
\end{equation}
Since $P$, $C$ and $G$ are multiplicative quantum numbers these relations are especially 
restrictive for mesons that are eigenstates of $C$ and $G$, e.g. for neutral non-strange 
mesons. For example, for two identical neutral non-strange pseudoscalars (e.g. 
$\pi^0\pi^0$) with $S=0$, $C=+1$,  Eq. (\ref{6}) implies that $L$ is even and then Eq. 
(\ref{5}) requires $\ell$ to be odd (annihilation from P-states only). Equation (\ref{8}) 
further requires with $G=+1$ that $I=0$ 
and hence with Eq. (\ref{79}) annihilation from the ($I=0$) $0^{++}$ or $2^{++}$ atomic 
states.

For a pair of non-identical neutral pseudoscalars (e.g. $\pi^0\eta$) $L$ may be odd and 
hence the possible quantum numbers are $0^{++}$, $1^{-+}$, $2^{++}$, $3^{-+}$, etc. 
However, $1^{-+}$ and $3^{-+}$ do not couple to $\pbarp$ since Eqs. (\ref{5}) and 
(\ref{80}) require $\ell$ even and $s=0$ and hence $J$ even. In fact these ``exotic" 
quantum numbers do not couple to any fermion-antifermion pair and are, in particular,  
excluded for $\qqbar$ mesons.  

\subsection{Annihilation into two neutral mesons}
\label{sec:ann}
Crystal Barrel has measured the branching ratios for $\pbarp$ 
annihilation into two neutral light mesons 
from about $10^7$ annihilations into 0-prong (Amsler, 1993b).
These data have been collected by vetoing 
charged particles with the PWC's and the internal layers of the JDC. 
The lowest $\gamma$-multiplicity was four (e.g. 
$\pi^0\pi^0, \pi^0\eta$) and the highest nine 
(e.g. $\eta\omega$, with $\eta\rightarrow 3\pi^0$ and 
$\omega\rightarrow\pi^0\gamma$). To control systematic errors in 
the detection efficiency, some of these branching ratios  
have been determined from different final state multiplicities.  
For example, $\eta$ decays to $2\gamma$ and $3\pi^0$ and hence 
$\eta\eta$ is accessible from $4\gamma$ and $8\gamma$ events.

Figure \ref{gaga} shows a scatterplot of $2\gamma$-invariant masses 
for events with $4\gamma$. Events have been selected by 
requiring four clusters in the barrel and applying momentum and 
energy conservation (4C fit).  Signals from $\pi^0\pi^0$, 
$\pi^0\eta$, $\eta\eta$ and even $\pi^0\etap$ are clearly visible. 
The dark diagonal band at the edge is due to wrong combinations. The 
detection and reconstruction efficiency was typically 
40\% for 4$\gamma$ events, obtained by Monte Carlo simulation 
with GEANT. As discussed  
above, two neutral pseudoscalars couple only to atomic P-states and 
are therefore suppressed in liquid hydrogen. On the 
other hand, the channels $\pi^0\omega$ and $\eta\omega$  
couple to $^3S_1$ and hence have a larger branching ratio. In 
spite of the good detection efficiency of the detector one therefore 
observes the background signals from $\pi^0\omega$ and 
$\eta\omega$, where $\omega$ decays to $\pi^0\gamma$ with a 
missing (undetected) photon.

Figure \ref{ninega} shows the $\pi^0\gamma$ momentum distribution for 
$\pbarp\rightarrow 4\pi^0\gamma$ events (8C fit requiring $4\pi^0$). 
The peak at 657 MeV/c is due to 
the channel $\pbarp\rightarrow\eta\omega$. For these $9\gamma$ 
events the detection efficiency was 10\%.

The branching ratios are given in Table 
\ref{BR}. They are always  corrected for the 
unobserved (but known) decay modes of the final state mesons (Barnett, 1996).  
For Crystal Barrel data 
the absolute normalization was provided by comparison with 
$\pi^0\pi^0$ which has been measured with minimum bias data (Eq. (\ref{30})). 

Signals for $\omega\pi^0$ and $\omega\eta $ have also been 
observed for $\omega$ decaying to $\pi^+\pi^-\pi^0$, leading to 
$\pi^+\pi^-4\gamma$ (Schmid, 1991). Figure \ref{Schmid} shows the 
$\pi^+\pi^-\pi^0$ invariant mass spectrum for  $\pi^+\pi^-\pi^0\eta$ 
events. The branching ratio for $\omega\pi^0$ and $\omega\eta $ 
are in excellent agreement with the ones from 0-prong (Table \ref{BR}). 

The angular distribution in the $\omega$ rest frame contains information on the initial 
atomic state. The distribution of the angle between the normal to 
the plane spanned by the three pions 
and the direction of the recoiling $\eta$ is plotted in the inset of Fig. 
\ref{Schmid}. Using the method described in section \ref{sec:spa} one 
predicts the distribution sin$^2\theta$ for annihilation from $^3S_1$ while the 
distribution should be isotropic for annihilation from $^1P_1$. The fit (curve) 
allows (12 $\pm$ 4)\% P-wave. Figure \ref{Strassb} shows the angular distribution of the 
$\gamma$ in the $\omega$ rest frame for $\omega\eta 
(\omega\rightarrow\pi^0\gamma)$. The 
predicted distribution is ($1+\cos^2\theta$) from $^3S_1$ and is again isotropic for $^1P_1$. 
The fit (curve) allows (9 $\pm$ 3)\% P-wave. However, these results 
assume that the relative angular momentum between $\eta$ and $\omega$  is  
$L=0$ from $^1P_1$, thus neglecting $L=2$. Without this assumption, the fraction 
of P-wave cannot be determined from the angular distributions due to the unknown 
interference between the $L=0$ and $L=2$ amplitudes.

Some of the branching ratios for two-neutral mesons have been 
measured earlier (Adiels, 1989; Chiba, 1988) by detecting and 
reconstructing $\pi^0$'s or $\eta$'s with small solid angle detectors 
and observing peaks in the $\pi^0$  or $\eta$ inclusive momentum 
spectra. Since the branching ratios are small, these early data are 
often statistically weak or subject to uncertainties in the baseline 
subtraction from the inclusive spectra. In fact most of the Crystal Barrel 
results disagree with these measurements which should not 
be used anymore. Table \ref{BR} therefore updates Table 1 in 
Amsler and Myhrer (1991). 

\subsection{The annihilation mechanism}
There is currently no model which completely and satisfactorily 
describes the measured two-body branching ratios listed in Table 
\ref{BR} (for a review of annihilation models and references, see Amsler and 
Myhrer (1991). Since the proton and the antiproton wavefunctions 
overlap one 
expects quarks to play an important role in the annihilation 
dynamics. For instance, $\pbarp$ annihilation into two mesons can 
be described by the annihilation of two $\qqbar$ pairs and the 
creation of a new pair (annihilation graph A) or by the 
annihilation of one $\qqbar$ pair and the rearrangement of the 
other two pairs (rearrangement graph R), see Fig.~\ref{AR}. At 
low energies there is, however, no consensus as to which 
operator should be used to describe the emission and absorption 
of gluons.  In a first approach one assumes that only the 
flavor flow between initial and final states is important 
(Genz, 1983; Hartmann, 1988). The Quark Line Rule (QLR) states that 
annihilation into  $\uubar$ and $\ddbar$  is excluded if 
the A graph dominates, while annihilation into two $\ddbar$ 
mesons is forbidden if R dominates (see Fig. \ref{AR}). 
The OZI-rule (Okubo, 1963) is a special case of the QLR: Annihilation into one or 
more $\ssbar$ mesons is forbidden. We shall confront these simple 
rules below with Crystal Barrel data.

Another approach, which we shall use, is the nearest 
threshold dominance model which describes  reasonably well the 
observed final state multiplicity as a function of $\pbar$ 
momentum (Vandermeulen, 1988).  The branching ratio for 
annihilation into two mesons with masses $m_a$ und $m_b$ is given by
\begin{equation}
W = p C_0 C_{ab}\exp(-A\sqrt{s-(m_a + m_b)^2}),
\label{10}
\end{equation}
where $p$ is the meson momentum in the $\pbarp$ center of mass 
system with total energy $\sqrt{s}$, $C_0$ a normalization constant  
and $C_{ab}$  a multiplicity 
factor depending on spin and isospin. The constant $A$= 1.2 GeV$^{-1}$
has been fitted to the cross section for $\pbarp$ annihilation 
into $\pi^+\pi^-$ as a function of $\pbar$ momentum. For 
annihilation into kaons the fit to kaonic channels requires the additional normalization 
factor $C_1/C_0$ = 0.15. Thus annihilation into the heaviest possible meson pair is 
enhanced 
with respect to phase space $p$ by the exponential form factor in 
Eq. (\ref{10}). This is natural in the framework of baryon exchange 
models which prefer small momentum transfers at the 
baryon-meson vertices. 

In more refined models the branching 
ratios for annihilation at rest depend on the atomic wave function distorted by strong 
interaction at short distances (Carbonell, 1989). Predictions for the 
branching ratios therefore depend on models for 
the meson exchange potential which are uncertain below 1 fm. Also,  
the  quark description has to be complemented by baryon and 
meson exchanges to take the finite size of the emitted mesons into account. 

In the absence of strong interaction the $\pbarp$ atomic state is an  
equal superposition of isospin $I=0$ and 1 states. Naively one 
would therefore expect half the protonium states to annihilate 
into a final state of given isospin. However, $\pbarp$ to $\nbarn$ 
transitions occuring at short distances may modify the population 
of $I=0$ and $I=1$ states and therefore enhance or reduce the 
annihilation rate to a  final state of given isospin (Klempt, 1990; 
Jaenicke, 1991). Nonetheless, one expects that predictions for 
ratios of branching ratios for channels with the same isospin and 
proceeding from the same atomic states are less sensitive to model 
dependence. We shall therefore compare predictions from the QLR with ratios of 
branching ratios from Table \ref{BR}. 

The flavor content of the $\eta$ and $\etap$ mesons is given by
\begin{eqnarray}
|\eta\rangle & = & \frac{1}{\sqrt{2}}(|\uubar\rangle + 
|\ddbar\rangle)\sin(\theta_i -\theta_p) - 
|\ssbar\rangle \cos(\theta_i -\theta_p),\nonumber\\
|\etap\rangle & = & \frac{1}{\sqrt{2}}(|\uubar\rangle + 
|\ddbar\rangle)\cos(\theta_i -\theta_p) +
|\ssbar\rangle \sin(\theta_i -\theta_p),
\label{31}
\end{eqnarray}
where $\theta_i = 35.3^{\circ}$ is the ideal mixing angle. The 
flavor wave functions of the $\pi^0$ and $\rho^0$ are
\begin{equation}
|\pi^0\rangle, |\rho^0\rangle = \frac{1}{\sqrt{2}}(|\ddbar\rangle - 
|\uubar\rangle),
\label{32}
\end{equation}
and those of $\omega$ and $\phi$, assuming ideal mixing in the 
vector nonet,
\begin{equation}
|\omega\rangle = \frac{1}{\sqrt{2}}(|\uubar\rangle + 
|\ddbar\rangle), \  \  |\phi\rangle = -|\ssbar\rangle.
\label{33}
\end{equation}
The branching ratio for annihilation into two neutral mesons is then
given by $B = \tilde{B}\cdot W$  with 
\begin{equation}
\tilde{B} = |\langle\pbarp|T|M_1M_2\rangle|^2 = |\sum_{i,j} 
T([q_i\overline{q_i}]_1, [q_j\overline{q_j}]_2)\langle q_i\overline{q_i}| 
M_1\rangle\langle q_j\overline{q_j}|M_2\rangle|^2,
\label{34}
\end{equation}
and \footnote{The theoretical prediction $\tilde{B}$ has to be multiplied by two for a pair 
of non-identical mesons.} $q_i=u$ or $d$ (Genz, 1985). In the 
absence of $\ssbar$ pairs in the nucleon, the QLR forbids the 
production of $\ssbar$ mesons and therefore the $\ssbar$ 
components in $M_1$ and $M_2$ can be ignored in Eq. (\ref{34}). 
The predicted ratios of branching ratios are given by the first four rows in Table 
\ref{QLR} for various channels from the same atomic states. They depend only on the 
pseudoscalar mixing angle $\theta_p$. To extract $\theta_p$ the measured branching 
ratios from Table \ref{BR} must be first divided by $W$ (Eq. (\ref{10})), ignoring $C_0$ 
and $C_{ab}$ which cancel in the ratio.

The pseudoscalar mixing angle has been measured in various 
meson decays (e.g.  $\eta$ and $\etap$ radiative  decays, $J/\psi$ 
radiative decays to $\eta$ and $\etap$) and is known to be close to 
-20$^{\circ}$ (Gilman and Kauffmann, 1987). The agreement with 
our simple model of annihilation is amazing (third column of Table 
\ref{QLR}). We emphasize that the predictions in the upper four rows of 
Table \ref{QLR} are valid independently of the relative 
contributions from the A and R graphs. 

Conversely, one can assume the validity of the model and extract 
from the first four rows in Table \ref{QLR} the average
\begin{equation}
\theta_p = (-19.4 \pm 0.9)^{\circ}.
\label{81}
\end{equation}
Leaving the constant $A$ in Eq. (\ref{10}) as a free fit parameter one obtains $\theta_p = 
(-17.3 \pm 1.8)^{\circ}$ from early Crystal Barrel data (Amsler, 1992b). Assuming now 
dominance of the planar graph A, the amplitudes  $T([\uubar]_1,[\ddbar]_2)$ and 
$T([\ddbar]_1,$ $[\uubar]_2)$ vanish and one obtains the 
predictions in the lower part of Table \ref{QLR}. The measurements lead in general to 
incorrect values for $\theta_p$, presumably due to the contribution of the R graph. Also, 
for 
$\rho^0\rho^0$ and $\omega\omega$ one expects 
from A dominance
\begin{equation}
\tilde{B}(\rho^0\rho^0) = \tilde{B}(\omega\omega),
\end{equation}
in violent disagreement with data (Table \ref{BR}).
On the other hand, if R dominates the amplitude  
$T([\ddbar]_1,[\ddbar]_2)$ vanishes and one predicts from Eq. (\ref{34}) 
with $a^2\equiv\sin^2(\theta_i-\theta_p)$ the inequality (Genz, 1990)
\begin{equation}
|a^2\sqrt{2\tilde{B}(\pi^0\pi^0)}-\sqrt{2\tilde{B}(\eta\eta)}|^2 \leq
4a^2\tilde{B}(\pi^0\eta) \leq 
|a^2\sqrt{2\tilde{B}(\pi^0\pi^0)}+\sqrt{2\tilde{B}(\eta\eta)}|^2, 
\end{equation}
which is fulfilled by data. 

There is, however, a caveat: the predictions (\ref{34}) have been compared to the 
measured branching ratios corrected by  $W$. As pointed out earlier,  Eq. (\ref{10}) 
provides a good fit to the mutiplicity  distribution in low energy $\pbarp$ annihilation as 
a function of $\pbar$ momentum. Other correcting factors can, however, be found in the 
literature. In section \ref{sec:spectro} we shall use the phase space factor
\begin{equation}
W = p F_L^2(p)
\label{39}
\end{equation}
where $F_L(p)$ is the Blatt-Weisskopf damping factor which suppresses high angular 
momenta $L$ for small $p$. This factor is determined by the range of the 
interaction, usually chosen as 1 fm  ($p_R$ =197 MeV/c). Convenient expressions for 
$F_L(p)$ 
are given in Table \ref{Blatt}. For $p$ 
much larger than $p_R$, $F_L(p)\simeq 1$ and for $p$ much 
smaller than $p_R$ 
\begin{equation}
F_L(p) \simeq p^L. 
\end{equation}
This last prescription provides a reasonable agreement when comparing the measured 
decay branching ratios of mesons, especially tensors, with predictions from SU(3), as we 
shall discuss in section \ref{sec:new}. These alternative phase space factors may also be 
used to determine the pseudoscalar mixing angle. However, they do not lead to consistent 
values for $\theta_p$ (Amsler, 1992b). Agreement is achieved with prescription (\ref{10}), 
which we shall also employ in the next section.

In conclusion, the naive quark model assuming only the QLR and 
two-body threshold dominance reproduces the correct pseudoscalar 
mixing angle from the measured two-meson final states. This is a clear indication for 
quark dynamics in the annihilation process. The 
relative contribution from R and A cannot be extracted but the 
non-planar graph R must contribute substantially to the annihilation process.

\section{Electromagnetic Processes}
\label{sec:em}
\subsection{Radiative annihilation}
For annihilations leading to  
direct photons the  isospin $I=0$ and 1 amplitudes from the 
same $\pbarp$ atomic state interfere since isospin is not conserved in electromagnetic 
processes. Radiative annihilation  $\pbarp\rightarrow\gamma X$, where $X$ stands for 
any neutral meson, involves the annihilation of a 
$\qqbar$ pair into a photon. The branching ratios can be calculated 
from the Vector Dominance Model (VDM) which relates  
$\gamma$ emission  to the emission of $\rho$, $\omega$ and 
$\phi$ mesons (Delcourt, 1984). Assuming ideal mixing in the 
vector nonet one may actually neglect $\phi$ production which is forbidden 
by the QLR. The amplitude for $\pbarp\rightarrow \gamma X$ is 
then given by the coherent sum of the two $I=0$ and $I=1$ 
amplitudes  with unknown relative phase 
$\beta$ (fig. \ref{Isospin}). According to VDM, the $\gamma\rho$ coupling 
$g_{\rho\gamma}$ is a factor 
of three stronger than the $\gamma\omega$ coupling. The 
branching ratio is then\footnote{For two identical particles, e.g. 
$\rho^0\rho^0$ or $\omega\omega$, the measured branching ratios divided by phase 
space, $\tilde{B}$, have to be 
multiplied by two.}
\begin{equation}
\tilde{B}(\gamma X) = A^2 [\tilde{B}(\rho X) + 
\frac{1}{9}\tilde{B}(\omega X) + \frac{2}{3}\sqrt{\tilde{B}(\rho 
X)\tilde{B}(\omega X)}\cos\beta],
\label{35}
\end{equation}
with
\begin{equation}
A = \frac{eg_{\rho\gamma}}{m_\rho^2} = 0.055.
\end{equation}
Equation (\ref{35}) provides lower and upper limits (cos$\beta = \pm$ 1) for branching 
ratios.

Radiative annihilation has not been observed so far with the 
exception of $\pi^0\gamma$ (Adiels, 1987).
Crystal Barrel has measured the rates for 
$\pi^0\gamma$, $\eta\gamma$, $\omega\gamma$ and has 
obtained upper limits for $\etap\gamma$ and $\gamma\gamma$ 
(Amsler, 1993c). Annihilation into $\phi\gamma$ (Amsler, 1995c) is treated in section 
\ref{sec:phi}. 

The starting data sample consisted of $4.5\times 10^6$ 0-prong 
events. Figure \ref{pi0gam} shows a typical $\pi^0\gamma$ event 
leading to three detected photons. The main background to $\pi^0\gamma$ stems 
from annihilations into $\pi^0\pi^0$ for which one of the photons 
from $\pi^0$ decay has not been detected, mainly because its energy 
lies below detection threshold (10 MeV). 
The $\pi^0$ momentum for $\pi^0\gamma$ is slightly higher (5 
MeV/c)  than for $\pi^0\pi^0$. The small downward shift of the 
$\pi^0$ momentum peak due to the $\pi^0\pi^0$ contamination 
could be observed thanks to the good energy resolution of the 
detector and could be used to estimate the feedthrough from 
$\pi^0\pi^0$:  (29 $\pm$ 8) \%, in agreement with  Monte Carlo 
simulations. The result for the $\pi^0\gamma$ branching ratio is 
given in Table \ref{BRG}. It disagrees with the one obtained 
earlier from the $\pi^0$ inclusive momentum spectrum: (1.74 $\pm 0.22)\times 10^{-5}$ 
(Adiels, 1987).  

Results for $\eta\gamma, \omega\gamma$ and $\etap\gamma$ are also given in Table 
\ref{BRG}. The $\omega$ was detected in its 
$\pi^0\gamma$ and the $\etap$ searched for in its $2\gamma$ decay mode. The 
main contaminants were $\eta\pi^0$, $\omega\pi^0$ and 
$\etap\pi^0$, respectively, with one photon escaping detection. For 
$\pbarp\rightarrow\gamma\gamma$, 98 $\pm$ 10 events were observed of which 70 
$\pm$ 8 were expected feedthrough from $\pi^0\gamma $ and 
$\pi^0\pi^0$. This corresponds to a branching ratio of (3.3 $\pm 
1.5) \times 10^{-7}$ which the collaboration prefers to quote as an 
upper limit (Table \ref{BRG}).

The branching ratios, divided by $W$ (Eq. (\ref{10})), are 
compared in Table \ref{BRG} with the range allowed by Eq. 
(\ref{35}). Apart from $\phi\gamma$ to which we shall return later, the results agree 
with predictions from VDM. For 
$\pi^0\gamma$ and $\eta\gamma$ (from $^3S_1$) the isospin 
amplitudes interfere destructively (cos$\beta\sim$ -0.3) while for 
$\omega\gamma$ (from $^1S_0$) they interfere constructively 
(cos$\beta\sim$ 0.13). 
We emphasize that these conclusions depend 
on the prescription for the phase space correction. With a phase 
space factor of the form $p^3$ one finds strongly destructive amplitudes (Amsler, 1993c), 
see also Locher  (1994)  and Markushin (1997). 

No prediction can be made from VDM for $\gamma\gamma$  due to 
the contribution of three amplitudes with unknown relative phases: $\rho^0\omega$ from 
$I=1$, $\rho^0\rho^0$ and $\omega\omega$ from $I=0$. Also, the branching ratio for 
$\rho^0\rho^0$ is poorly known (Table \ref{BR}).

\subsection{Search for light gauge bosons in pseudoscalar meson decays}
Extensions of the standard model allow additional gauge bosons, some of which could be 
light enough to be produced in the decay of pseudoscalar mesons (Dobroliubov and 
Ignatiev, 1988). Radiative decays $\pi^0, \eta, \etap\rightarrow\gamma X$ are particularly 
suitable since they are only sensitive to gauge bosons $X$ with quantum numbers $J^P=1^-
$.   Branching ratios are predicted to lie  in the range $10^{-7}$ to $10^{-3}$ (Dobroliubov, 
1990). Experimental upper limits for $\pi^0\rightarrow\gamma X$ are of the order  
$5\times 10^{-4}$ for long lived gauge bosons with lifetime $\tau > 10^{-7}$ s (Atiya, 1992).
Short lived gauge bosons decaying subsequently to $e^+e^-$ are not observed with an 
upper limit of $4\times 10^{-6}$ (Meijer Drees, 1992). 

Crystal Barrel has searched for radiative decays
where $X$ is a long lived weakly interacting gauge boson 
escaping from the detector without 
interaction, or decaying to $\nu\overline{\nu}$. The search was performed using the 
reactions $\pbarp\rightarrow 3\pi^0$, $\pi^0\pi^0\eta$ and $\pi^0\pi^0\etap$ at rest 
(Amsler, 1994a, 1996a) which 
occur with a sufficiently high probability (see Table \ref{TBBR} below) and are 
kinematically 
well 
constrained. Events with five photons were selected from a sample of 15 million 
annihilations into neutral final states (18 million for 
$\etap$ decays). Since 
the branching ratio for 0-prong annihilation is about 4\%, the  
data sample corresponds to some 400 million $\pbarp$ 
annihilations in liquid hydrogen.

Events consistent with $\pi^0\pi^0$ decays and a single 
(unpaired) $\gamma$ were then 
selected by requiring energy and momentum conservation for 
$\pbarp$ annihilation into $\pi^0\pi^0\pi^0$, $\pi^0\pi^0\eta$ or $\pi^0\pi^0\etap$
with a missing $\pi^0$, $\eta$ or $\etap$. Thus 3C kinematic fits 
were applied, ignoring the remaining fifth photon. The measured energy of the latter was 
then transformed into the rest frame of the missing pseudoscalar. 
In this frame a missing state $X$ with mass $m_X$ would produce  
a peak in the $\gamma$-energy distribution at 
\begin{equation}
E_{\gamma}^* = \frac{m}{2}\left(1 - \frac{m_X^2}{m^2}\right),
\end{equation}
with width 
determined by the experimental resolution,  where $m$ is the mass of the missing 
pseudoscalar. 
Thus, if $X$ is simply a missing (undetected) $\gamma$ from 
$\pi^0, \eta$ or $\etap$ decay, one finds with $m_X = 
0$ that $E_{\gamma}^* = m/2$, as expected. 

The main source of background is annihilation into three pseudoscalars for which one  
of the photons escaped detection. This occurs for (i)  photons with energies below 
detection 
threshold ($E< 20$ 
MeV) or (ii)  for photons emitted into insensitive areas of the detector. The latter 
background can be reduced by rejecting events for which the missing $\gamma$ could 
have been emitted e.g. in the holes along the beam pipe. The high efficiency and large 
angular coverage of the Crystal Barrel are therefore crucial in this analysis.
For the $3\pi^0$ channel an important background also arose from 
$\pbarp\rightarrow\pi^0\omega$ with $\omega\rightarrow\pi^0\gamma$, leading to 
5$\gamma$. This background could be reduced by rejecting events which satisfy the 
$\pi^0\omega(\rightarrow\pi^0\gamma)$ kinematics. For $\pi^0\pi^0\eta$, the 
background channel $K_S(\rightarrow 2\pi^0)K_L$ with an interacting $K_L$ faking a 
missing $\eta$ could  be eliminated with  
appropriate kinematic cuts. 

The $E_{\gamma}^*$-energy distribution for $\pi^0$ decay 
is shown in Fig. ~\ref{pi0X}. The broad peak around 70 
MeV is due to $\pi^0$ decays into $2\gamma$ where one 
$\gamma$ has escaped detection, or from residual $\pi^0\omega$ 
events. The fit to the distribution (full line) agrees with the 
simulated rate of  background from $3\pi^0$ and $\pi^0\omega$. The dotted 
line shows the expected signal for a state with mass $m_X=120$ MeV, produced in $\pi^0$ 
decay with a branching ratio of $5\times 10^{-4}$. The corresponding distributions for 
$\eta$ and $\etap$ decays can be found in Amsler (1996a).

Upper limits for radiative decays are given in Fig.~\ref{upplim} as a 
function of  $m_X$. The upper limit for $\pi^0$ decay is an order  of magnitude lower than 
from previous experiments (Atiya, 1992). For $\eta$ and $\etap$ decays no limits were 
available previously. Light gauge bosons are therefore not observed in radiative 
pseudoscalar decays at a level of $10^{-4}$ to $10^{-5}$. 

\subsection{$\eta\rightarrow 3\pi$}
The $3\pi$ decay of the $\eta$ plays an important role in testing low energy QCD 
predictions. This isospin breaking decay is mainly due to the mass difference between $u$ 
and $d$ quarks. Crystal Barrel has measured the relative branching ratios for 
$\eta\rightarrow\pi^+\pi^-\pi^0 $, $\eta\rightarrow 3\pi^0 $ and $\eta\rightarrow 
2\gamma$ from  samples of annihilations into $2\pi^+2\pi^-\pi^0$, $\pi^+\pi^-3\pi^0$ 
and $\pi^+\pi^-2\gamma$, respectively (Amsler, 1995d). The ratios of partial widths are
\begin{eqnarray}
r_1 & \equiv  & \frac{\Gamma(\eta\rightarrow 3\pi^0)}{\Gamma(\eta\rightarrow 
\pi^+\pi^-
\pi^0)} = 1.44 \pm 0.13,\nonumber\\
r_2 & \equiv & \frac{\Gamma(\eta\rightarrow 2\gamma)}{\Gamma(\eta\rightarrow 
\pi^+\pi^-\pi^0)} = 1.78 \pm 0.16.
\end{eqnarray}
The result for $r_1$ is in good agreement with chiral perturbation theory: 1.43 $\pm$ 0.03 
(Gasser and Leutwyler, 1985) and 1.40 $\pm$ 0.03 when taking unitarity corrections  into 
account 
(Kambor, 1996). With the known $2\gamma$ partial width (Barnett, 1996) one 
can calculate from $r_2$ the  partial width $\Gamma(\pi^+\pi^-\pi^0)$ = 
258 $\pm$ 32 eV, in accord with chiral perturbation theory ($\Gamma$ = 230 eV), taking 
into account corrections to the $u-d$ mass difference (Donoghue, 1992). In good 
approximation, the $\eta\rightarrow\pi^+\pi^-\pi^0$ Dalitz plot may be described by the 
matrix element squared
\begin{equation}
|M(x,y)|^2  \propto  1 + ay + by^2
\end{equation}
with 
\begin{equation}
y \equiv \frac{3T_0}{m(\eta)-m(\pi^0)-2m(\pi^\pm)} - 1,
\end{equation}
where $T_0$ is the kinetic energy of the neutral pion. 
The parameters $a$ and $b$ were determined in Amsler (1995d), but more accurate values 
are now available from the annihilation channel $\pbarp\rightarrow\pi^0\pi^0\eta$. 
Abele (1997b) finds  
\begin{equation}
a = -1.19 \pm 0.07, \ \  b = 0.19 \pm 0.11,
\end{equation}
in reasonable agreement with chiral perturbation calculations which predict $a$ = -1.3 
and  $b$ = 0.38 (Gasser and Leutwyler, 1985). 

The matrix element for $\eta$ decay to $3\pi^0$ is directly connected to the matrix element 
for the charged mode because $3\pi^0$ is an $I=1$ state. The matrix element squared for 
$\eta$ decay to $3\pi^0$ is given by
\begin{equation}
|M(z)|^2 = 1 + 2 \alpha z ,
\label{72}
\end{equation}
where $z$ is the distance from the center of the $\eta\rightarrow 3\pi^0$ Dalitz plot, 
\begin{equation}
z = \frac{2}{3}\sum_{i=1}^{3}\left[\frac{3E_i-m(\eta)}{m(\eta)-3m(\pi^0)}\right]^2,
\end{equation}
and where $E_i$ are the total energies of the pions. 
Chiral perturbation theory up to next-to-leading order predict $\alpha$ to be zero (Gasser 
and Leutwyler, 1985) leading to a homogeneously populated Dalitz plot. Taking unitarity  
corrections into account, Kambor (1996)  predicts $\alpha\sim -0.01$. Experiments 
have so far reported values for $\alpha$  compatible with zero, e.g.  Alde (1984) finds  
$-0.022 \pm 0.023$.

Crystal Barrel has analyzed the $3\pi^0$ Dalitz plot with 98,000 $\eta$ decays from the 
annihilation channel $\pi^0\pi^0\eta$, leading to 10 detected photons (Abele, 1997c). The 
background and acceptance corrected matrix element is shown in Fig. \ref{zplot} as a 
function of $z$. The slope is clearly negative: 
\begin{equation}
\alpha = -0.052 \pm 0.020.
\end{equation}

\subsection{$\etap\rightarrow\pi^+\pi^-\gamma$}
The $\pi^+\pi^-\gamma$ decay mode of the $\etap$ is generally believed  to proceed 
through the $\rho(770)\gamma$ intermediate state (Barnett, 1996). However, the $\rho$ 
mass extracted from a fit to the $\pi^+\pi^-$ mass spectrum appears to lie some 20 MeV 
higher than for $\rho$ production in $e^+e^-$ annihilations. This effect is due to the 
contribution of the direct decay into $\pi^+\pi^-\gamma$ (Bityukov, 1991) through the so-
called box anomaly (Benayoun, 1993). Crystal Barrel has studied the 
$\etap\rightarrow\pi^+\pi^-
\gamma$ channel where the $\etap$ is produced from the annihilation channels 
$\pi^0\pi^0\etap$, $\pi^+\pi^-\etap$ and $\omega\etap$ (Abele, 1997i). Evidence for the 
direct decay was confirmed at the $4\sigma$ level by fitting the $\pi^+\pi^-$ mass 
spectrum from a sample of 7,392 $\etap$ decays. Including contributions from the box 
anomaly, the $\rho$ mass turns out to be consistent with the standard value from $e^+e^-$ 
annihilation. Using the known two-photon decay widths of $\eta$ and $\etap$ and the 
$\eta\rightarrow\pi^+\pi^-\gamma $ decay spectrum from Layter (1973) the collaboration 
derived  the pseudoscalar nonet parameters $f_\pi/f_1$=0.91 $\pm$ 0.02, $f_8/f_\pi$=0.90 
$\pm$ 0.05 and the pseudoscalar mixing angle $\theta_p = (-16.44 \pm 1.20)^\circ$.

\subsection{Radiative $\omega$ decays}
The rates for radiative meson decays can be calculated from 
the naive quark model using SU(3) and the OZI rule (O$'$Donnell, 1981). Assuming ideal 
mixing in the vector nonet one finds, neglecting the small difference 
between $u$ and $d$ quark masses:
\begin{equation}
\frac{B(\oeg)}{B(\opg)} =
\frac{1}{9} \frac{p_{\eta}^3}{p_{\pi^0}^3} \cos^2 (54.7^\circ +\theta_p) = 0.010,
\label{76}
\end{equation}
where $p_\pi $ = 379 MeV/c  and $p_\eta$ = 199 MeV/c are the decay momenta in the
$\omega$ rest frame and $\theta_p$ is the pseudoscalar mixing angle
(Amsler, 1992b). 
However, the production and decay of the $\omega$ and $\rho$ mesons are coupled by the 
isospin breaking $\omega$ to $\rho$ transition since these mesons overlap (for 
references on $\ro$ mixing, see O$'$Connell (1995)).  Since the width of $\rho$ is much 
larger than the width of $\omega$, the effect of $\ro$ mixing is essential in processes 
where $\rho$ production is larger than $\omega$
production, for example in $e^+e^-$ annihilation where $\omega$
and $\rho$ are produced with a relative rate of 1/9. The
determination of the branching ratio for $\oeg$ varies by a factor of
five depending on whether the interference between $\oeg$ and
$\reg$ is constructive or destructive (Dolinsky, 1989). A similar
effect is observed in photoproduction (Andrews, 1977).
The GAMS collaboration has determined the $\oeg$ decay
branching ratio, $(8.3 \pm 2.1) \times 10^{-4}$,
using the reaction $\pi^-p\rightarrow\omega n$ at large
momentum transfers, thus suppressing $\rho$ production
(Alde, 1994). 

In $\pbarp$ annihilation at rest the branching ratio for $\omega\pi^0$ is much smaller 
than the branching ratio for $\rho^0\pi^0$ while the converse is true for $\eta\omega$ 
and $\eta\rho^0$ (Table \ref{BR}). We therefore expect that a determination of the $\oeg$ 
branching ratio from $\pi^0\omega$, neglecting $\ro$ mixing, will lead 
to a larger value than from $\eta\omega$. However, a simultaneous analysis of both 
branching ratios, including $\ro$ mixing, should lead to consistent results and allow a 
determination of the relative phase between the two amplitudes.

In Abele (1997d) the channels $\po$ and $\eo$ ($\oeg$) were
reconstructed from  15.5 million 0-prong events with five detected $\gamma$'s. The events 
were submitted to a 6C kinematic 
fit assuming total energy and total momentum conservation, at least one 
$\pi^0\rightarrow 2\gamma$ (or one $\eta\rightarrow
2\gamma$) and $\otg$. The $\otg$ Dalitz plots for $\po$
and $\eo$ are shown in Fig. \ref{Omega}, using the variables
\begin{equation}
\label{eqn:DalitzVariables}
x = \frac{T_2-T_1}{\sqrt{3}Q}, \ y = \frac{T_3}{Q} - \frac{1}{3},
\end{equation}
where $T_1, T_2$ and $T_3$ are the kinetic energies of the $\gamma$'s
in the $\omega$ rest frame and  $Q=T_1 + T_2 + T_3$. The prominent
bands along the boundaries are due to $\opg$ and the weaker
bands around the center to $\oeg$.

The main background contributions arose from
$6\gamma$ events ($\pbarp\rightarrow 3\pi^0, 2\pi^0\eta,
2\eta\pi^0$) with a missing photon. This background (10\% in the $\pi^0\omega$ and 5\% 
in the $\eta\omega$ Dalitz plots) was simulated using the 3-pseudoscalar distributions 
discussed in section \ref{sec:three} 
and could be reduced with appropriate cuts (Pietra, 1996). After removal of the $\pi^0$ 
bands 147 $\pm$ 25 $\oeg$ events were found in the $\eta$ bands of Fig. \ref{Omega}(a) 
and 
123 $\pm$ 19 events in the $\eta$ bands of Fig. \ref{Omega}(b).

The branching ratio for $\oeg$ was derived by normalizing
on the known branching ratio for $\opg$, $(8.5 \pm 0.5)$ \% (Barnett, 1996). 
Correcting for the reconstruction efficiency Abele (1997d) finds a branching ratio
of  $(13.1 \pm  2.4) \times 10^{-4}$ from $\po$ and $(6.5 \pm  1.1) \times 10^{-4}$ from 
$\eo$,  hence a much larger signal from $\pi^0\omega$. 

Consider now the isospin breaking electromagnetic $\ro$ transition. The amplitude $S$ for 
the reaction $\pbarp \rightarrow X(\rho - \omega) \rightarrow X \eta \gamma$ is, up to 
an arbitrary phase factor (Goldhaber, 1969):
\begin{eqnarray}
S & = & \frac{|A_{\rho}| |T_{\rho}|}{P_{\rho}}
\left(1 - \frac{|A_{\omega}|}{|A_{\rho}|}
\frac{e^{i\alpha} \delta}{P_{\omega}} \right)\nonumber\\
& + & e^{i(\alpha + \phi)} \frac{|A_{\omega}|
|T_{\omega}|}{P_{\omega}}
\left(1 - \frac{|A_{\rho}|}{|A_{\omega}|}
\frac{e^{-i\alpha}\delta}{P_{\rho}} \right),
\label{74}
\end{eqnarray}
where $A$ is the production and $T$
the decay amplitude of the two mesons and $P_{\rho} \equiv m -
m_{\rho} + i\Gamma_{\rho}/2$, $P_{\omega} \equiv m -
m_{\omega} + i\Gamma_{\omega}/2$. The parameter $\delta$ was determined from 
$\omega,\rho\rightarrow\pi^+\pi^-$: $\delta$ = (2.48 $\pm$ 0.17) MeV  (Weidenauer, 
1993). The relative phase between the production amplitudes $A_{\rho}$ and 
$A_{\omega}$
is $\alpha$ while the relative phase between the decay amplitudes
$T_{\rho}$ and $T_{\omega}$ is $\phi$.  In the absence of $\ro$
interference ($\delta = 0$) Eq. (\ref{74})
reduces to a sum of two Breit-Wigner functions with relative phase
$\alpha + \phi$. The magnitudes of the amplitudes $A$ and $T$ 
are proportional to the production branching ratios and the partial decay widths, 
respectively.

The production phase $\alpha$ can be determined from $\rho,\omega\rightarrow\pi^+ 
\pi^-$ since the isospin violating decay amplitude  $T(\omega \rightarrow \pi^+ \pi^-)$ 
may be neglected, leaving only the first term in Eq. (\ref{74}). A value for $\alpha$ 
consistent with zero, $(-5.4 \pm 4.3)^\circ$, was measured by Crystal Barrel, 
using the channel $\pbarp\rightarrow \eta\pi^+\pi^-$ where $\ro$ interference is 
observed directly (Abele, 1997a). This phase is indeed predicted to be zero in 
$e^+e^-$ annihilation, in photoproduction and also in $\pbarp$ annihilation (Achasov and 
Shestakov, 1978).

With the branching ratios for $\omega\pi^0$, $\omega\eta $, $\pi^0\rho^0$ and 
$\eta\rho^0$ given in Table \ref{BR} the intensity $|S|^2$ was fitted to the number 
of observed $\oeg$ events in $\omega\pi^0$ and $\omega\eta $, using Monte Carlo 
simulation. Both channels lead to consistent results for 
\begin{equation}
B(\oeg)  =  (6.6 \pm 1.7) \times 10^{-4},
\label{75}
\end{equation}
in agreement with the branching ratio from $\eo$, obtained by neglecting $\ro$ 
interference. The phase $\phi =  (-20^{+70}_{-50})^{\circ}$ leads to constructive 
interference. The result 
Eq. (\ref{75}) is in excellent agreement with  Alde (1994) and with the constructive 
interference solution in 
$e^+e^-$, $(7.3 \pm 2.9) \times 10^{-4}$ (Dolinsky, 1989). This then solves the longstanding 
ambiguity in $e^+e^-$ annihilation between the constructive ($\phi=0$) and destructive 
($\phi=\pi$) interference solutions. The branching ratio  for $\reg$, $(12.2 \pm 10.6) 
\times 10^{-4}$, is not competitive but 
agrees with results from $e^+e^-$,  $(3.8 \pm 0.7) \times 10^{-4}$ for 
constructive interference (Dolinsky, 1989; Andrews, 1977). Using  Eq. (\ref{76}) one 
then finds
\begin{equation}
\frac{B(\oeg)}{B(\opg)} = (7.8 \pm 2.1) \times 10^{-3},
\end{equation}
in agreement with SU(3).

The $\omega\rightarrow 3\gamma$ Dalitz plot 
is also useful to search for the direct process $\otg$ which  is similar to the decay of 
($^3S_1$) orthopositronium into 3$\gamma$  and has not been observed so far. By analogy,
the population in the $\otg$ Dalitz plot is expected to be almost
homogeneous except for a slight increase close to its boundaries
(Ore and Powell, 1949). Using the central region in Fig. \ref{Omega}(a)
which contains only one event (6 entries) one obtains the upper limit
$B(\opg) = 1.9 \times 10^{-4}$ at 95\% confidence level. This is somewhat more precise that 
the previous upper limit,  $2 \times 10^{-4}$ at 90\% confidence level (Prokoshkin and 
Samoilenko, 1995).

\section{Production of $\phi$ Mesons}
\label{seq:phi}
It has been known for some time that $\phi$ production is 
enhanced beyond expectation from the OZI rule in various 
hadronic reactions (Cooper, 1978).  Let us return to Eq. 
(\ref{31}) and replace $\eta$ by $\phi$ and 
$\etap$ by $\omega$. The mixing angle becomes the mixing angle 
$\theta_v$ in the vector nonet.  According to the OZI rule, $\phi$ 
and $\omega$ can only be produced through their $\uubar + 
\ddbar$ components. Hence $\phi$ production should vanish for an 
ideally mixed vector nonet ($\theta_v = \theta_i$) in which $\phi$ 
is purely $\ssbar$. Since $\phi$ also decays to $3\pi$ this is not 
quite the case and we find for the ratio of branching ratios with a recoiling meson $X$ 
and apart from phase space corrections, 
\begin{equation}
\tilde{R}_X = \frac{\tilde{B}(X\phi)}{\tilde{B}(X\omega)} = 
\tan^2  (\theta_i - \theta_v) = 4.2 \times 10^{-3} \ {\rm or} \ 1.5 
\times 10^{-4}, 
\label{eq:tgv}
\end{equation}
for the quadratic ($\theta_v = 39^{\circ}$) or linear ($\theta_v = 
36^{\circ}$) Gell-Mann-Okubo mass formula (Barnett, 1996).   

The branching ratios for $\pbarn$ annihilation into $\pi^-\phi$ and $\pi^-\omega$
have been measured in deuterium bubble chambers. The ratio $\tilde{R}_{\pi^-}$ lies in 
the range 0.07 to 0.22 indicating a strong violation of the OZI rule (for a review, see 
Dover and Fishbane (1989)). The Asterix experiment at LEAR 
has measured $\phi$ production in $\pbarp$ annihilation into 
$\pi^0\phi $, $\eta\phi $, $\rho^0\phi$ and $\omega\phi $  in 
gaseous hydrogen at NTP (50\% P-wave annihilation) and in 
coincidence with atomic L X-rays (P-wave annihilation). The 
branching ratios for pure S-wave were then obtained indirectly by 
linear extrapolation (Reifenr\"other, 1991). With the corresponding 
$\omega$ branching ratios, then available 
from literature, the authors reported a strong violation of the OZI rule, especially for 
$\pi^0\phi$. Some of the  branching ratios 
from Crystal Barrel  are, however, in 
disagreement with previous results. We shall therefore review the 
direct measurement of $\phi$ production in liquid from Crystal 
Barrel and then reexamine the evidence for OZI violation with the 
two-body branching ratios listed in Table \ref{BR}.

\subsection{Annihilation into $\pi^0\phi$, $\eta\phi$ and 
$\gamma\phi$}
\label{sec:phi}
Crystal Barrel has studied the channels
\begin{equation}
\pbarp  \rightarrow  K_SK_L\pi^0 \ {\rm and}  \ K_SK_L\eta , \\
\end{equation}
where $K_S$ decays to $\pi^0\pi^0$ and $\eta$ to 
$\gamma\gamma$, leading to six photons and a missing 
(undetected) $K_L$ (Amsler, 1993d). The starting data sample 
consisted of $4.5\times 10^6$ 0-prong annihilations. By 
imposing energy and momentum conservation, the masses of the 
three reconstructed pseudoscalars and the $K_S$ mass,  a (5C) 
kinematic fit was applied leading to 2,834 $K_SK_L\pi^0$ 
and 72 $K_SK_L\eta$ events with an estimated background of 4\%, 
respectively 36\%.

The $K_SK_L\pi^0$ Dalitz plot 
is shown in Fig.~\ref{kskl}. One observes the production of 
$K^*(892)$ $(\rightarrow K\pi)$ and $\phi(\rightarrow K_SK_L)$. 
A Dalitz plot analysis was performed with the method  described in 
section \ref{sec:spa}. Since the $C$-parity of $K_SK_L$ is negative\footnote{Note that 
$K^0\overline{K^0}$ recoiling against $\pi^0$  appears as $K_SK_S + K_LK_L$ ($J^{PC}$ = 
even $^{++}$) from $^1S_0$ 
and as $K_SK_L$ ($J^{PC}$ = odd$^{--}$) from $^3S_1$.} the  contributing 
initial atomic S-state is $^3S_1$. One 
obtains a good fit to the Dalitz plot with only two amplitudes, one 
for $K^*\overline{K}$ (and its charge conjugated $K\overline{K}^*$ 
which interferes constructively) and one for $\pi^0\phi$ with a relative 
contribution to the $K_SK_L\pi^0$ channel of
\begin{equation}
\frac{ K^*\overline{K} + K\overline{K}^*}{\pi^0\phi}  = 2.04 \pm 
0.21. 
\label{40}
\end{equation}

The final state $K_SK_L\eta$ is much simpler since only $\eta\phi$ contributes 
(Amsler, 1993d). One obtains by comparing the 
intensities for $\pi^0\phi$ and $\eta\phi$
\begin{equation}
\frac{B(\pi^0\phi)}{B(\eta\phi)} = 8.3 \pm 2.1,
\label{36}
\end{equation}
taking into account the unobserved decay modes of the $\eta$ 
meson.

Events with a $K_L$ interacting in the CsI crystals have been 
removed by the selection procedure which required exactly six 
clusters in the barrel. The last two results therefore assume that 
the interaction probability for $K_L$ in CsI does not vary 
significantly with $K_L$ momentum. Therefore, this interaction 
probability needs to 
be determined to derive absolute branching ratios for $\pi^0\phi$ 
and $\eta\phi$. This number cannot be obtained directly by Monte 
Carlo simulation due to the lack of data for low energy $K_L$ 
interacting with nuclear matter. With monoenergetic (795 MeV/c) 
$K_L$ from the channel $\pbarp\rightarrow K_SK_L$, Amsler 
(1995c) finds an interaction probability of (57 $\pm$ 3) \% in the CsI 
barrel.  This leads to a branching ratio of 
(9.0 $\pm$  0.6) $\times 10^{-4}$ for $K_SK_L$, in agreement with 
bubble chamber data (Table \ref{BR}). 

An average 
interaction probability of (54 $\pm$ 4)\% was measured with the kinematically well 
constrained 
annihilation channel $K_S(\rightarrow\pi^+\pi^-)K_L\pi^0$ (Abele, 1997e). However, in 
Amsler (1995c) a somewhat lower probability was used. Updating their $\pi^0\phi$ 
branching ratio  one finds together with their compatible result from 
$\pi^0(\phi\rightarrow K^+K^-$): 
\begin{equation}
B(\pbarp\rightarrow\pi^0\phi) = (6.5 \pm 0.6) \times 10^{-4}. 
\end{equation}
From Eq. (\ref{36}) one then obtains
\begin{equation}
B(\pbarp\rightarrow\eta\phi) = (7.8 \pm 2.1) \times 10^{-5}.
\end{equation}
Both numbers are slightly higher than from indirect data  in gaseous hydrogen 
which were extrapolated to pure S-wave annihilation 
(Reifenr\"other, 1991), see Table \ref{BRR}.

Radiative annihilation into $\phi$ mesons should be suppressed by 
both the OZI rule and the electromagnetic coupling.  Crystal Barrel has studied the 
channel $\gamma\phi$ with the reactions 
\begin{equation}
\pbarp \rightarrow K_SK_L\gamma \ {\rm and} \  K^+K^-\gamma,
\label{38}
\end{equation}
(Amsler, 1995c). In the first reaction $K_S$ decays to $\pi^0\pi^0$, $K_L$ is not 
detected and thus the final state consists of five photons. The $K_SK_L\gamma$ final state 
was selected from $8.7 \times 10^6$ 0-prong annihilations by performing a 4C fit, imposing 
energy and momentum 
conservation, the masses of the two pions and the $K_S$ mass.
The background reaction $\pbarp\rightarrow K_SK_L$ with an 
interacting $K_L$ faking the fifth photon could easily be suppressed 
since $K_L$ and $K_S$ are emitted back-to-back. The experimental 
$K_SK_L\gamma$ Dalitz plot is dominated by background from $K_SK_L\pi^0$ with a 
missing (undetected) low energy 
$\gamma$ from $\pi^0$ decay and is therefore similar to the one 
shown in Fig. \ref{kskl}. The background contribution to 
$\gamma\phi $, mainly from $\pi^0\phi$ with a missing photon, 
was estimated by Monte Carlo simulation and by varying the 
photon detection threshold. This led to 211 $\pm$ 41 
$\gamma\phi $ events corresponding to a branching ratio 
$B(\gamma\phi)$ = $(2.0 \pm 0.5) \times 10^{-5}$, after correcting 
for the (updated) $K_L$ interaction probability.

A sample of $1.6 \times 10^6$ 2-prong annihilations was used to 
select the second reaction in (\ref{38}). After a cut on energy and 
momentum conservation (assuming kaons) the measured ionization 
loss in the JDC was used to separate kaons from pions. Again, the 
dominating background to $\gamma\phi $ 
arose from $\pi^0\phi$ with a missing $\gamma$. This background 
was subtracted by varying the $\gamma$ detection threshold 
and by keeping only $\gamma$'s with energies around  661 MeV, as 
required by two-body kinematics. The  signal of 29  events led to 
a branching ratio 
$B(\gamma\phi)$ = $(1.9 \pm 0.7) \times 10^{-5}$ which is less 
precise but in good agreement with data from the neutral 
mode. The average is then
\begin{equation}
B(\pbarp\rightarrow\phi\gamma) = 
(2.0 \pm 0.4) \times 10^{-5},
\end{equation}
which updates the result from Amsler (1995c).

\subsection{$\phi/\omega$ ratio}
Table \ref{R_X} and Fig. \ref{figR_X} show the phase space 
corrected ratios $\tilde{R}_X$ (Eq. (\ref{eq:tgv})).  The nearest threshold 
dominance factors (Eq. (\ref{10})) have been used but the measured 
ratios do not differ significantly from $\tilde{R}_X$. Phase space factors of the type 
(\ref{39}) lead to even larger ratios.  For $X=\gamma, \pi^0, \eta$ we 
used Crystal Barrel data. For $X =\omega$ we used for  
$\omega\omega$ the branching ratio from Crystal Barrel 
(multiplied by two for identical particles) and for $\omega\phi $ the 
branching ratio from Bizzarri (1971) (Table \ref{BR}). 
For completeness we also list the result for $X=\rho$ from 
Reifenr\"other (1991) and Bizzarri (1969) and the recent Obelix 
data for $R_{\pi^-}$ (Ableev, 1995) and $R_{\pi^+}$ (Ableev, 1994) in $\pbarn$ and 
$\overline{n}p$ annihilation.  

Annihilation into $\omega\pi^0\pi^0$ and $\phi(\rightarrow K_L 
K_S)\pi^0\pi^0$ can be used to extract the ratio $\tilde{R}_\sigma$ 
(Spanier, 1997),  where $\sigma$ stands for the low energy 
$(\pi\pi)$ S-wave up to 900 MeV (section \ref{sec:coupled}). Finally, the ratio 
$R_{\pi^+\pi^-}$ was also measured in $\pbarp$ at rest (Bertin, 
1996). Table \ref{R_X} and Fig. \ref{figR_X} show their result for 
$\pi^+\pi^-$ masses between 300 and 500 MeV.

The production of $\phi$ mesons is enhanced in all channels except $\eta\phi $ 
and is  especially dramatic in $\gamma\phi\ $ (from $^1S_0 $) and $\pi^0\phi$ (from 
$^3S_1$). Several 
explanations for this effect have been proposed:

(i) High energy reactions reveal the presence of sea  
$\ssbar$ pairs in the nucleon at high momentum transfers but valence $\ssbar$ pairs 
could enhance the production of $\phi$ mesons already at 
small momentum transfers. Figure \ref{OZIphi} shows the OZI 
allowed production of $\ssbar$ mesons through the shake-out 
mechanism (a) and through the OZI allowed 
rearrangement process (b) (Ellis, 1995). The fraction of $\ssbar$ pairs in the 
nucleon required to explain the measured $\pi^0\phi$ rate lies 
between 1 and 19\%. Deep 
inelastic muon scattering data indicate an $\ssbar$ polarization 
opposite to the spin of the nucleon (Ellis, 1995).  For 
annihilation from the $^3S_1$ state the wave function of the 
($^3S_1$) $\ssbar$ would match the wave function of the 
$\phi$ in the rearrangement process of Fig. \ref{OZIphi}(b), leading 
to an enhanced production of $\phi$ mesons. The absence of
enhancement in $\eta\phi$ could be due to destructive interference 
between additional graphs arising from the $\ssbar$ content of 
$\eta$. However, this model does not explain the large 
branching ratios for the two-vector channels $\rho^0\phi $, $\omega\phi$ and especially 
$\gamma\phi$ which proceed from the $^1S_0$ atomic state. 

In the tensor nonet, the mainly $\ssbar$ meson is $f_2'(1525)$ and $f_2(1270)$ is the 
mainly $\uubar + \ddbar$. Using annihilation into $K_LK_L\pi^0$ and $3\pi^0$ (section 
\ref{sec:three}) Crystal Barrel has measured the ratio of  $f_2'(1525)\pi^0$ to 
$f_2(1270)\pi^0$ from $^1S_0$. After dividing by $W$  (Eq. (\ref{10})) one finds with the 
most recent branching ratio for  $f_2'(1525)$ decay to $\KKbar$  from Barnett (1996), see 
Table \ref{BRTB} below:
\begin{equation}
\frac{\tilde{B}(\pbarp\rightarrow f_2'(1525)\pi^0)}{\tilde{B}(\pbarp\rightarrow 
f_2(1270)\pi^0)} = \tan^2(\theta_i-\theta_t) = (2.6 \pm 1.0) 
\times 10^{-2}.
\end{equation}
The mixing angle $\theta_t$ in the $2^{++}$ nonet  is found to be 
$(26.1^{+2.0}_{-1.6})^\circ$, in good agreement with the linear 
(26$^\circ$) or quadratic (28$^\circ$) mass formulae (Barnett, 1996). There is therefore no 
OZI violating $\ssbar$ enhancement from $f_2'(1525)\pi^0$ in liquid hydrogen.

(ii) Dover and Fishbane (1989) suggest that the 
$\pi^0\phi$ enhancement is due to mixing with a four-quark 
state ($\ssbar\qqbar$) with mass below $2m_p$ 
(Fig.~\ref{OZIphi}(c)). This exotic meson would then have the quantum 
numbers of the $\pbarp$ initial state ($^3S_1 = 1^{--}$ with $I = 1$). This 
would also explain why $\eta\phi$ ($I = 0$) is not enhanced and 
why $\pi^0\phi$ is weak from  $^1P_1$ (Reifenr\"other, 1991). A $1^{--}$ state, 
$C(1480)\rightarrow\pi\phi $,  has in fact been reported in  $\pi^- 
p\rightarrow\pi^0\phi n$ (Bityukov, 1987). This isovector    
cannot be $\qqbar$ since it would decay 
mostly to $\pi\omega$, which is not observed. Indeed, Crystal 
Barrel does not observe any $\pi^0\omega$ signal in this mass 
region in $\pbarp\rightarrow\pi^0\pi^0\omega$ at rest (Amsler, 
1993a).  Also, $C(1480)$ has not been observed in 
$\pbarp$ annihilation at rest into $\pi^+\pi^-\phi$ (Reifenr\"other, 
1991) nor into $\pi^0\pi^0\phi$ (Abele, 1997f). In any case the large $\gamma\phi $ 
signal with the ``wrong" quantum numbers $^1S_0$ remains 
unexplained by this model.

(iii) We have seen (Eq.~(\ref{40})) that the $\KKbar\pi$ final 
state is dominated by $K^*\overline{K}$ production which proceeds dominantly from the 
$I = 1$ $^3S_1$ state (see section \ref{sec:KLKP}). The 
$\phi$ enhancement could then 
be due to $K^*\overline{K}$ and 
$\rho\rho$ rescattering (Fig.\ref{OZIphi}(d)). In Gortchakov (1996) 
$K^*\overline{K}$ and $\rho\rho$ interfere constructively to produce a 
$\pi^0\phi $ branching ratio as high as $4.6 \times 10^{-4}$, nearly in 
agreement with experimental data. In Locher (1994) and 
Markushin (1997) the large $\gamma\phi $ branching ratio simply 
arises from VDM: The channels $\rho^0\omega $ and 
$\omega\omega$ interfere destructively in Eq. (\ref{35}) (thereby lowering the 
branching ratio for $\gamma\omega$) while 
$\rho^0\phi $ and $\omega\phi $ interfere constructively (thus 
increasing the branching ratio for $\gamma\phi$). This 
conclusion, however, depends strongly on the phase space correction:  
Prescription (\ref{10}) leads to a $\gamma\phi $ branching ratio 
which exceeds the OZI prediction by a factor of ten (Table 
\ref{BRG}). Also, as pointed out by Markushin (1997), the large 
$\rho^0\phi$ and $\omega\phi$ rates remain unexplained but could 
perhaps be accommodated within a two-step mechanism similar to 
the one shown in Fig. \ref{OZIphi}(d). 

The origin of the $\phi$ enhancement is therefore not clear. If $I=1$ 
also dominates $\pbarp$ annihilation into  $K^*\overline{K}$ from 
P-states in gaseous hydrogen, then the rescattering model would presumably conflict 
with the weak $\pi\phi $ production observed from $^1P_1$ states 
(Reifenr\"other, 1991). With strange quarks in the nucleon, 
the $^3S_1$ contribution and hence the contribution from $\ssbar$ 
pairs in the nucleon will be diluted by the large number of partial 
waves at higher $\pbar$ momenta. Hence $\phi$ production should 
decrease with increasing momentum. Also, the small $\eta\phi $ 
rate, possibly due to destructive interference, implies 
that $\etap\phi$ should be abnormally large (Ellis, 1995). 
Finally, since $f_2'(1525)$ is a spin triplet meson one would expect a strong production of 
$f_2'(1525)\pi^0$ from triplet $\pbarp$ states at rest, hence $^3P_1$ (Ellis, 1995).
The analysis of Crystal Barrel data in gaseous hydrogen and in 
flight will hopefully contribute to a better understanding of the 
$\phi$ enhancement in hadronic reactions.  

\section{Meson Spectroscopy}
\label{sec:spectro}
\subsection{Introduction}
Mesons made of light quarks $u, d, s$  are classified within the $\qqbar$ nonets of  
SU(3)-flavor. The ground states (angular momentum $L=0$) pseudoscalars ($0^{-+}$) and 
vectors ($1^{--}$) are well established. Among the first orbital excitations ($L=1$), 
consisting of the four nonets  $0^{++}$, $1^{++}$, $2^{++}$, $1^{+-}$, only the tensor 
($2^{++}$) nonet is complete and unambiguous with the well established $a_2(1320)$, 
$f_2(1270)$, $f_2'(1525)$ and $K_2^*(1430)$ but another tensor, $f_2(1565)$ was discovered 
at LEAR in the 1500 MeV mass range (May, 1989). 

Before Crystal Barrel three scalar ($0^{++}$) states were already well established: 
$a_0(980)$, $f_0(980)$ and $K^*_0(1430)$. Further candidates have been reported and we 
shall discuss the scalars in more details below. In the 
$1^{++}$ nonet two states compete for the $\ssbar$ assignment, $f_1(1420)$ and 
$f_1(1510)$. In the 
$1^{+-}$ nonet the $\ssbar$ meson is not established although  a 
candidate, $h_1(1380)$, has been reported (Aston, 1988a; Abele 1997f).  Many  of the radial 
and higher orbital excitations are still missing. Recent experimental reviews on light 
quark mesons 
have been written by Bl\"um (1996) and Landua (1996) and theoretical 
predictions for the mass spectrum can be found in Godfrey and Isgur (1985). 

Only overall color-neutral $\qqbar$ configurations are allowed by QCD but 
additional colorless states are possible, among them 
multiquark mesons ($\twoqqbars$, $\threeqqbars$) and mesons 
made of $\qqbar$ pairs bound by an excited gluon $g$, the hybrid states (Isgur and Paton, 
1985; Close and Page, 1995). The $2^{-+}$ state 
$\eta_2(1870)$ has been reported by Crystal Barrel (Adomeit, 1996) with decay rates to 
$a_2^0(1320)\pi^0$ and $f_2(1270)\eta$ compatible with predictions for a hybrid state 
(Close and Page, 1995). Hybrids may have exotic quantum numbers, e.g.  $1^{-+}$,  
which do not couple to $\qqbar$. An isovector state, $\hat\rho(1405)$,  with quantum 
numbers $1^{-+}$ has been reported (Alde, 1988a; Thompson, 1997). However, lattice QCD 
predicts the lightest hybrid, a $1^{+-}$, around 2000 MeV (Lacock, 1997).

A striking prediction of QCD is the existence of isoscalar 
mesons which contain only gluons, the glueballs (for a recent experimental review, see 
Spanier (1996)). They are a consequence of the non-abelian structure of QCD which 
requires that gluons couple and hence may bind. The models predict low-mass glueballs 
with quantum numbers $0^{++}$, $2^{++}$ and $0^{-+}$ (Szczepaniak, 1996).
The ground state glueball, a $0^{++}$ meson, is expected by lattice 
gauge theories to lie in the mass range 1500 to 1700 MeV. The mass of the pure gluonium 
state is calculated at 1550 $\pm$ 50 MeV by Bali (1993) while Sexton (1995) predicts a 
slightly higher mass of 1707 $\pm$ 64 MeV. The first excited state, a $2^{++}$, is expected 
around 2300 MeV (Bali, 1993). 

Since the mass spectra of $\qqbar$ and glueballs overlap, the latter are easily 
confused with ordinary $\qqbar$ states. This is presumably the reason why they have not 
yet been identified unambiguously. For pure gluonium one expects couplings of similar 
strengths to $\ssbar$ and $\uubar$ + $\ddbar$ 
mesons since gluons are flavor-blind. In contrast, $\ssbar$ mesons decay mainly to kaons 
and $\uubar$ + $\ddbar$ mesons mainly to pions. Hence decay rates to $\pi\pi$, 
$\KKbar$, $\eta\eta$ and $\eta\etap$ can be used to distinguish glueballs from ordinary 
mesons. However, mixing with nearby $\qqbar$ states may modify the decay branching 
ratios (Amsler and Close, 1996) and obscure the nature of the observed state. 
Nevertheless, the existence of a scalar gluonium state, whether pure or mixed with 
$\qqbar$, is signalled by a 
third isoscalar meson in the $0^{++}$ nonet. It is therefore essential to complete the SU(3) 
nonets  in the 1500 - 2000 MeV region and to identify supernumerary states. The most 
pressing questions to be addressed are:
\begin{enumerate}
\item
What are the ground state scalar mesons, in particular is $f_0(980)$ the $\ssbar$ state and 
is $a_0(980)$ the isovector or are these states $\KKbar$ molecules (Weinstein and Isgur, 
1990; Close, 1993) in which case the nonet members still need to be identified? Where 
are the first radial excitations and is there a supernumerary  $I=0$ scalar in the 1500 MeV 
region? Is  $f_J(1710)$ scalar or tensor? 
\item
In the $0^{-+}$ sector, are $\eta(1295)$ and $\eta(1440)$ the two isoscalar radial excitations 
of $\eta$ and $\etap$ or is $\eta(1440)$ a structure  containing 
several states (Bai, 1990; Bertin, 1995), in particular a non-$\qqbar$ state around 1400 
MeV?
\item
Where are the hybrid states? Is $\eta_2(1870)$ a hybrid and does $\hat\rho(1405)$ really 
exist?
\end{enumerate}
Before reviewing the new mesons discovered by Crystal Barrel and providing clues to some 
of these issues, we shall recall 
the mathematical tools used to 
extract the mass, width, spin and parity of 
intermediate resonances in $\pbarp$ annihilation at rest.

\subsection{Spin-parity analysis}
\label{sec:spa}
The Crystal Barrel data have been analyzed using the isobar model in which the $\pbarp$ 
system annihilates into N ``stable" particles ($\pi^{\pm}, 
K^{\pm}, K^0, \pi^0, \eta, \etap$)  
through  intermediate resonances. The decay chain is assumed 
to be a succession of two-body decays $a\rightarrow bc$ followed 
by $b\rightarrow b_1b_2$ and $c\rightarrow c_1c_2$, etc. Final state rescattering is 
ignored. We shall calculate from the $N$ momentum vectors the probability $w_D$ that the 
final state proceeds through a given cascade of resonances. The final state may be from 
real data or from phase space distributed Monte Carlo events to be weighted by $w_D$.

The spins and parities of intermediate resonances are determined using the helicity 
formalism developed by Jacob and Wick (1959) or the  equivalent method of  Zemach 
tensors (Zemach, 1964, 1965). Here we describe briefly the helicity formalism. Suppose that 
a mother  
resonance with mass $m_0$ and spin $J$ decays into two 
daughters (spins $S_1$ and $S_2$) with 
total spin $S$ and relative angular momentum $L$. As 
quantization axis we choose the flight direction of the mother. The 
decay amplitude is given by the matrix (Amsler and Bizot, 1983)
\begin{equation}
A_{\lambda_1,\lambda_2;M} = D^J_{\lambda 
M}(\theta,\phi)\langle J\lambda|LS0\lambda\rangle\langle 
S\lambda|S_1S_2\lambda
_1,-\lambda_2\rangle\times BW_L(m),
\label{54}
\end{equation}
where the row index  $\lambda = \lambda_1-\lambda_2$ runs 
over the $(2S_1+1)(2S_2+1)$ helicity states and the column index 
M over the $2J+1$ magnetic substates; $\theta$ and $\phi$ refer to 
the decay angles in the mother rest frame.  $BW_L(m)$ is the 
Breit-Wigner amplitude\footnote{ The phase space 
factor $\sqrt{\rho} = \sqrt{2p/m}$ should be dropped in Eq. (\ref{55}) when events 
are drawn by Monte Carlo simulation,  already assuming phase space 
distribution.}  
\begin{equation}
BW_L(m) = \frac{m_0\Gamma_0}{m_0^2-m^2-
im_0\Gamma(m)} F_L(p) \sqrt{\rho}, 
\label{55}
\end{equation}
where
\begin{equation}
\Gamma(m) = 
\Gamma_0\frac{m_0}{m}\frac{p}{p_0}
\frac{F^2_L(p)}{F^2_L(p_0)}.
\label{83}
\end{equation}
The mass and width of the 
resonance are $m_0$ and $\Gamma_0$,  $p$ is the two-body decay momentum and $p_0$ 
the decay momentum for $m=m_0$. The damping factors $F_L(p)$ are given in Table 
\ref{Blatt}. The matrix $D$ is given by
\begin{equation}
D^J_{\lambda M} (\theta,\phi) = e^{iM\phi} 
d^J_{\lambda   M}(\theta)
\label{58}
\end{equation}
where the matrix $d^J_{\lambda 
M}(\theta)$ is the usual representation of a rotation 
around the $y$-axis, see for example Barnett (1996). 

We shall describe the annihilation to 
the observed final state by a product of matrices $A$ 
for successive decays in the cascade. Hence we first calculate from the $N$ final state 
momentum vectors 
the angles $\theta$ and $\phi$ for all resonances through a series of 
Lorentz boosts, apply Eq. (\ref{54}) to each decay and obtain 
the total amplitude through matrix multiplications, for example 
\begin{equation}
A = [A(c\rightarrow c_1c_2) \otimes A(b\rightarrow b_1b_2)]   
A(\pbarp\rightarrow bc) ,
\label{59}
\end{equation}
where $\otimes$ denotes a tensor product. The matrix $A$ has as many rows as the total 
final state spin multiplicity and has $2J+1$ columns, where $J$ is the total spin of the 
$\pbarp$ system. We define the 
quantization axis as the direction of one of the daughters in the 
first decay (the annihilating $\pbarp$ atom) for which we choose 
$\theta$ = 0 ,  $\phi=0$ and $BW_L(2m_p)=F_L(p)\sqrt{p}$.

Several decay chains of intermediate resonances may lead to 
the same observed final state of $N$ stable particles. The transition 
probability $w_D$ for chains starting from the {\it same} atomic 
state is given by the coherent sum 
\begin{equation}
w_D =  w\times\epsilon\times {\rm Tr}\  [(\sum_{j}^{}\alpha_jA_j) \tilde{\rho}  
(\sum_{k}^{}\alpha_k^*A_k^{\dagger})] = w\times\epsilon\times {\rm Tr}\   
| \sum_{j}^{}\alpha_jA_j|^2, 
\label{56}
\end{equation}
where the sums extend over all decay chains labelled by the matrices $A_j$. We have 
assumed that the initial spin-density matrix $\tilde{\rho}$ is unity since 
the $\pbarp$ system is unpolarized. The phase space  $w$ and the detection probability 
$\epsilon$ will be ignored for Monte Carlo events drawn according to phase space and 
submitted to the detector simulation, since $w=1$ and $\epsilon = 1$ or 0 for every Monte 
Carlo event. The  
parameters $\alpha_j = a_j\exp (-i\phi_j)$ are unknown constants to be fitted and one 
phase, say $\phi_0$, is arbitrary and set to zero. 
For chains decaying into the same resonances but with different 
electric charges (e.g. $\rho^+\pi^-, \rho^-\pi^+, \rho^0\pi^0$ 
$\rightarrow\pi^+\pi^-\pi^0$) these constants are given by isospin
relations.  The contributions from 
{\it different} atomic states are given by incoherent sums, i.e. by 
summing weights $w_D$ of the form (\ref{56}). 

As an example, let us derive the weight $w_D$ for the annihilation 
channel $\pbarp\rightarrow\rho^0\rho^0\rightarrow\ 2\pi^+2\pi^-$ 
from the atomic state  $^1S_0$ ($J^{PC}=0^{-+}$). Parity, $C$-parity and total angular 
momentum 
conservation require for $\rho^0\rho^0$ that $L=1$ and $S=1$ (see section \ref{sec:ann}). 
The first Clebsch-Gordan 
coefficient in (\ref{54}) is, apart from a constant, 
\begin{equation}
\langle 0\lambda|110\lambda\rangle\langle 1\lambda|11
\lambda_1,-\lambda_2\rangle\ = 
\lambda_1\delta_{\lambda_1\lambda_2},
\end{equation}
and hence the amplitude vanishes unless the $\rho$'s are emitted with the same helicity 
$\lambda_1=\lambda_2\neq 0$.
With $J=0$, the matrices (\ref{58}) are unity and therefore  
$A(\pbarp)$ is a column-vector with 9 rows
\begin{eqnarray}
A(\pbarp) = \left( \begin{array}{c} 1 \\ 0 \\  . \\  .  \\ . \\ 0 \\ - 
1 \end{array} \right) F_1(p)\sqrt{p}.
\end{eqnarray}
For $\rho\rightarrow\pi^+\pi^-$ one finds with $S=0$ and $J=1$ that $L=1$ and the  
Clebsch-Gordan coefficients are unity. Hence we get with $\lambda=0$ the 
3-dimensional row-vector
\begin{equation}
A(\rho) = [D^1_{01}(\theta, \phi), D^1_{00}(\theta, \phi), 
D^1_{0-1}(\theta, \phi)] BW_1 (m).
\end{equation}
With Eq. (\ref{59}) one then obtains
\begin{eqnarray}
A & = & [A(\rho_1) \otimes A(\rho_2)] A(\pbarp) \nonumber\\
& = & [D^1_{01}(\theta_1, \phi_1) D^1_{01}(\theta_2, \phi_2) -
D^1_{0-1}(\theta_1, \phi_1) D^1_{0-1}(\theta_2, \phi_2)]\nonumber\\
& & \times BW_1(m_1) BW_1(m_2) F_1(p)\sqrt{p}, 
\nonumber\\
& = & i\sin\theta_1\sin\theta_2\sin(\phi_1+\phi_2)
BW_1(m_1) BW_1(m_2) F_1(p)\sqrt{p}, 
\end{eqnarray}
and therefore
\begin{equation}
w_D = \sin^2\theta_1\sin^2\theta_2\sin^2(\phi_1+\phi_2)|BW_1(m_1)  
BW_1(m_2)|^2  F_1^2(p)p.
\end{equation}
The angles refer to the directions of the pions in the $\rho$ rest 
frames, with respect to the flight direction of the $\rho's$. Therefore the most probable 
angle between the planes spanned by the two dipions is 90$^{\circ}$. This angular 
dependence is familiar in parapositronium ($0^{-+}$) annihilation or $\pi^0$ decay where 
the $\gamma$ polarizations are preferably orthogonal ($\phi_1+\phi_2=90^\circ$). 
However, there are two ways to combine four pions into 
$\rho^0\rho^0$ 
and therefore the final weight is actually given by the coherent sum (\ref{56}) of  two 
decay chains with $\alpha_1=\alpha_2$.

As another example of symmetrization let us consider $\pbarp$ annihilation into 
$\pi\KKbar$ which will be  discussed in detail below. The amplitudes for 
annihilation through the intermediate $K^*$ are related through isospin  
Clebsch-Gordan coefficients (see for example Conforto (1967) or Barash (1965)). In 
general,  annihilation may occur from $^1S_0$ or $^3S_1$ with isospin 
$I=0$ or 1. For example, $\pi^0K^+K^-$ proceeds through $K^{*+}\rightarrow 
\pi^0K^+ $ or $K^{*-}\rightarrow\pi^0K^-$ with coefficients $\alpha_1$ and $\alpha_2$ 
equal in absolute 
magnitude and the two chains interfere.  Table \ref{symm} gives the relative sign 
between $\alpha_1$ and $\alpha_2$. Note that for $^3S_1$  the matrix   
(\ref{58}) flips the sign so that the {\it observed} interference pattern is the same for 
$^1S_0$ or $^3S_1$, namely constructive 
in $\pi^0K^+K^-$,  $\pi^0K^0\overline{K^0}$ and ($I=0$) 
$\pi^{\pm} K^{\mp} K^0$, and destructive in ($I=1$) $\pi^{\pm} K^{\mp} K^0$.
The signs given in Table \ref{symm} also apply to $\KKbar$ intermediate 
states with isospin $i=1$ from $I(\pbarp) = 0$ and  $I(\pbarp) = 1$ (the latter 
only contributing to $\pi^{\pm} K^{\mp} K^0$) and for states with isospin $i=0$
from $I(\pbarp) = 1$ (to which $\pi^{\pm} K^{\mp} K^0$ does not contribute).    

The procedure to analyze data is as follows: Phase space distributed Monte Carlo events are 
generated, tracked through the detector simulation and 
submitted to the reconstruction program. As already mentioned, this procedure 
automatically takes care of the
factors $\rho$, $w$ and $\epsilon$. For $\pbarp$ annihilation into three
 stable particles, there are only 
two independent kinematic variables. One usually chooses the 
invariant masses squared $m_{12}^2$ and $m_{13}^2$. The 
two-dimensional distribution (Dalitz plot) is then  uniformly populated 
for phase space distributed events. The procedure consists in 
calculating $w_D$ for each Monte Carlo event and to vary the constants 
$\alpha_i$, the widths, masses and spin-parity assignments of the 
resonances until a good fit to the observed Dalitz plot 
density is achieved. Resonances with spins larger than 2 are too heavy to be produced in 
$\pbarp$ annihilation at rest and are therefore ignored.

For the Dalitz plot fits it is  convenient to factorize $w_D$ in 
terms of the real constants $a_j$ and $\phi_j$:
\begin{equation}
w_D = \sum_{i} a_i^2 Q_{ii}  + 2 \sum_{i<j} a_ia_j \mbox{\rm Re} 
[Q_{ij}] \cos(\phi_i-\phi_j) + 2 \sum_{i<j} a_ia_j \mbox{\rm Im}  
[Q_{ij}] \sin(\phi_i-\phi_j),
\label{87}
\end{equation}
where
\begin{equation}
Q_{ij} = {\rm Tr}\  [A_iA_j^{\dagger}] .
\end{equation}
Dalitz plots weighted by $Q_{ii}$, Re$[Q_{ij}]$ and Im$[Q_{ij}]$are 
then produced for 
each pair of chains $i, j$ ($i\leq j$). The  $Q_{ij}$,  and 
correspondingly the weights $w_D$, are normalized to the 
total number of real events $N_T$:
\begin{equation}
Q_{ij}\rightarrow Q_{ij}/\sqrt{f_if_j}, 
\end{equation}
with
\begin{equation}
f_i = \sum Q_{ii}/ N_T,
\end{equation}
where the sum runs over all Monte Carlo 
events. One then divides the Dalitz plots into cells, adds them according to Eq. (\ref{87}) 
and builds the $\chi^2$
\begin{equation}
\chi^2 = \sum \frac{(n - w_D)^2}{n + w_D^2/n_{MC}}
\end{equation}
where the sum extends over all cells. The number of real events 
in each cell is denoted by $n$ and the number of Monte Carlo 
events by $n_{MC}$.

With limited statistics or for more than two degrees of freedom (final states with more 
than three stable particles) the $\chi^2$ minimization may be replaced by a likelihood 
maximization. One minimizes 
the quantity $S$ = - 2 ln$L$ or 
\begin{equation}
S  = 2 N_T {\rm ln} \left( \sum_{i=1}^{M_T}w_i[MC]\right) - 2 
\sum_{i=1}^{N_T}{\rm ln}w_i[DAT], 
\label{60}
\end{equation}
where $w_i[MC]$ and $w_i[DAT]$ are weights $w_D$ calculated for 
Monte Carlo and data events, respectively. The sums run over 
$N_T$ 
data events and $M_T$ Monte Carlo events. 

From the best fit the fractional contributions 
of the resonances in chain $i$ are given by
\begin{equation}
r_i \equiv \frac{a_i^2}{\sum_{i}^{} a_i^2},
\label{57}  
\end{equation}
where, obviously, $\sum r_i$ = 1. This is a somewhat arbitrary 
definition which may not agree with the directly visible Dalitz plot densities, 
because interferences beween the chains are neglected in 
Eq. (\ref{57}). One may define alternatively
\begin{equation}
r_i\equiv a_i^2,
\label{63}
\end{equation}
but then $\sum r_i$ may differ significantly from unity in the 
presence of strong interferences. Hence decay branching fractions for broad interfering 
resonances are not measurable unambiguously. This is an unavoidable caveat to keep in 
mind when extracting the internal structure of broad states from 
their decay branching ratios. 

\subsection{$K$-matrix analysis}
The Breit-Wigner factors (\ref{55}) violate unitarity when two 
resonances with the 
same quantum numbers overlap and decay into the same final 
state. Also, they do not describe distortions  in the mass spectrum 
that occur around kinematical thresholds. For example, the 
$f_0(980)\rightarrow\pi\pi$ appears  as a  dip  rather than a peak 
in the $\pi\pi$ mass spectrum of elastic $\pi\pi$ scattering, due to 
the opening of the decay channel $f_0(980)\rightarrow \KKbar$ (Au, 1987).

This behaviour can be described with the $K$-matrix formalism. A 
detailed description can be found in Chung (1995) and I shall only 
recall the formulae used in the analysis of Crystal Barrel 
data.  Consider for instance the four scattering reactions 
\begin{eqnarray}
\left( \begin{array}{c c}
\pi\pi\rightarrow\pi\pi & \pi\pi\rightarrow\KKbar\\
\KKbar\rightarrow\pi\pi & \KKbar\rightarrow\KKbar
\end{array} \right).
\end{eqnarray}
The transition amplitude $T$ for a given partial wave is described 
by the $2\times 2$ $K$-matrix 
\begin{equation}
T = (1 - i K \rho)^{-1} K
\label{42}
\end{equation}
with the real and symmetric matrix
\begin{equation}
K_{ij}(m) = \sum_{\alpha} 
\frac{\gamma_{\alpha_i}\gamma_{\alpha_j}m_\alpha
\Gamma'_\alpha}{m_\alpha^2-m^2} B_{\alpha_i}(m)B_{\alpha_j}(m) 
+ c_{ij}.
\label{43}
\end{equation}
The sum runs over all resonances with $K$-matrix poles $m_\alpha$ 
decaying to $\pi\pi$ and $\KKbar$ with (real) coupling constants 
$\gamma_{\alpha_1}$ and  
$\gamma_{\alpha_2}$, respectively,  where
\begin{equation}
\gamma_{\alpha_1}^2 + \gamma_{\alpha_2}^2 = 1.
\end{equation}
The factors $B_{\alpha_i}$ are ratios of Blatt-Weisskopf damping 
factors (Table \ref{Blatt})
\begin{equation}
B_{\alpha_i}(m) = \frac{F_L (p_i)}{F_L (p_{\alpha_i})},
\end{equation}
where $L$ is the decay angular momentum, $p$ the $\pi$ or $K$ 
momenta and $p_{\alpha_i}$ their momenta at 
the pole mass $m_\alpha$. The optional real constants $c_{ij}$ allow for a 
background (non-resonating) amplitude\footnote{For the $\pi\pi$ S-wave a factor ($m^2-
2m_\pi^2)/m^2$ is multiplied to the $K$-matrix to ensure a smooth behaviour near  
threshold.}.
In Eq. (\ref{42}) the matrix $\rho(m)$ describes the two-body phase 
space and is diagonal with $\rho_{11}\equiv\rho_1=2p_\pi/m$ and
$\rho_{22}\equiv\rho_2=2p_K/m$. For masses far above 
kinematical threshold $\rho_i\sim 1$ and below $\KKbar$ 
threshold $\rho_2$ becomes imaginary. 

The ($K$-matrix) partial width of resonance $\alpha$ to decay into 
channel $i$ is defined as
\begin{equation}
\Gamma_{\alpha_i}(m_\alpha) = 
\gamma^2_{\alpha_i}\Gamma'_\alpha\rho_i(m_\alpha),
\end{equation}
and the ($K$-matrix) total width as
\begin{equation}
\Gamma_\alpha = \sum_i \Gamma_{\alpha_i}.
\end{equation}
For a resonance with mass far above kinematic thresholds one 
obtains the partial and total widths
\begin{equation}
\Gamma_{\alpha_i} = 
\gamma^2_{\alpha_i}\Gamma'_\alpha, \  \  \ \Gamma_\alpha = 
\Gamma'_\alpha,
\end{equation}
respectively. For one resonance and one channel (elastic scattering) the $K$-matrix 
reduces to 
\begin{equation}
K = \frac{m_0\Gamma'_0 B^2(m)}{m_0^2-m^2},
\end{equation}
and $T$ (Eq. (\ref{42})) reduces to the relativistic Breit-Wigner
\begin{equation}
T = \frac{m_0\Gamma_0B^2(m)/\rho(m_0)}
{m_0^2 - m^2 - im_0\Gamma(m)},
\label{48}
\end{equation}
with
\begin{equation}
\Gamma(m) = \Gamma_0\frac{\rho(m)}{\rho(m_0)} B^2(m).
\end{equation}
For a resonance far above threshold and with $\Gamma_0 << m_0$ 
we get the familiar expression
\begin{equation}
T = \frac{\Gamma_0/2}
{m_0 - m - i\Gamma_0/2}.
\label{82}
\end{equation}
Normally, the mass $m_R$ and width $\Gamma_R$ of a resonance 
are obtained from the poles of the $T$-matrix.  Extending the mass 
$m$ to complex values we find from (\ref{82}) the poles  at
\begin{equation}
m_P = m_R - i\frac{\Gamma_R}{2}, 
\label{45}
\end{equation}
with $m_R=m_0$ and $\Gamma_R=\Gamma_0$. In general, 
however, $m_R$ does not 
coincide with the pole of the $K$-matrix  and $\Gamma_R$ 
is different from the $K$-matrix width. For example, for 
two non-overlapping resonances far 
above threshold, the $K$-matrix, $T$-matrix and 
Breit-Wigner 
poles coincide. As the resonance tails begin to overlap the 
two $T$-matrix poles move towards one another (for an example, see Chung, 
(1995)). 

The $\pi\pi$ S-wave scattering amplitude is related to the $\pi\pi$ phase 
shift $\delta$ and inelasticity $\eta$ through the relation
\begin{equation}
\rho_1(m) T_{11}(m) = \frac{\eta(m)\exp[2i\delta(m)] - 1}{2i}.
\label{50}
\end{equation}
According to Eq. (\ref{42}) the corresponding $K$-matrix then 
reads for 
pure  elastic $\pi\pi$-scattering ($\eta\equiv 1$) 
\begin{equation}
K_{11}(m) = \frac{\tan\delta(m)}{\rho_1(m)}  
\end{equation}
and becomes infinite at $m=m_0$, when $\delta$ 
passes through 90$^\circ$. However, the amplitude $T$ does not,  in general, reach a 
resonance when 
$\delta=\pi/2$. As an example, consider the $\pi\pi$ S-wave 
scattering amplitude described by the amplitude (\ref{50}) in the 
complex plane (Argand diagram):  The intensity $|T|^2$ reaches its 
maximum value around 850 MeV ($\delta$ = $90^{\circ}$), loops 
back and passes rapidly through the $\KKbar$ threshold 
(see Fig. \ref{Argand} below and Au (1987)). At 
$\sim$1000 MeV $|T|^2$ reaches its minimum value ($\delta = 
180^{\circ}$) and then starts a new (inelastic) loop. The 
$f_0(980)$ then appears as a hole in the $\pi\pi$ intensity 
distribution. We shall return to the S-wave Argand diagram when 
discussing the fits to Crystal Barrel data.

Consider now the production of a resonance $\alpha$ in $\pbarp$ 
annihilation. In the isobar model, the  resonance is assumed  not to interact with the 
recoiling system. The coupling strength to $\pbarp$ is 
denoted by the (complex)  constant  $\beta_\alpha$ while 
$\gamma_{\alpha_i}$ describes  its decay strength into channel  
$i$ (say $\pi\pi$ for $i=1$ and $\KKbar$ for $i=2$). 
Following Aitchison (1972) the amplitudes  are given by the components of the vector 
\begin{equation}
{\cal T}  = (1 - i K \rho)^{-1} P.
\label{46}
\end{equation}
The $K$-matrix now describes the propagation of the channel  
$ i$ through the resonances $\alpha$ while the vector $P$ 
describes their production. $P$ and $K$ share the common poles 
$m_\alpha$ so that ${\cal T}$ remains finite at the poles. The vector 
$P$ is given by
\begin{equation}
P_j(m) = \sum_{\alpha}^{} \frac{\beta_{\alpha} \gamma_{\alpha j} 
m_{\alpha} \Gamma'_{\alpha} B_{\alpha_j}(m)}{m_{\alpha}^2-
m^2},
\label{47}
\end{equation}
where the sum runs over all resonances. For a single resonance 
feeding only one decay channel we again obtain from Eq. (\ref{46})  
a Breit-Wigner distribution of the form (\ref{48}) with coupling 
strength $\beta$: 
\begin{equation}
{\cal T} = \frac{\beta m_0\Gamma_0B (m)/\rho(m_0)}
{m_0^2 - m^2 - im_0\Gamma(m)}.
\end{equation}

Let us now assume  a series of resonances with the same quantum numbers 
decaying into two final states. The amplitude for the first final state 
is given by Eq. (\ref{46}):
\begin{equation}
{\cal T}_1  = \frac{(1 - i K_{22}\rho_2)P_1 + i K_{12}\rho_2P_2}{1 - 
\rho_1\rho_2 D - i (\rho_1K_{11}+\rho_2K_{22})},
\label{49}
\end{equation}
with
\begin{equation}
D \equiv K_{11}K_{22} - K_{12}^2.
\end{equation}
As an example, consider a single resonance, say $a_0(980)$ 
decaying to $\eta\pi$ and $\KKbar$. In this case $D\equiv 0$ and 
$B(m)\equiv 1$ (S-wave). We then obtain from Eq. (\ref{49}) the 
formula (Flatt\'e, 1976)
\begin{equation}
{\cal T} (\eta\pi) = \frac{bg_1}{m_0^2-m^2-i(\rho_1g_1^2+\rho_2g_2^2)}, 
\label{53}
\end{equation}
with
\begin{equation}
g_i \equiv \gamma_i\sqrt{m_0\Gamma'_0}  \ \Rightarrow  \  
\sum_{i=1}^2  g_i^2 
= m_0 \Gamma'_0, 
\label{52}
\end{equation}
and
\begin{equation}
b \equiv \beta\sqrt{m_0\Gamma'_0}.
\label{51}
\end{equation}
The phase space factors are
\begin{equation}
\rho_1(m) = \frac{2}{m} p_\eta \ \ {\rm and} \ \ 
\rho_2(m) = \frac{2}{m} p_K =  \sqrt{1-\frac{4m_K^2}{m^2}}.
\end{equation}
The  ${\cal T}(\KKbar)$ amplitude is obtained 
by interchanging the labels 1 and 2 in Eq. (\ref{53}). Below 
$\KKbar$ 
threshold $\rho_2$ becomes imaginary. Compared to pure 
$\eta\pi$ decay this leads to a shift of 
the resonance peak and to a narrower and asymmetric distribution  
of the observed signal in the $\eta\pi$  channel. 
This is shown in Fig. \ref{Flatte} for $g_1$ = 0.324 GeV, 
$g_2$ = 0.279 GeV (hence $\Gamma_0'$ = 0.43 GeV). These 
parameters have been extracted from the $a_0(980)$ contribution to  
$\pbarp\rightarrow\eta\pi^0\pi^0$ and $\KKbar\pi$ (section \ref{sec:KLKP}). A width 
of  54.12 $\pm$ 0.36 MeV was determined directly from the $a_0(980)\rightarrow\eta\pi$ 
signal in the annihilation channel $\omega\eta\pi^0$ (Amsler, 1994c), in good agreement 
with the observed width in Fig. \ref{Flatte}. Also shown in Fig. \ref{Flatte} is 
the expected distribution for  $\Gamma_0'=0.43$ 
GeV,  assuming no $\KKbar$ decay. The observed width 
$\Gamma_0 = \Gamma_0'\rho_1(m_0)$ increases to  0.28 GeV.

The standard procedure in the analysis of Crystal Barrel data is to 
replace the Breit-Wigner function (\ref{55}) by ${\cal T}$ (Eq. (\ref{46})) and to fit the 
parameters $g_{\alpha_i}$, $\beta_\alpha$ and $m_\alpha$. The 
resonance parameters $m_R$ and $\Gamma_R$ are then extracted 
by searching for the complex poles (Eq. (\ref{45})) of the matrix ${\cal T}$. 
A one-channel resonance appears as a pole in the second 
Riemann-sheet and a two-channel resonance manifests itself as a pole in the 
second or third Riemann-sheet (Badalyan, 1982). 

Some of the Crystal Barrel Dalitz plots have also been analyzed using the N/D formalism 
(Chew and Mandelstam, 1960) which takes into account the direct production of three 
mesons and also the final state interaction. In this formalism the amplitude ${\cal T}$  has 
the same denominator as, e.g., in Eq. (\ref{49}), but the numerator allows for additional 
degrees of freedom (Bugg, 1994).

\section{Annihilation at Rest into Three Pseudoscalars}
\label{sec:three}
Proton-antiproton annihilation at rest into three pseudoscalars  
is the simplest process to search for scalar resonances $0^{+}\rightarrow  0^{-}0^{-}$, the 
recoiling third pseudoscalar removing the excess energy. The annihilation rates for these 
processes in liquid hydrogen ($^1S_0$ atomic state) are reasonably large since no angular 
momentum barrier is involved. Channels with three pseudoscalars have 
been studied earlier, but essentially in the 2-prong configuration and with limited 
statistics, e.g  $\pi^+\pi^-\pi^0$  with 3,838 events (Foster, 1968b), 
$\pi^+\pi^-\eta$  with 459 events (Espigat, 1972) and $\pi^+\pi^-\etap$ with 104 events  
(Foster, 1968a). The samples collected by the Asterix collaboration at LEAR (May, 1989; 
Weidenauer, 1990) are larger but were collected from 
atomic P-states. The $\pi^+\pi^-
\pi^0$ final state revealed the existence of a tensor meson, $f_2(1565)$, produced from 
P-states (May, 1989, 1990). In the kaonic sector, data were collected in bubble chambers  
for the final states $\pi^{\pm}K^{\mp}K_S$ (2,851 events) and $\pi^0K_SK_S$ (546 events) 
in the experiments of Conforto (1967) and Barash (1965). Branching ratios for 
annihilation into three mesons are listed in Table \ref{TBBR}.

Annihilation with charged pions is dominated  by $\rho(770)$ production which 
complicates the spin-parity analysis of underlying scalar resonances in the 
$\pi\pi$ S-wave.  Also, both $^1S_0$ and $^3S_1$ atomic states contribute. All-neutral 
(0-prong) channels are therefore simpler to analyze but more complex to select due to the 
large $\gamma$-multiplicity. The channel $\pi^0\pi^0\pi^0$ with 2,100 events 
has been reconstructed earlier with optical spark chambers (Devons, 1973). The existence 
of a scalar resonance decaying to $\pi\pi$ with mass 1527 and width 101 MeV was 
suggested in the $3\pi^0$ channel  and in its $\pi^-\pi^-\pi^+$ counterpart in $\pbarn$ 
annihilation in deuterium (Gray, 1983).  This was actually the first sighting of $f_0(1500)$ 
which will be discussed  below.

The sizes of the data samples have been vastly increased by Crystal Barrel. We shall first   
review annihilation into three neutral non-strange mesons. We start from 
$6\gamma$ final states and select the channels $\pbarp\rightarrow \pi^0\eta\eta, 3\pi^0, 
\pi^0\pi^0\eta$,  $\pi^0\eta\etap$ and $\pi^0\pi^0\etap$ by assuming total energy and 
total momentum conservation and constraining the $2\gamma$ masses to $\pi^0, \eta$ and 
$\etap$ decays (7C fits), excluding any other possible  
configuration: Events are accepted if the kinematic fit satisfies the assumed 
three-pseudoscalar hypothesis with a confidence level typically larger than 10\%. 
Background from the other 
$6\gamma$ channels  is suppressed by rejecting those events that also satisfy 
any other $6\gamma$ final state hypothesis (including the strong 
$\omega\omega$, $\omega\rightarrow\pi^0\gamma$) with a  
confidence level of at least 1\%. The absolute branching ratios for the 
$6\gamma$ channels are determined by normalyzing on the branching ratio 
for $\pbarp\rightarrow\omega\omega$. These three-pseudoscalar channels have all been 
analyzed and we now review the salient features in their Dalitz plots. Results on kaonic 
channels are appended to the next sections.

\subsection{$\pbarp\rightarrow\pi^0\eta\eta$}
The first evidence for two $I=0$ scalars in the 1400 MeV mass region, now 
called $f_0(1370)$  and $f_0(1500)$, was obtained from a 
reduced sample of 
$2.3\times 10^4$ $\pi^0\eta\eta$ events (Amsler, 1992c).  The  invariant mass 
distributions are shown in Fig.~\ref{etaetapi}. The two scalars decaying to 
$\eta\eta$  are also observed when one $\eta$ decays to $3\pi^0$ ($10\gamma$ 
final state), a channel with entirely different systematics (Fig.~\ref{etaetapi}(b)). The 
distributions in Fig.~\ref{etaetapi}(a) and (b) are nearly identical. 
 
An amplitude analysis of the Dalitz plot distribution for the $6\gamma$ final 
state was  performed with the method outlined in the 
previous section. However, Breit-Wigner functions of the form 
(\ref{55}) were used to describe the resonances. The fit required 
$J^{PC}=0^{++}$ for both $\eta\eta$ resonances. The (Breit-Wigner) masses and 
widths were $m$ =1430, $\Gamma$ = 250 and  $m$ = 1560,  $\Gamma$ = 245 MeV, 
respectively.  Note that the width 
of the upper state appears smaller in Fig.~\ref{etaetapi}(a,b),  due to interference 
effects. This state may be identical to the scalar meson observed earlier by the GAMS 
collaboration at 1590 MeV in the  $\eta\eta$ and $\eta\etap$ mass spectrum of 
high energy $\pi N$-interactions (Alde, 1988b; Binon, 1983, 1984).

The final analysis of this channel was performed with a tenfold increase in 
statistics, namely $3.1\times 10^4$ $\pi^0\eta\eta$ events from 0-prong data 
and $1.67\times 10^5$ $\pi^0\eta\eta$ events from a triggered data sample  
requiring online one $\pi^0$ and two $\eta$ mesons (Amsler, 1995e).  
The Dalitz plot is shown in Fig.~\ref{Dalitz}(a). The horizontal and vertical bands 
are due to  $a_0(980)$ decaying to $\eta\pi^0$. One also observes diagonal 
bands which correspond to the two  states decaying to $\eta\eta$. A residual 
incoherent flat background of 5\%, mainly due to $\pi^0\pi^0\omega\rightarrow 
7\gamma$ with a missing photon, has been subtracted from the Dalitz plot 
before applying the amplitude analysis, this time with the 
full $K$-matrix formalism. 

Since S-wave dominates in liquid, the channel $\pi^0\eta\eta$ proceeds mainly 
through the $^1S_0$ atomic state. The $\eta\pi$ S-wave was parametrized by a 
2$\times 2$ $K$-matrix with poles from 
$a_0(980)$ and  $a_0(1450)$. The parameters were taken from the 
$\pi^0\pi^0\eta$ analysis (section \ref{secppe}), leaving the production 
constants $\beta$ free. A contribution from $a_2(1320)$ ($\eta\pi$ D-wave) with fixed 
mass and 
width was also offered to the fit. The  $\eta\eta$ S- and D-waves were described 
by one-channel $K$-matrices. Annihilation from atomic P-states was not 
included in the fit except for tensor mesons (e.g. the expected 
$f_2(1565)$) decaying to $\eta\eta$. In fact the fit demands a contribution 
from a tensor meson with mass $\sim$1494 and width $\sim$155 MeV, produced mainly 
from P-states.

The best fit was obtained with two poles for the $\eta\eta$ S-wave. The 
resonance parameters ($T$-matrix poles) are:
\begin{eqnarray}
f_0(1370): m & = & 1360 \pm 35, \  \Gamma = 300 - 600 \ {\rm MeV},\nonumber\\
f_0(1500): m & = & 1505 \pm 15, \  \Gamma = 120 \pm 30 \ {\rm MeV}.
\end{eqnarray}
The $K$-matrix mass and width of $f_0(1500)$ are 1569 and 
191 MeV, respectively, in accord with the Breit-Wigner parameters of the 
GAMS resonance (Binon, 1983). This state, previously called $f_0(1590)$, 
and $f_0(1500)$ are therefore assumed to be identical. The branching ratio for 
$\pi^0\eta\eta$ is given in Table \ref{TBBR} and the products of  resonance production 
and decay  branching ratios  are listed in Table \ref{DPBR}.

\subsection{$\pbarp\rightarrow\pi^0\pi^0\eta$}
\label{secppe}
This channel is relevant to  search for 
isovector $0^{++}$ states decaying to $\eta\pi$.
The $\pi^0\pi^0\eta$ Dalitz plot ($2.8\times 10^5$ events) is 
shown in Fig.~\ref{Dalitz}(b). Qualitatively, one observes 
$a_0(980)$ and $a_2(1320)$ decaying to $\eta\pi$ and $f_0(980)$ decaying 
to $\pi\pi$. The strong interference patterns point to coherent contributions from a 
single $\pbarp$ atomic state ($^1S_0$).

An amplitude analysis based on the $K$-matrix formalism (and, alternatively, 
the N/D formalism) has been  performed, 
assuming pure S-wave annihilation (Amsler, 1994b). The $\pi\pi$ S-wave was 
described by 
two poles, one for $f_0(980)$, coupling to $\pi\pi$ and $\KKbar$, 
and one for $f_0(1370)$. Elastic $\pi\pi$-scattering data (Grayer, 1974; 
Rosselet, 1977) were included in the fit. The $\pi\pi$ D-wave 
($f_2(1270)$) was also introduced but turned out to be negligible. The $\eta\pi$ D-wave was 
described by one pole for $a_2(1320)$. 
The $\eta\pi$ S-wave was described by a one-pole $2\times 2$ $K$-matrix 
for $a_0(980)$ with couplings $g_1$ to $\eta\pi$ and $g_2$ to $\KKbar$. Since decay to 
$\KKbar$ was not measured, $g_2$ was obtained indirectly from the $\eta\pi$ line shape. 
The fit yielded $g_1$ = 0.353 GeV and
\begin{equation}
\frac{g_2}{g_1} \sim 0.88.
\label{88}
\end{equation}
These amplitudes were, however, not sufficient to 
describe the data.  A satisfactory fit was obtained by adding (i) a second pole 
to the $\eta\pi$ S-wave,  (ii) a second pole to the 
$\eta\pi$ D-wave and (iii) an $\eta\pi$ P-wave. Branching ratios are 
given in Table \ref{DPBR}. The branching ratio for $\pi^0\pi^0\eta$ (Table \ref{TBBR}) is 
in excellent agreement with the one derived from the channel 
$\pi^0\pi^0\eta\rightarrow\pi^+\pi^- 3\pi^0$ (Amsler, 1994d).

The main result was the observation of a new
isovector scalar resonance in the $\eta\pi$ S-wave:
\begin{equation}
a_0(1450):  m = 1450 \pm 40,  \  \Gamma = 270 \pm 40 \ {\rm MeV}.
\end{equation}
This resonance manifests itself as a depletion in the bottom right (or top left)
corner of the Dalitz plot (Fig. \ref{Dalitz} (b)). Evidence for the $a_0(1450)$ decaying to 
$\eta\pi$ is also reported in an analysis of the channel $\pi^+\pi^-\eta$ which requires, 
in addition, an amplitude for $\rho\eta$ production from $^3S_1$ (Abele, 1997a).

The $\eta\pi$ D-wave contribution 
corresponds to a $2^{++}$ resonance around 1650 MeV, called $a_2'(1650)$ in Table 
\ref{DPBR}, with a width of about 200 MeV. This state could be the radial excitation of 
$a_2(1320)$. The mass and width of the structure in the $\eta\pi$ P-wave (exotic $1^{-+}$)  
are  not  well defined. They vary from 1200 to 1600 MeV and from 
400 to 1000 MeV, respectively, without significant changes in the $\chi^2$. The $1^{-+}$  
$\hat\rho(1405)$ reported by Alde 
(1988a) is therefore not confirmed by the present data.  

\subsection{$\pbarp\rightarrow\pi^0\pi^0\pi^0$}
\label{sec:threepi0}
The first analysis of this channel used a sample of only $5.5\times 
10^4$ events and reported an isoscalar $2^{++}$ meson at 1515 $\pm$ 10  MeV with width 
120 $\pm$ 10 MeV, decaying to $\pi^0\pi^0$ (Aker, 1991). This state was identified with  
$f_2(1565)$ that had been observed before 
in the final state $\pi^+\pi^-\pi^0$ in hydrogen gas  (May, 1989, 1990). P-wave 
annihilation  from 
$^3P_1$ and $^3P_2$ was therefore allowed when fitting the $3\pi^0$ channel.
Resonances in the $\pi\pi$ 
S-wave were described by the $\pi\pi$ elastic scattering amplitude, replacing the Breit-
Wigner amplitude by 
\begin{equation}
BW_0(m) = \frac{m}{p}\left(\frac{\eta(m)\exp[2i\delta(m)] - 1}{2i}\right),
\label{61}
\end{equation}
according to Eq. (\ref{50}), where $\delta$ and $\eta$ were 
taken from the Argand diagram of Au (1987). This is an  
approximation assuming equal production strengths for all 
resonances in the $\pi\pi$ S-wave, which is reasonable for the $3\pi^0$ channel, as I will 
show below.

A statistical sample an order of magnitude larger then revealed a new feature (Amsler, 
1994e) which was consolidated by a reanalysis of the 
early Crystal Barrel  data (Anisovich, 1994): the presence in the 
Dalitz plot of a narrow homogeneously populated band from a scalar resonance, 
$f_0(1500)$,  decaying to $2\pi^0$. The $3\pi^0$ Dalitz plot  is shown in Fig. 
\ref{Dalitz}(c) and the $2\pi^0$ mass projection in Fig.~\ref{pipipi}.
  
Qualitatively, one observes the following features: the population along the $\pi\pi$ mass 
band  marked  $f_2(1270)$  increases  at the edges of the Dalitz plot indicating that one 
decay $\pi^0$ is preferably emitted along  the flight direction of the 
resonance. This is typical of a spin 2 resonance decaying with the 
angular distribution  $(3\cos^2\theta -1)^2$ from $^1S_0$ or $(1+3\cos^2\theta)$ 
from $^3P_1$. The blobs labelled $f_2(1565)$ at the 
corners correspond to an angular distribution sin$^2\theta$ from another 
spin 2 resonance  produced from $^3P_2$, together with constructive interference from 
the two $\pi\pi$ S-waves. The $f_0(980)$ appears as a narrow dip in the $\pi\pi$ S-wave. 
The new feature is the homogeneous narrow band marked  $f_0(1500)$ which must be due 
to a spin 0 state.

The analysis of the full data sample was performed with the $K$-matrix 
formalism (Amsler, 1995f). A 2 $\times$ 2 $K$-matrix with three poles was 
sufficient to describe the $\pi\pi$ S-wave. Elastic  $\pi\pi$ scattering data up 
to 1200 MeV from Grayer (1974) and Rosselet (1977) were also included in the 
fit. The contributing scalar resonances are $f_0(980)$ and
\begin{eqnarray}
f_0(1370): \  m & \simeq &  1330, \ \Gamma  \sim  760 \ {\rm 
MeV},\nonumber\\
f_0(1500): \  m & = &  1500 \pm15, \ \Gamma  = 120 \pm 25 \ {\rm 
MeV}.
\end{eqnarray}
A 4-pole $K$-matrix helps to constrain the $f_0(1370)$ parameters, giving  
\begin{equation}
f_0(1370): \ m =  1330 \pm 50, \ \Gamma = 300 \pm 80 \ {\rm 
MeV},
\end{equation}
and the 4th pole corresponds to  a 600 MeV broad structure around 1100 MeV (called 
$f_0(400-1200)$ by the Particle Data Group (Barnett, 1996)) and also reported by a recent 
reanalysis of $\pi\pi$ S-wave data (Morgan and Pennington, 1993). The data are therefore 
compatible with 3 or 4 poles and it is not obvious that $f_0(1370)$ and $f_0(400-1200)$ are 
not part of the one and the same object. 

A one-dimensional $K$-matrix describes the $\pi\pi$ D-wave. In  addition to $f_2(1270)$ 
one finds
\begin{equation}
f_2(1565): \  m \sim 1530, \ \Gamma  \sim 135 \ {\rm 
MeV}.
\end{equation}
The fractional contribution of P-waves is 46\%. Without P-waves the fit deteriorates 
markedly but the $f_0(1370)$ and $f_0(1500)$ parameters remain stable. We shall return to 
the $f_2(1565)$ in the discussion  below. Branching ratios are given in Table \ref{DPBR}.

\subsection{Coupled channel analysis}
\label{sec:coupled}
A simultaneous fit was performed to the channels
$\pi^0\eta\eta$, $\pi^0\pi^0\eta$ and $3\pi^0$ (Amsler, 1995g) using the full data samples
presented in the previous sections with, in addition, the $\pi\pi$-scattering 
data up to 1200 MeV. However, pure S-wave annihilation was 
assumed.  A 3 $\times$ 3 $K$-matrix with 4 poles was used to describe the $\pi\pi$ S-wave 
coupling to $\pi\pi$, $\eta\eta$ and the at that time unmeasured $\KKbar$ through the 
resonances $f_0(980)$, $f_0(1370)$, $f_0(1500)$, taking also into account the 
broad structure around 1100 MeV. Common $\beta_\alpha$ parameters (Eq. (\ref{47})) 
were introduced to describe the production of resonances  associated 
with the same recoiling particle. For example, $f_0(1500)$ recoiling against $\pi^0$ is 
produced with the same 
strength in $\pi^0\eta\eta$ and $3\pi^0$. The $\eta\pi$ S-wave was 
described by a 2 $\times$ 2 $K$-matrix for $a_0(980)$ and $a_0(1450)$. The 
$\pi\pi$, $\eta\eta$ and $\eta\pi$ D-waves were treated with one dimensional 
$K$-matrices, the latter including $a_2(1320)$ and $a_2'(1650)$ with pole parameters 
taken from the $\pi^0\pi^0\eta$ analysis of section \ref{secppe}.

The branching ratios are given in Table \ref{DPBR}. Note that the $\pi\pi$ S-wave 
includes contributions from $f_0(400-1200)$, $f_0(980)$ and $f_0(1370)$ (but excluding 
$f_0(1500)$) which cannot be disentangled due to interferences.  
The fit is in good agreement with the single channel analyses and constrains the 
resonance parameters.  Hence a 
consistent description of all three annihilation channels was achieved with the 
following main features:
\begin{enumerate}
\item
The data demand three scalar resonances in the 1300 - 1600 MeV region:
\begin{eqnarray}
a_0(1450): \  m & = &  1470 \pm 25, \ \Gamma  = 265 \pm 30 \ {\rm 
MeV},\nonumber\\
f_0(1370): \  m & = &  1390 \pm 30, \ \Gamma  = 380 \pm 80 \ 
{\rm MeV},\nonumber\\
f_0(1500): \  m & = &  1500 \pm 10, \ \Gamma  = 154 \pm 30 \ {\rm 
MeV}.
\end{eqnarray}
\item
The broad scalar structure around 1100 MeV ($f_0(400-1200)$) has very different pole 
positions in sheets II and III, making a resonance interpretation of this state difficult. 
\item
The production data demand a larger width ($\simeq$ 100 MeV) for $f_0(980)$ 
than the $\pi\pi$ scattering data alone ($\simeq$ 50 MeV, according to 
Morgan and Pennington (1993)).
\item
A tensor meson is observed in the $\pi\pi$  D-wave with mass 1552 
and width 142 MeV, in accord with May (1889, 1990) notwithstanding the absence of 
atomic P-waves in the present analysis.  A structure is also required in this mass range in 
the $\eta\eta$ D-wave. This state is not compatible with $f_2^\prime (1525)$ which is 
produced with a much lower rate, as we shall see in section \ref{sec:KLKL}.
\end{enumerate}

The $\pi\pi$ scattering amplitude $T$ (Eq. (\ref{50})) obtained from the fit to the 
elastic scattering data and the Crystal Barrel data is shown in Fig. 
\ref{Argand}. Note that Crystal Barrel $\KKbar$ data are not yet included and therefore 
the third $K$-matrix channel also contains by default all other unmeasured inelasticities. 
Figure \ref{Intense} shows the $I=0$ S-wave production intensity  
$|{\cal T}|^2$ (Eq. (\ref{46})) for the three annihilation channels. The maxima 
around 1300 and 1550 MeV correspond to the $K$-matrix poles for $f_0(1370)$ and 
$f_0(1500)$. 
It is instructive to compare the dip around 1000 MeV in the $\pi\pi$ S-wave for the 
$3\pi^0$ channel  to the peak in the  $\pi^0\pi^0\eta$ channel, both due to $f_0(980)$. This 
is produced by interferences between the $\pi\pi$ S-waves in $3\pi^0$ which have the 
opposite sign to the interference between the $\pi\pi$ and the $\eta\pi$ S-waves in 
$\eta\pi^0\pi^0$. The $\pi\pi$ 
S-wave in $\pi\pi$ scattering  exhibits qualitatively the 
same behaviour as in $\pbarp$ annihilation into $3\pi^0$, namely sharp minima around 
1000 and 1450 MeV. The ansatz (\ref{61}) used in several 
Crystal Barrel analyses (e.g. in Aker (1991)) for the $\pi\pi$ S-wave is 
therefore a rough but reasonable approximation.  

\subsection{$\pbarp\rightarrow\pi^0\eta\etap$}
Another piece of evidence for $f_0(1500)$ stems from $\pi^0\eta\etap$ 
(Amsler, 1994f). This channel was also  reconstructed from the 6$\gamma$ 
final state. Since $\etap$ decays to $\gamma\gamma$ with a branching ratio of 
only 2.1\% the data sample is small (977 events). Most of these events  
were collected with the online trigger requiring one $\pi^0$, one $\eta$ and 
one $\etap$. The $\pi^0\eta\etap$ Dalitz  plot shows an accumulation of events at small 
$\eta\etap$ masses. Figure 
\ref{peep} shows the $\eta\etap$ mass projection. The $\eta\etap$ mass spectrum from the 
same annihilation channel, but with $\etap\rightarrow\eta\pi^+\pi^-$, has  
entirely different systematics but is identical (inset in Fig. \ref{peep}).
The enhancement at low masses is due to a scalar resonance since the angular distribution 
in the $\eta\etap$ system is isotropic. A maximum likelihood fit was performed to the 
$6\gamma$ channel using a (damped) Breit-Wigner according to Eqs. (\ref{55}),  
(\ref{83}) and a flat incoherent background. The resonance parameters are
\begin{equation}
f_0(1500):  \ m =  1545 \pm 25, \ \Gamma =  100 \pm 40 \ {\rm MeV}.
\label{78}
\end{equation}
The branching ratio is given in Table \ref{DPBR}. The $f_0(1500)$ mass is somewhat 
larger than for $3\pi^0$ and $\pi^0\eta\eta$.  However, Abele (1996a) points out that a 
constant width in the denominator of the Breit-Wigner function yields a mass around 1500 
MeV. This is a more realistic procedure since the total width at the $\eta\etap$ threshold 
remains finite due to the channels $\pi\pi$ and $\eta\eta$. However, this does not 
modify the branching ratio for $f_0(1500)\rightarrow\eta\etap$. We shall therefore 
ignore Eq. (\ref{78}), when averaging below the $f_0(1500)$ mass and width.
 
\subsection{$\pbarp\rightarrow\pi^0\pi^0\etap$}
With 0-prong data this final state is accessible through $\etap\rightarrow 
2\gamma$ or $\etap\rightarrow\eta\pi^0\pi^0$, leading to 6, respectively 10 
photons (Abele, 1997g). The branching ratios from both final state configurations agree 
(Table \ref{TBBR}). A sample of  8,230 10$\gamma$ events were kinematically fitted to 
$4\pi^0\eta$.  The $\pi^0\pi^0\eta$ mass distribution (Fig. \ref{fourpe}) shows a sharp 
signal from $\etap$ and a shoulder around 1400 MeV due to the $E$ meson  that will be 
discussed in section \ref{seciota}.

The $\pi^0\pi^0\etap$ Dalitz plot was obtained by selecting events in the $\etap$ peak and 
subtracting background Dalitz plots from either sides of the peak. It shows a broad 
accumulation of events in its center which can be described by a dominating $\pi\pi$  
S-wave and small contributions from $a_2(1320)$ and $a_0(1450)$ with branching ratios 
given in Table \ref{DPBR}. These resonances  are included in the fit with fixed mass and 
width. The ratios of rates for $a_2(1320)$ and 
$a_0(1450)$ decays into $\eta\pi$ and  $\etap\pi$ can be predicted from SU(3) and 
compared with measurements. This will be discussed in section 
\ref{sec:new}. The analysis of the 6$\gamma$ Dalitz plot (3,559 events) leads to similar 
results. Figure \ref{ppetpfit} shows that $a_0(1450)$ is required for a satisfactory 
description of this annihilation channel.

\subsection{$\pbarp\rightarrow\pi^0K_LK_L$}
\label{sec:KLKL}
The isoscalar $f_0(1500)$ has been observed to decay into $\pi^0\pi^0$, 
$\eta\eta$ and $\eta\etap$. To clarify its internal structure it was essential to also 
search for its $\KKbar$ decay mode. In a previous bubble chamber experiment 
(Gray, 1983) no $\KKbar$ signal had been observed in the 1500 MeV region in 
$\pbarp$ annihilation into $\KKbar\pi$, leading to the conclusion that the $f_0(1500)$ 
coupling to $\KKbar$ is suppressed (Amsler and Close, 1996). However,
no partial wave analysis was performed due to limited statistics.
  
Crystal Barrel has therefore searched for $f_0(1500)$ in the annihilation 
channel $\pi^0K_LK_L$ (Abele, 1996b; Dombrowski, 1996) which 
proceeds from the $^1S_0$ atomic state.   All-neutral events were selected with three 
energy clusters in the barrel and the channel $\pbarp\rightarrow\pi^0K_LK_L$ could be 
reconstructed by measuring the $\pi^0(\rightarrow 2\gamma)$ momentum and 
the direction of one of the $K_L$ which interacts hadronically in the 
CsI barrel. The main background contribution was due to events with a 
reconstructed $\pi^0$ and one additional $\gamma$ which happen to fulfil the 
$\pi^0K_LK_L$ kinematics but for which one or more $\gamma$'s have 
escaped detection. This background ($\sim$ 17\%) could be removed  
by subtracting a Dalitz plot constructed from $2\gamma$ pairs with invariant masses just 
below or above the $\pi^0$ mass. 

Further background contributions were due to 
$\pbarp\rightarrow\omega K_{L}K_{L}$ where $\omega$ decays to 
$\pi^{0}\gamma$ and both $K_{L}$ are not detected, and 
$\pbarp\rightarrow K_{S}K_{L}$ where one photon from 
$K_{S}\rightarrow\pi^{0}\pi^{0}\rightarrow 4\gamma$ and the 
$K_{L}$ are undetected. These events can be removed with 
appropriate mass cuts. Background from final states like 
$\omega\eta$, $\omega\pi^0$ and $3\pi^{0}$ were studied by Monte Carlo 
simulation. The total residual background contamination was $(3.4\pm 0.5)\%$.

The background subtracted Dalitz plot is shown in Fig. \ref{Dalitz}(d).  This plot has not 
been symmetrized with respect to the diagonal axis since one $K_L$ is detected (vertical 
axis) whilst the other is missing (horizontal axis). The interaction probability as a 
function of $K_L$ momentum was measured by comparing the intensities along 
the two $K^*$ bands, $K^*\rightarrow K_L\pi^0$. The interaction probability was found to 
be flat with $K_L$ momentum, but increasing slowly below 300 MeV/c (Dombrowski, 1996).
The $\pi^0K_LK_L$ Dalitz plot shown in Fig. \ref{Dalitz}(d) is already corrected for 
detection efficiency and for $K_L$ decay between the production vertex and 
the crystals. Its symmetry along the diagonal axis is nearly 
perfect, indicating that acceptance and backgrounds have been 
taken into account properly.

One observes signals from $K^*(892)$ decaying to $K\pi$ and 
$a_2(1320)/ f_2(1270)$ decaying to $\KKbar$. The accumulation of events at 
the edge for small $\KKbar$ masses is due to the tensor $f_2'(1525)$   
which is observed for the first time in $\overline{p}p$ annihilation at rest.
These resonances and the broad $K_0^*(1430)$ were introduced in a first 
attempt to fit the Dalitz plot\footnote{The 
absence of threshold enhancement from $a_0(980)$ or $f_0(980)$ at the upper 
right border of the Dalitz plot could be due to destructive interference between these states 
and/or to the loss of acceptance close to the 
$\KKbar$ threshold.}. The $K$-matrix for the $K\pi$ ($I$ = 1/2) S-wave 
was written as
\begin{equation}
K = \frac{am}{2+abp^2} + \frac{m_0\Gamma_0/\rho_0}{m_0^2-m^2},
\label{62}
\end{equation}
where the first term describes the low energy $K\pi$ scattering ($a$ is the 
scattering length and $b$ the effective range) and the second term describes 
the  $K_0^*(1430)$ resonance.  The parameters $m_0, \Gamma_0, a$ and 
$b$ were determined by fitting the scattering amplitude $T$ (Eq. (\ref{42})) to the $K\pi$ 
phase shifts (Aston, 1988b). The fit is shown in Fig. \ref{LASS}. Note that resonance occurs 
at $\delta\simeq 120^{\circ}$.  The corresponding mass and width ($T$-matrix pole) for 
$K_0^*(1430)$ are $m = (1423 \pm 10)$ MeV and $(277 \pm 17)$ MeV, in close 
agreement with Aston (1988b) who used a different parametrization and found $m_0 = 1429 
\pm 6$ and $\Gamma_0 = 287 \pm 23$ MeV.

A one-pole $K$-matrix for a scalar resonance 
was added for the peak around 1450 MeV in the $K\overline{K}$ mass distribution (Fig. 
\ref{KLKLp}). The fit clearly fails to describe the $K\overline{K}$ mass spectrum (dashed 
line in Fig. \ref{KLKLp}). The second attempt assumed a $K$-matrix with two scalar 
resonances, 
$f_0(1370)$ and  $f_0(1500)$, in the $K\overline{K}$ amplitude. The fit now 
provided a satisfactory description of the Dalitz plot and the $\KKbar$ mass 
spectrum (full line in Fig. \ref{KLKLp}). However, the isovector 
$a_0(1450)$ is also expected to decay to $K_LK_L$ and one cannot 
distinguish between isovectors and isoscalars from the $\pi^0K_LK_L$ data 
alone. Therefore a Breit-Wigner was added for the isovector $a_0(1450)$ with 
fixed mass and width taken from the coupled channel analysis (section \ref{sec:coupled}). 
The $f_0(1370)$ and  
$f_0(1500)$ poles are stable, 
nearly independent of $a_0(1450)$ contribution. One finds
\begin{eqnarray}
f_0(1370): \ m & = &  1380 \pm 40, \ \Gamma  = 360 \pm 50 \ 
{\rm MeV},\nonumber\\
f_0(1500): \ m & = &  1515 \pm 20, \ \Gamma  = 105 \pm 15 \ {\rm 
MeV},
\end{eqnarray}
in agreement with the resonance parameters measured in the other decay 
channels. 

Due to uncertainties in the $K_L$ interaction probability the 
branching ratio for $\pi^0K_LK_L$  could not be determined accurately. The 
branching ratios given in Table \ref{DPBRK} (Abele, 1996b; Dombrowski, 1996) are 
therefore normalized to the known branching ratio for $\pi^0K_SK_S$ (Armenteros, 1965; 
Barash, 1965)\footnote{The isospin contributions from 
$^1S_0$ to the $K^*\overline{K}$  system cannot be determined in this channel since the 
$K^*$ bands interfere constructively for both $I=0$ and $I=1$.}.

The branching ratios for $\pbarp\rightarrow f_0(1370)$ and  
$f_0(1500)\rightarrow K_LK_L$ also vary with $a_0(1450)$ contribution
(Fig. \ref{correl}). In Table \ref{DPBRK} the branching ratios 
for $f_0(1370)$ and $f_0(1500)$ are therefore derived from Fig. \ref{correl}  assuming an 
$a_0(1450)$ contribution $r_0$ to the $\pi^0K_LK_L$ final state, derived from  its measured 
contribution $r = (10.8 \pm 2.0) \%$ to $\pi^{\pm}K^{\mp}K_L$ (next section):
\begin{equation}
r_0 = \frac{r}{4}\frac{B(\pbarp\rightarrow 
\pi^{\pm}K^{\mp}K_S)}{B(\pbarp\rightarrow \pi^0K_SK_S)} = (9.8 \pm 1.9) \%,
\end{equation}
where we have used the branching ratios from Armenteros (1965) and Barash (1965).

\subsection{$\pbarp\rightarrow\pi^{\pm}K^{\mp}K_L$}
\label{sec:KLKP}
This channel selects only isospin 1 resonances decaying to $\KKbar$ and 
therefore permits a direct measurement of the contribution from isovectors to 
$\KKbar\pi$, in particular from $a_0(1450)$.  Crystal Barrel has studied the reaction 
$\pbarp\rightarrow \pi^{\pm}K^{\mp}K_L $ with a non-interacting $K_L$
(Heinzelmann, 1996; Abele 1997e). This channel is selected from 7.7$\times 10^6$ triggered 
2-prong data by requiring exactly  two clusters in the barrel from $\pi^{\pm}$ and 
$K^{\mp}$.  
Particle identification is achieved by measuring the  ionisation in the drift 
chamber (Fig. \ref{dedx}) and a (1C) kinematic fit ensures momentum and 
energy conservation. The background contribution, mainly from $\pi^+\pi^-\pi^0$, is 
about 2\%. The Dalitz plot (Fig. \ref{KKp}) has been corrected for 
background and acceptance, and for the $K_L$ interaction probability. The latter was 
determined by reconstructing the channel $\pi^0K_S(\rightarrow\pi^+\pi^-)K_L$ with 
and without missing $K_L$. The branching ratio 
\begin{equation}
B(\pbarp\rightarrow \pi^{\pm}K^{\mp}K_L) = (2.91 \pm 0.34)\times 10^{-3}
\end{equation}
is in excellent agreement with the one given in Table \ref{DPBRK} for 
$\pi^{\pm}K^{\mp}K_S$ from bubble chamber experiments (Armenteros, 1965; Barash, 
1965).

The Dalitz plot shows clear signals from $K^*(892)$, $a_2(1320)$ and $a_0(980)$.
The $K^*(892)$ and $a_2(1320)$ were parametrized by Breit-Wigner functions.
The $a_0(980)$ was described by a 2 $\times$ 2 $K$-matrix (Flatt\'e formula 
(\ref{53})). The 
$K\pi$ S-wave  ($K^*_0(1430)$) was treated using the data from Aston (1988b),  
as described in the previous section. In contrast to $\pi^0K_LK_L$, both atomic 
states $^1S_0$ and $^3S_1$ contribute. The $I=0$ and $I=1$ contributions to 
$K^*(892)\overline{K}$ can be determined from the (destructive) interference pattern 
around the crossing point of the $K^*$ bands in Fig. \ref{KKp}. 

The fit, however, did not provide a satisfactory description of the Dalitz plot and the fitted 
$a_2(1320)$ mass, 1290 MeV, was significantly lower than the table value, a problem that 
had been noticed earlier in bubble chamber data (Conforto, 1967). A 
substantial improvement in the $\chi^2$ (Fig. \ref{a0toKK}) was obtained when 
introducing the $a_0(1450)$ as a second pole in the $K$-matrix, together with $a_0(980)$,  
leading to the resonance parameters ($T$-matrix pole)
\begin{equation}
a_0(1450): m = 1480 \pm 30, \ \ \ \Gamma = 265 \pm 15  \ {\rm  MeV},
\end{equation}
in agreement with the $\eta\pi$ decay mode. The $a_2(1320)$ mass now became compatible 
with the table value (Barnett, 1996). We shall show below that the contribution to 
$\pi^{\pm}K^{\mp}K_L $ of  (10.8 $\pm$ 2.0) \% (Fig. \ref{a0toKK}) is consistent with 
predictions from the $\pi^0\pi^0\eta$ channel, using SU(3). 

An even better fit was achieved by adding a broad structure in the $\KKbar$ P-wave, 
presumably from the radial excitations $\rho(1450)$ and $\rho(1700)$, which could not, 
however, be disentangled by the fit.

For $a_0(980)$ one finds for the $T$-matrix pole in the second Riemann sheet the mass 987 
$\pm$ 3 MeV and the width 86 $\pm$ 7 MeV. Using the $\eta\pi$ decay branching ratio 
from the $\pi^0\pi^0\eta$ analysis (Table \ref{DPBR}) one also obtains the ratio
\begin{equation}
\frac{B(a_0(980)\rightarrow\KKbar)}{B(a_0(980)\rightarrow\eta\pi)} = 0.24 \pm 0.06.
\label{77}
\end{equation}
With the coupling $g_1$ = 0.324 GeV to $\eta\pi$ (Amsler, 1994b) the coupling 
$g_2$ to $\KKbar$ can be tuned to satisfy Eq. (\ref{77}) by integrating the mass 
distributions over the $a_0(980)$ distribution (Fig. \ref{Flatte}). One obtains $g_2$ = 0.279 
MeV and the ratio $g_2/g_1$ = 0.86, is in excellent agreement with the estimate from the 
line shape in the  $\pi^0\pi^0\eta$ channel (Eq. (\ref{88})).

The branching ratios are given in Table \ref{DPBRK}. The intermediate $K^*\overline{K}$ 
is largest in the (I=1) $^3S_1$ channel, a feature that was noticed before (Barash, 1965; 
Conforto, 1967) and that we have used in section \ref{sec:phi} as a possible explanation for 
the $\pi\phi$ enhancement. The 
branching ratios are in fair agreement 
with those from Conforto (1967) and those for $\pi^0K_LK_L$, except for the much larger 
$K\pi$ S-wave in $\pi^{\pm}K^{\mp}K_L$.

\section{Annihilation at Rest into $5\pi$}
Resonances in $\rho^+\rho^-$ 
have been reported in $\pbarn$ annihilation in deuterium: A $2^{++}$ state was observed 
in bubble chamber exposures in liquid deuterium (Bridges, 
1987). An enhancement was also seen around 1500 MeV by the Asterix collaboration at 
LEAR in gaseous deuterium but no spin-parity analysis was performed (Weidenauer, 1993). 
This state was interpreted as $f_2(1565)$ in its $\rho\rho$ decay mode, the slightly  lower 
mass being due to $\pi$-rescattering with the recoil proton exciting the $\Delta$ 
resonance (Kolybashov, 1989). The $2^{++}$ assignment was, however, disputed in favour 
of $0^{++}$ by a reanalysis of the bubble chamber data (Gaspero, 1993).

The Crystal Barrel has also searched for scalar mesons decaying to $4\pi$.
The $4\pi^0$ decay mode was investigated using $\pbarp$ annihilation into $5\pi^0$,  
leading to 10 detected photons (Abele, 1996c). The branching ratio for annihilation into 
$5\pi^0$ was found to be $(7.1 \pm 1.4)\times 10^{-3}$. After removing the 
$\eta\rightarrow 3\pi^0$ contribution they performed a maximum likelihood analysis of a 
sample of 25,000 $5\pi^0$ events. The data demand contributions from 
$\pi(1330)\rightarrow 3\pi^0$ and from two scalar resonances decaying to $4\pi^0$. The 
mass and width of the  lower state, presumably $f_0(1370)$, could not be determined 
precisely.

The upper state has mass $\sim$ 1505 MeV and width $\sim$  169 MeV and decays into 
two $\pi\pi$ S-wave pairs and $\pi^0(1300)\pi^0$ with approximately equal rates. The 
branching ratio for $\pbarp\rightarrow f_0(1500)\pi^0$$\rightarrow 5\pi^0$ is (9.0 
$\pm$ 1.4) $\times 10^{-4}$. Using the $2\pi$ 
branching ratio from the coupled channel analysis one finds
\begin{eqnarray}
\frac{B(f_0(1500)\rightarrow 4\pi)}{B(f_0(1500)\rightarrow 2\pi)} = 2.1 \pm 
0.6, \label{69} \\
\frac{B(f_0(1500)\rightarrow 4\pi^0)}{B(f_0(1500)\rightarrow\eta\eta)} = 1.5 \pm 0.5.
\label{68} 
\end{eqnarray}
The $4\pi^0$ ($2\pi^0$) contribution in (\ref{69}) has been multiplied by 9 (3) 
to take into account the unobserved charged pions. The ratio (\ref{69}) is, in principle, a 
lower limit which does not include $\rho\rho$. 
However, a reanalysis of the Mark III data on $J/\psi\rightarrow\gamma 2\pi^+2\pi^-
$ finds evidence for $f_0(1500)$ decaying to $4\pi$ through two S-wave 
dipions with negligible $\rho\rho$ contribution (Bugg, 1995). The ratio (\ref{68}) is in 
agreement with the result for the former $f_0(1590)$ from Alde (1987) in $\pi^-
p\rightarrow 4\pi^0 n$: 0.8 $\pm$ 0.3. 

A scalar resonance with mass 1374 $\pm$ 38 and width 375 $\pm$ 61 MeV decaying to 
$\pi^+\pi^-2\pi^0$ was also reported by Crystal Barrel in the 
annihilation channel $\pi^+\pi^-3\pi^0$ (Amsler, 1994d). The 
branching ratio for $\pi^+\pi^-3\pi^0$ was measured to be (9.7 $\pm$ 0.6) \%. The $4\pi$ 
decay mode of the resonance is five times larger than the $2\pi$, 
indicating a large inelasticity in the $2\pi$ channel. The relative decay ratio 
to $\rho\rho$ and two $\pi\pi$ S-waves is 1.6 $\pm$ 0.2. However, the data do not exclude the 
admixture of a  $f_0(1500)$ contribution.

\section{The New Mesons}
It is instructive to first check the consistency within Crystal Barrel data and also 
compare with previous measurements. The squares of the isospin Clebsch-Gordan 
coefficients determine the total branching ratios, including all charge modes. Note that 
two neutral isovectors (e.g. $a_2^0(1320)\pi^0$) do not contribute to $I=1$. Table 
\ref{iweight} gives the corresponding weights to $\KKbar$. For $K^*(892), K^*_0(1430)$, 
$a_2(1320)$ and $a_0(1450)$ contributions to $\KKbar\pi$ the $\pi^0K_LK_L$ ratios in 
Table \ref{DPBRK} must be multiplied by 12 while those for $f_2(1270)$, $f_0(1370)$,  
$f_0(1500)$ and $f_2'(1525)$ must be multiplied by 4 (since $K_S$ is not observed). For  
$\pi^{\pm}K^{\mp}K_L$ the resonance contributions in Table \ref{DPBRK} have to be 
multiplied by 3 for $K^*(892)$, $K^*_0(1430)$, by 3 for $a_0(980)$,  $a_2(1320)$, 
$\rho(1450/1700)$ from $I=0$ and by 2 for $a_2(1320)$, $\rho(1450/1700)$ from  $I=1$.

Furthermore, the branching ratios must be corrected for all decay modes of the 
intermediate resonances to obtain the two-body branching 
ratios in Table \ref{BRTB}. We have used the following decay branching ratios (Barnett, 
1996): (28.2$\pm$0.6)\% for $f_2(1270)\rightarrow\pi^0\pi^0$, (4.6$\pm$0.5)\% for 
$f_2(1270)\rightarrow\KKbar$, (14.5$\pm$1.2)\% 
for $a_2(1320)\rightarrow\eta\pi$, (4.9$\pm$0.8)\% 
for $a_2(1320)\rightarrow\KKbar$, (0.57$\pm$0.11)\% for 
$a_2(1320)\rightarrow\etap\pi$ and 
(88.8$\pm$3.1)\% for $f_2'(1525)\rightarrow\KKbar$. There are indications that the 
$a_2(1320)$ contribution to $\pi\pi\eta$ is somewhat too large. Otherwise the consistency 
is in general quite good and Crystal Barrel results also agree with previous data. This gives 
confidence in the following discussion on the new mesons. 
\subsection{$a_0(1450)$}
\label{sec:new}
We begin with the isovector 
$a_0(1450)$ which has been observed in its $\eta\pi$, $\etap\pi$ and $\KKbar$ decay 
modes. Averaging mass and width from the coupled channel  and the $\KKbar$ analyses 
one finds:
\begin{equation}
a_0(1450): \ m = 1474 \pm 19, \ \ \ \Gamma = 265 \pm 13 \ {\rm MeV}.
\end{equation}
The $a_0(1450)$ decay rates are related by SU(3)-flavor which can be tested with Crystal 
Barrel data. Following  Amsler and Close (1996) we  
shall write for a quarkonium state 
\begin{equation}
|\qqbar\rangle = {\rm cos}\alpha |n\bar{n}\rangle -
{\rm sin}\alpha |s\bar{s}\rangle ,
\end{equation}
where
\begin{equation}
|n\bar{n}\rangle \equiv (u\bar{u} + d\bar{d})/\sqrt{2}.
\end{equation}
The mixing angle $\alpha$ is related to the usual nonet mixing
angle $\theta$ (Barnett, 1996)  by the relation
\begin{equation}
\alpha = 54.7^{\circ} + \theta.
\end{equation}
Ideal mixing occurs for $\theta=35.3^{\circ}$ (-54.7$^{\circ}$)
for which the quarkonium state becomes pure $\ssbar$
($\nnbar$).

The flavor content of $\eta$ and $\etap$ are then given by the superposition (see also Eq. 
(\ref{31}))
\begin{eqnarray}
|\eta\rangle & = & {\rm cos}\phi |n\bar{n}\rangle - {\rm sin}\phi
|s\bar{s}\rangle,\nonumber\\
|\etap\rangle & = & {\rm sin} \phi |n\bar{n}\rangle +
{\rm cos} \phi |s\bar{s}\rangle,
\end{eqnarray}
with $\phi = 54.7^{\circ} +\theta_p$, where $\theta_p$ is the
pseudoscalar mixing angle which we take as $\theta_p$ = $(-17.3\pm 1.8)^{\circ}$ (Amsler, 
1992b). 

The partial decay width of a scalar (or tensor) quarkonium into a pair of pseudoscalars 
$M_1$ and $M_2$ is given by
\begin{equation}
\Gamma(M_1, M_2) = \gamma^2 (M_1, M_2)  f_L(p) p.
\label{64}
\end{equation}
The couplings $\gamma$ can be derived from SU(3)-flavor. The two-body decay 
momentum is denoted by $p$ and the relative angular momentum by $L$. The form factor
\begin{equation}
f_L(p) = p^{2L} \exp(-\frac{p^2}{8\beta^2})
\label{65}
\end{equation}
provides a good fit to the decay branching ratios of the well known ground state $2^{++}$ 
mesons if $\beta$ is chosen $\geq$ 0.5 GeV/c (Amsler and Close, 1996). We choose 
$\beta$ = 0.5 GeV/c although the exponential factor can be ignored 
($\beta\rightarrow\infty$) without altering the forthcoming conclusions. 
Replacing $f_L(p)p$ by prescription (\ref{39}) also leads to a good description of decay 
branching ratios provided that $p_R > 500$ MeV/c, corresponding to an interaction radius 
of less than 0.4 fm (Abele, 1997g).

The decay of quarkonium into a pair of mesons
involves the creation of $\qqbar$ pair from the vacuum.
We shall assume for the ratio of the matrix elements for the creation
of $s\bar{s}$ versus $u\bar{u}$ (or $d\bar{d}$) that 
\begin{equation}
\rho \equiv \frac{\langle 0|V|s\bar{s}\rangle}{\langle
0|V|u\bar{u}\rangle} \simeq 1.
\end{equation}
This assumption is reasonable since from the measured decay branching ratios of tensor 
mesons one finds $\rho$ = 0.96 $\pm$ 0.04 (Amsler and Close, 1996). Similar conclusions 
are reached by Peters and Klempt (1995).

Let us now compare the Crystal Barrel branching ratios for $a_2(1320)$ decays to 
$\KKbar$, $\eta\pi$ and $\etap\pi$ with predictions from SU(3).  One predicts for an 
isovector with the coupling constants $\gamma$ given in the appendix of 
Amsler and Close (1996):
\begin{eqnarray}
\frac{\Gamma(a^\pm_2(1320)\rightarrow K^\pm K^0)}
{\Gamma(a_2^0(1320)\rightarrow\eta\pi^0)} & = & \frac{1}{2\cos^2\phi} 
\frac{f_2(p_K)p_K}{f_2(p_\eta) p_\eta} = 0.295 \pm 0.013,
\label{84}\\
\frac{\Gamma(a_2^0(1320)\rightarrow \etap\pi^0)}
{\Gamma(a_2^0(1320)\rightarrow\eta\pi^0)} & = & 
\tan^2\phi\frac{f_2(p_K)p_K}{f_2(p_\eta) 
p_\eta} = 0.029 \pm 0.004.
\label{85}
\end{eqnarray}
Both ratios are in excellent agreement with world data (Barnett, 1996). The 
Crystal Barrel numbers to be compared with are taken from Tables \ref{DPBR} and 
\ref{DPBRK} for $I=0$. One finds the ratios of branching ratios
\begin{eqnarray}
\frac{B(a^\pm_2(1320)\rightarrow K^\pm K^0)}
{B(a_2^0(1320)\rightarrow\eta\pi^0)} & = & 0.21 ^{+0.04}_{-0.06}, \\
\frac{B(a_2^0(1320)\rightarrow \etap\pi^0)}
{B(a_2^0(1320)\rightarrow\eta\pi^0)} & = & 0.034 \pm 0.009,
\end{eqnarray}
in fair agreement with SU(3) predictions. We now compare predictions and data for  
$a_0(1450)$. From SU(3) one expects, using Eq. (\ref{84}) and (\ref{85}) with $L=0$: 
\begin{eqnarray}
\frac{\Gamma(a^\pm_0(1450)\rightarrow K^\pm K^0)}
{\Gamma(a_0^0(1450)\rightarrow\eta\pi^0)} & = &   0.72 \pm 0.03, \\
\frac{\Gamma(a_0^0(1450)\rightarrow \etap\pi^0)}
{\Gamma(a_0^0(1450)\rightarrow\eta\pi^0)} & = &   0.43 \pm 0.06,
\end{eqnarray}
in agreement with the experimental results from Tables \ref{DPBR} and \ref{DPBRK}:  
\begin{eqnarray}
\frac{B(a^\pm_0(1450)\rightarrow K^\pm K^0)}
{B(a_0^0(1450)\rightarrow\eta\pi^0)} & = & 0.87 \pm 0.23, \\
\frac{B(a_0^0(1450)\rightarrow \etap\pi^0)}
{B(a_0^0(1450)\rightarrow\eta\pi^0)} & = & 0.34 \pm 0.15.
\end{eqnarray}
This then establishes $a_0(1450)$ as a $\qqbar$ isovector in the scalar 
nonet. 

The existence 
of $a_0(1450)$ adds further evidence for $a_0(980)$ being a non-$\qqbar$ state. 
The $f_0(980)$ and $a_0(980)$ have been assumed to
be $\KKbar$ molecules (Weinstein, 1990; Close, 1993). This is motivated by
their strong couplings to $\KKbar$ - in spite of their
masses close to the $\KKbar$ threshold - and their small
$\gamma\gamma$ partial widths. For $f_0(980)$ the $2\gamma$ partial width is 
$\Gamma_{\gamma\gamma}$ = (0.56 $\pm$ 0.11) keV (Barnett, 1996).
The relative ratio for $a_0(980)$ decay to $\KKbar$ and $\eta\pi$ has been determined by 
Crystal Barrel (Eq. (\ref{77})). Using 
\begin{equation}
\Gamma_{\gamma\gamma} B(a_0(980)\rightarrow\eta\pi) = 0.24 \pm 0.08 \ {\rm keV} 
\end{equation}
(Barnett, 1996), one then derives the partial width $\Gamma_{\gamma\gamma}$ = (0.30  
$\pm$ 0.10) keV. Thus the $2\gamma $ 
partial widths for $f_0(980)$ and $a_0(980)$ appear to be similar, close to predictions 
for $\KKbar$ molecules (0.6 keV) and much smaller than for $\qqbar$ states
(Barnes, 1985). However, not everybody agrees: In T\"ornqvist (1995) the 
$f_0(980)$/$f_0(1370)$ and the $a_0(980)$/$a_0(1450)$ are different manifestations of the 
same uniterized $\ssbar$ and $\overline{u}d$ states, while the broad structure around 1100 
MeV is the $\uubar + \ddbar$ state (T\"ornqvist and Roos, 1996). This then leaves 
$f_0(1500)$ as an extra state.

\subsection{$f_0(1370)$ and $f_0(1500)$}
From the single channel analyses and the $\KKbar$ decay mode we find for 
$f_0(1370)$ and $f_0(1500)$ the average masses and widths:
\begin{eqnarray}
f_0(1370):  m & = & 1360 \pm 23 \ {\rm MeV}, \ \Gamma = 351 \pm 41 \ {\rm 
MeV},\nonumber\\
f_0(1500):  m & = & 1505 \pm 9 \ {\rm MeV}, \ \Gamma = 111 \pm 12 \ {\rm MeV}.
\end{eqnarray}
The closeness of $a_0(1450)$ and $f_0(1500)$ or even $f_0(1370)$ masses is 
conspicuous and points to a close to  ideally mixed scalar nonet, one of the latter mesons 
being one of the $\qqbar$ isoscalars. However, $f_0(1500)$ with 
a width of about 100 MeV is 
much narrower than $a_0(1450)$, $f_0(1370)$ and 
$K^*_0(1430)$ with widths of typically 300 MeV. Theoretical predictions for the widths of 
scalar $\qqbar$ mesons, based on the $^3P_0$ model, agree that scalar $\qqbar$ mesons 
have widths of at least 250 MeV (for a discussion and references see Amsler and Close 
(1996)). We therefore tentatively assign $f_0(1370)$ to the ground state scalar nonet. 

If $f_0(980)$ is indeed a molecule then the (mainly) $\ssbar$ member of 
the scalar nonet still needs to be identified. We now show from their decay branching 
ratios that neither $f_0(1370)$ nor $f_0(1500)$ are likely candidates.  
To investigate the quark content of $f_0(1500)$ we calculate its relative couplings 
to $\eta\eta$, $\eta\etap$ and $\KKbar$ and search for a common value of the scalar 
mixing angle $\alpha$. The ratios of couplings for a pseudoscalar  mixing angle $\phi$ are 
(Amsler and Close, 1996):
\begin{eqnarray}
R_1 & \equiv & \frac{\gamma^2(\eta\eta)}{\gamma^2(\pi\pi)} = \frac{(\cos^2\phi -
\sqrt{2}\tan\alpha\sin^2\phi)^2}{3},\nonumber\\
R_2 & \equiv & \frac{\gamma^2(\eta\etap)}{\gamma^2(\pi\pi)} = 
\frac{2(\cos\phi\sin\phi[1 + \sqrt{2}\tan\alpha])^2}{3},\nonumber\\
R_3 & \equiv & \frac{\gamma^2(\KKbar)}{\gamma^2(\pi\pi)} = 
\frac{(1 -\sqrt{2}\tan\alpha)^2}{3}.
\end{eqnarray}
For $\eta\eta$ and $\pi\pi$ we use the branching ratios from the coupled channel 
analysis (Table \ref{DPBR}) and multiply $\pi\pi$ by 3 to take into account the 
$\pi^+\pi^-$ decay mode. The branching ratio for $\KKbar$ is taken from Table 
\ref{DPBRK} and is multiplied by  4. The branching fractions are, including the $4\pi$ 
mode from Eq. (\ref{69}) and ignoring a possible small $\rho\rho$ contribution to $4\pi$:
\begin{eqnarray}
\begin{array}{l r c r l}
\pi\pi: &  (29.0 & \pm & 7.5) & \% \\
\eta\eta: &  (4.6 & \pm & 1.3) & \% \\
\eta\etap: & (1.2 & \pm & 0.3) & \% \\
\KKbar: & (3.5 & \pm & 0.3) & \% \\
4\pi:  &  (61.7 & \pm & 9.6) & \%.
\end{array}
\end{eqnarray}
After correcting for phase space and form factor (Eq. 
(\ref{65})) we obtain:
\begin{equation}
R_1 = 0.195 \pm 0.075, \ R_2 = 0.320 \pm 0.114, \ R_3 = 0.138 \pm 0.038.
\label{66}
\end{equation}
Since $f_0(1500)$ lies at the $\eta\etap$ threshold we have divided the branching ratios 
by the phase space factor $\rho$ integrated over the resonance and have neglected 
the form factor when calculating $R_2$.

Previously, the upper limit for $R_3$ was $ < 0.1$ (95\% confidence level) from Gray 
(1983), in which case no value for the mixing angle $\alpha$ could simultaneously fit 
$R_1$, $R_2$ and $R_3$ (Amsler and Close, 1996), therefore excluding $f_0(1500)$ as a 
$\qqbar$ state. The ($1\sigma$) allowed regions of tan$\alpha$ are shown in Fig. 
\ref{tanal} for the ratios (\ref{66}). The agreement between $R_1$ and 
$R_3$ is not particularly good. Remember, however, that branching ratios are 
sensitive to interference effects and therefore caution should be exercized in not  
overinterpreting the apparent discrepancy in Fig. \ref{tanal}. On the basis of the ratios 
(\ref{66}), one may conclude that $f_0(1500)$ 
is not incompatible with a mainly $\uubar + \ddbar$ meson ($\alpha=0$). For a pure 
$\ssbar$ state ($\alpha=90^\circ$) the ratios (\ref{66}) would, however,  become infinite. 
Therefore $f_0(1500)$ is not the missing $\ssbar$ scalar meson. 

Similar conclusions can reached for $f_0(1370)$ which has small  decay branching ratios 
to $\eta\eta$ and $\KKbar$. Precise ratios $R_i$ are difficult to obtain in this case 
since the branching ratio to $\pi\pi$ in Table \ref{DPBR} also includes the low energy 
$\pi\pi$ S-wave, in particular $f_0(400-1200)$. 

This analysis shows that both $f_0(1370)$ and $f_0(1500)$ are compatible with isoscalar  
$\uubar + \ddbar$ states, although the latter is much too narrow for the ground state 
scalar nonet. This then raises the question on whether $f_0(1500)$ could not be the first 
radial excitation of $f_0(1370)$. This is unlikely 
because (i) the splitting between ground state and first radial is expected to be around 700 
MeV (Godfrey and Isgur, 1985), (ii) the next $K^*_0$ lies at 1950 MeV (Barnett, 
1996) and, last but not least, first radials are expected to be quite broad (Barnes, 1997).

The most natural explanation is that $f_0(1500)$ is the ground state glueball predicted in 
this mass range by lattice gauge theories. However, a pure glueball should decay to 
$\pi\pi$, $\eta\eta$, $\eta\etap$ and $\KKbar$ with relative ratios 3 : 1 : 0 : 4, in 
contradiction with our ratios $R_i$. In the model of Amsler and Close (1996) the finite 
$\eta\etap$ and the 
small $\KKbar$ rates can be accommodated by mixing the pure glueball $G_0$ with the 
nearby two $\nnbar$ and $\ssbar$ states. Conversely, the two isoscalars in the $\qqbar$ 
nonet 
acquire a gluonic admixture. In first order perturbation one finds\footnote{We 
assume here that the quark-gluon coupling is flavor blind, see Amsler and Close (1996) 
for a generalization.}
\begin{equation}
|f_0(1500)\rangle =\frac{|G_0\rangle + \xi (\sqrt{2}|\nnbar\rangle + \omega 
|\ssbar\rangle)}
{\sqrt{1+\xi^2(2+\omega^2)}},
\label{71}
\end{equation}
where $\omega$ is the ratio of mass splittings
\begin{equation}
\omega = \frac{m(G_0) - m(\nnbar)}{m(G_0) - m(\ssbar)}.
\label{67}
\end{equation}
In the flux tube simulation of lattice QCD the pure gluonium $G_0$ does not decay to 
$\pi\pi$ nor to $\KKbar$ 
in first order and hence  $f_0(1500)$ 
decays to $\pi\pi$ and $\KKbar$ through its $\qqbar$ admixture in the wave function. 
If $G_0$ lies between the two $\qqbar$ states, $\omega$ is negative and the decay 
to $\KKbar$ is hindered by negative interference between the decay amplitudes of the 
$\nnbar$ and $\ssbar$ components in Eq. (\ref{71}). The ratio of couplings to $\KKbar$ 
and $\pi\pi$ is 
\begin{equation}
\frac{\gamma^2(\KKbar)}{\gamma^2(\pi\pi)} = \frac{(1+\omega)^2}{3}.
\end{equation}
The cancellation is perfect whenever $G_0$ lies exactly between $\nnbar$ and $\ssbar$ 
($\omega=-1$). We find with the measured $R_3$ two solutions, $\omega$ = - 0.36 or -1.64.  
Assuming that $f_0(1370)$ is essentially $\nnbar$ (with a small gluonic 
admixture) this leads to an $\ssbar$ state around 1900 MeV or 1600 MeV, respectively. 
Furthermore,  the ratio of $\pi\pi$ partial widths for $f_0(1500)$ and $f_0(1370)$, divided 
by phase-space and form factor,  
is given by
\begin{equation}
\frac{\tilde{\Gamma}_{2\pi}[f_0(1500)]} {\tilde{\Gamma}_{2\pi}[f_0(1370)]} =
\frac{2\xi^2(1+2\xi^2)}{1+\xi^2(2+\omega^2)} \sim 0.5,
\label{89}
\end{equation}
using for $f_0(1500)$  the branching ratios from Tables \ref{DPBR} and \ref{DPBRK} and 
the $4\pi/2\pi$ ratio (\ref{69}). There is, however, a large uncertainty in the  
ratio (\ref{89}) due to the branching ratios of $f_0(1370)$ which cannot easily be 
disentangled from $f_0(400-1200)$. This then leads to $|\xi|\sim 0.6$ and according to Eq. 
(\ref{71}) to about 30\% or about 60\% glue in $f_0(1500)$ for an $\ssbar$ state at 1600 
MeV or 1900 MeV, respectively. 

The $f_0(1500)$ has also been observed in $\pbarp$ annihilation at higher 
energies (Armstrong, 1993) and in other reactions, in particular in central production,  
decaying to $2\pi^+2\pi^-$ (Antinori, 1995). The VES experiment, studying $\pi^-p$ 
interactions on nuclei at 36 GeV/c, has reported a resonance, $\pi(1800)$, decaying to 
$\pi^-
\eta\etap$ (Beladidze, 1992) and $\pi^-\eta\eta$ (Amelin, 1996). The $\pi(1800)$ appears to 
decay into a resonance with mass 1460 
$\pm$ 20 MeV and width 100 $\pm$ 30 MeV - in agreement with $f_0(1500)$ - with a 
recoiling $\pi$.  They report an $\eta\etap/\eta\eta$ ratio of 0.29 $\pm$ 0.07 which is in 
excellent agreement with the Crystal Barrel ratio for $f_0(1500)$ decays, 0.27 $\pm$ 0.10 
(Table \ref{DPBR}). Note that if $\pi(1800)$ is indeed a $\qqbar g$ (hybrid), as advocated 
by Close and Page (1995), then decay into gluonium is favoured,.

A reanalysis of $J/\psi$ radiative decay to $2\pi^+2\pi^-$ finds evidence for $f_0(1500)$ 
decaying to two S-wave dipions  with a branching ratio in $J/\psi\rightarrow \gamma 
4\pi$ of $(5.7 \pm 0.8)\times 10^{-4}$ (Bugg, 1995). This leads to an expected branching 
ratio of $(2.7 \pm 0.9)\times 10^{-4}$ in $J/\psi\rightarrow \gamma 2\pi$, using the 
Crystal Barrel result (\ref{69}). It is interesting to compare this prediction with data 
on $J/\psi\rightarrow \gamma \pi^+\pi^-$ from Mark III (Baltrusaitis, 1987) where 
$f_2(1270)$ is observed together with a small  accumulation of events in the 1500 MeV 
region. 
Assuming that these are due to $f_0(1500)$, one finds by scaling to $f_2(1270)$ a 
branching ratio in $J/\psi\rightarrow \gamma 2\pi$ of $\simeq 2.9\times 10^{-4}$, in 
agreement with the above prediction. 

Summarizing, $f_0(1500)$  has been observed in $\pbarp$ annihilation in several decay 
modes, some with very high statististics ($\sim$ 150,000 decays into $\pi^0\pi^0)$ and also 
in other processes that are traditionally believed to enhance gluonium production, central 
production and $J/\psi$ radiative decay. The $K^*_0(1430)$ and $a_0(1450)$ define the 
mass scale of the $\qqbar$ scalar nonet. The $f_0(1500)$ is not the missing $\ssbar$ and is 
anyway too narrow for a scalar $\qqbar$ state. The most natural explanation for 
$f_0(1500)$ is the ground state glueball mixed with nearby scalars. 
The missing element in this jigsaw puzzle is the $\ssbar$ scalar expected between 1600 and 
2000 MeV. The analysis of in flight annihilation data will hopefully provide more 
information in this mass range. The spin of $f_J(1710)$ has not been determined 
unambiguously. If $J=0$ is confirmed then  $f_J(1710)$ could be this state or, alternatively,  
become a challenger for the ground state glueball (Sexton, 1995). A more detailed 
discussion on $f_0(1500)$ and $f_J(1710)$ can be found in Close (1997).

\subsection{$f_2(1565)$}
The $f_2(1565)$ with mass 1565 $\pm$ 20 and width 170 $\pm$ 40 MeV has been observed 
first by the Asterix collaboration at LEAR in the final state $\pi^+\pi^-\pi^0$ in hydrogen 
gas (May, 1989, 1990) and then by Aker (1991) in the $3\pi^0$ final state in liquid. The 
$3\pi^0$ analysis gave 60\% P-state contribution to $3\pi^0$ in liquid with roughly equal 
intensities 
from $f_2(1565)$ of 9\%, each from $^1S_0$, $^3P_1$ and $^3P_2$. The full $3\pi^0$ data 
sample demands 46\% P-state annihilation (Amsler, 1995f) and also requires a tensor 
around 1530 MeV 
(section \ref{sec:threepi0}). The $\ssbar$ tensor, $f_2'(1525)$, has been observed in its 
$\KKbar$ decay mode 
(section \ref{sec:KLKL}). From the observed rate (Table \ref{BRTB}) and the known 
$f_2'(1525)$ decay 
branching ratios (Barnett, 1996) one finds that $f_2'(1525)$ cannot account for 
much of the $2^{++}$ signal in $\pi\pi$ or $\eta\eta$. Hence 
$f_2(1565)$ is not $f_2'(1525)$. 

The large P-state fraction in the $3\pi^0$ channel in liquid is not too surprising: the 
corresponding 
channel $\pbarp\rightarrow\pi^+\pi^-\pi^0$ in liquid proceeds
mainly from the ($I$=0) $^3S_1$ atomic state while the ($I$=1) $^1S_0$ 
is suppressed by an order of magnitude (Foster, 1968b), as are normally P-waves 
in liquid. On the other hand, the channel $\pbarp\rightarrow 3\pi^0$ 
proceeds only through  the ($I$=1) $^1S_0$ atomic state while $^3S_1$ is forbidden. Indeed 
the branching ratios for $3\pi^0$ is an order of magnitude smaller than 
for $\pi^+\pi^-\pi^0$ (Table \ref{TBBR}). Hence for $3\pi^0$  S- and P-wave annihilations  
compete in liquid. 

A fraction of 50\% P-wave was also required in the Dalitz plot analysis of 
the $I=1$ final state $\pi^-\pi^0\pi^0$ at rest in liquid deuterium (Abele, 1997h) which 
shows evidence for $f_0(1500)$ and $f_2(1565)$ production and requires in addition the 
$\rho$-meson and two of its excitations, $\rho^-(1450)$ and $\rho^-(1700)$, decaying to 
$\pi^-\pi^0$.  

However, the coupled channel analysis described in section \ref{sec:coupled}, 
ignoring P-waves, still requires a tensor at 1552 MeV. Neglecting P-waves increases 
slightly the 
contribution from $f_0(1500)$ while decreasing the contribution from $f_2(1565)$,  
although the rates remain within errors (compare the two $\pi^0\pi^0$ branching ratios 
in Table \ref{DPBR} for the single and coupled channel analyses).

The alternative N/D analysis also reproduces the features of the $3\pi^0$ Dalitz plot 
without P-wave contributions, in particular the scalar state around 1500 MeV (Anisovich, 
1994). A tensor contribution with mass $\sim$ 1565 and width $\sim$ 165 MeV is, however, 
still required, but most of the blob structure in Fig. \ref{Dalitz}(c) is taken into account by 
interferences in the low energy $\pi\pi$ S-wave.

An N/D analysis of Crystal Barrel data, together with former data from 
other reactions, also reports a tensor with mass 1534 $\pm$ 20  and width 180 $\pm$ 60 
MeV (Abele, 1996a). They report strong $\rho\rho$ and $\omega\omega$ contributions 
and 
therefore assign this signal to $f_2(1640)$ discovered by GAMS in $\pi^-
p\rightarrow\omega\omega n$ (Alde 1990), also  reported to decay into $4\pi$ by the 
Obelix collaboration in $\nbar p\rightarrow 5\pi$ (Adamo, 1992). It should be emphasized, 
however, that in Abele (1996a) the inelasticity in the $K$-matrix is attributed to 
$\rho\rho$ and $\omega\omega$, although no $4\pi$ data are actually included in the fit. 
Given that  mass and width of the tensor agree with $f_2(1565)$, but disagree with 
$f_2(1640)$, it seems more 
natural to assign the $2^{++}$ signal to the former. It is interesting to note that a 
$2^{++}$ $\rho\rho$ 
molecule (T\"ornqvist, 1991) or a $2^{++}$ baryonium state (Dover, 1986) decaying strongly 
to $\rho\rho$ (Dover, 1991) are predicted in this mass range. The $f_2(1565)$ could be one 
of these states.

In conclusion, there is a certain amount of model dependence when extracting the precise 
production and decay rates of $f_2(1565)$. However, annihilation data 
require both a scalar and a tensor around 1500 MeV and the parameters and rates for 
$f_0(1500)$ are reasonably stable, independent of $f_2(1565)$ contribution. The Crystal 
Barrel data in gas will hopefully settle the issue of the fraction of P-wave in three-body 
annihilation.  

\section{$E/\iota$ Decay to $\eta\pi\pi$}
\label{seciota}
The $E$ meson, a  $0^{-+}$ state, was discovered in the sixties in the $\KKbar\pi$ mass 
spectrum of $\pbarp$ annihilation at rest into $(K_SK^{\pm}\pi^{\mp})\pi^+\pi^-$. Its 
mass and width were determined to be 1425 $\pm$ 7 and 80 $\pm$ 10 MeV (Baillon, 1967). 
Its quantum numbers have remained controversial since other groups have claimed a 
$0^{-+}$ state  (now called $\eta(1440)$) and a $1^{++}$ state (now called $f_1(1420)$) in 
this mass region from various hadronic reactions. A broad structure (previously called 
$\iota$), has also been observed in radiative $J/\psi$ decay to $\KKbar\pi$ 
(Scharre, 1980). Initially determined to be $0^{-+}$, the $E/\iota$ structure was then found 
to split into three states, the first ($0^{-+}$) at 1416 $\pm$ 10 MeV decaying to 
$a_0(980)\pi$, the second (presumably the $1^{++}$ $f_1(1420)$) at 1443 $\pm$ 8 MeV and 
the third ($0^{-+}$) at 1490 $\pm$ 16 MeV, both 
decaying to $K^*\overline{K}$ (Bai, 1990). The widths were not determined accurately. The 
Obelix collaboration has analyzed the $\KKbar\pi$ mass spectrum in $\pbarp$ 
annihilation at rest in liquid and has also reported a splitting of the $E$ meson into two 
pseudoscalar states at 1416 $\pm$ 2 MeV ($\Gamma$ = 50 $\pm$ 4 MeV) and 1460 $\pm$ 10 
MeV ($\Gamma$ = 105 $\pm$ 15 MeV) (Bertin, 1995). We shall refer to these pseudoscalars 
as $\eta(1410)$ and $\eta(1460)$. When using gaseous hydrogen, one expects the 
production of $1^{++}$ mesons from $^3P_1$ recoiling against an S-wave dipion: Bertin 
(1997) indeed observes three states in the $\KKbar\pi$ mass spectrum in gas: $\eta(1410)$, 
$f_1(1420)$ and $\eta(1460)$.

In $J/\psi$ radiative decay, $\iota$  decays to $\KKbar\pi$ through the intermediate 
$a_0(980)$  and hence a signal was also expected in the 
$a_0(980)\pi\rightarrow\eta\pi\pi$ mass spectrum. This has indeed been observed by 
Mark III and DM2:  Bolton (1992) reports a signal in $a_0^{\pm}\pi^{\mp}$ at 1400 $\pm$ 6 
MeV ($\Gamma$ = 47 $\pm$ 13 MeV) and Augustin (1990) in $\eta\pi^+\pi^-$ at 1398  $\pm$ 
6 MeV ($\Gamma$ = 53 $\pm$ 11 MeV). We shall tentatively assign these signals to  
$\eta(1410)$.
  
To clarify whether the structures observed in $J/\psi$ radiative decay and in $\pbarp$ 
annihilation are compatible and in particular to confirm the quantum numbers of $E$ 
($0^{-+}$ and not  $1^{++}$) Crystal Barrel has searched for the $\eta\pi\pi$ 
decay mode of the $E$ meson in the reaction 
$\pbarp\rightarrow(\eta\pi^+\pi^-)\pi^0\pi^0$ and 
$(\eta\pi^0\pi^0)\pi^+\pi^-$, leading to two charged particles and $6\gamma$ (Amsler, 
1995h; Urner, 1995). Since the rate for this reaction was expected to be rather small ($\sim 
10^{-3}$ of all annihilations), an online trigger required 8 clusters in the barrel and at 
least two $\pi^0$ and one $\eta$. A 7C kinematic fit then selected the channel $\pi^+\pi^-
2\pi^0\eta$ while suppressing $\pi^+\pi^-2\eta\pi^0$ and $\pi^+\pi^-3\pi^0$. The 
branching ratio for $\pi^+\pi^-2\pi^0\eta$ was found to be (2.09 $\pm$ 0.36) \%.

The final state $\pi^+\pi^-2\pi^0\eta$ includes a strong contribution from 
$\omega\eta\pi^0$ ($\omega\rightarrow\pi^+\pi^-\pi^0)$, a channel that has been 
studied 
in its $7\gamma$ decay mode (Amsler, 1994c). Events compatible with  $\omega\eta\pi^0$  
were removed, leaving a sample of about 127,000 $\pi^+\pi^-2\pi^0\eta$ events.  The 
evidence for $\eta(1410)$ decaying to $\eta\pi\pi$ is shown in the $\pi^0\pi^0\eta$ and 
$\pi^+\pi^-\eta$ mass distributions (Fig. \ref{Espec}). Some 9,000 $\eta(1410)$ decays into 
$\eta\pi\pi$  are observed in Fig. \ref{Espec}, an order of magnitude more than for $E$ 
to $\KKbar\pi$ in the seminal work of Baillon (1967).

A partial wave analysis was performed using a maximum likelihood optimization. The 
signal at 1400 MeV was described by the annihilation channels
\begin{eqnarray}
\pbarp & \rightarrow &\eta(1410)(\rightarrow\eta\sigma^0)
\sigma^{+-},\nonumber\\
& \rightarrow & \eta(1410) (\rightarrow a_0^0(980)\pi^0)
\sigma^{+-},\nonumber\\
& \rightarrow & \eta(1410) (\rightarrow\eta\sigma^{+-})
\sigma^0,\nonumber\\
& \rightarrow & \eta(1410) (\rightarrow a_0^\pm(980)\pi^\mp)
\sigma^0,
\end{eqnarray}
where $\sigma^0$ and  $\sigma^{+-}$ are shorthands for the $\pi^0\pi^0$ and 
$\pi^+\pi^-$ S-waves. The latter were described by prescriptions of the form (\ref{61}) 
which is reasonable since the $\pi\pi$ masses lie below 900 MeV. Background 
contributions, e.g. from $\eta\rho^0\sigma^0$ and $\etap\rho^0$, were also 
included in the fit. 
Figure \ref{Ange} shows for example the $a_0^\pm(980)$ angular distribution in the 
$\eta(1410)$ rest frame together with the best fit for a $0^{-+}$ state. The data 
exclude $1^{++}$,  hence $\eta(1410)$ is definitively pseudoscalar and is produced from the 
$^1S_0$ atomic state. It has mass and width
\begin{equation}
\eta(1410): \ m = 1409 \pm 3 \  {\rm MeV}, \ \ \Gamma = 86 \pm 10 \ {\rm MeV}.
\end{equation}
The width is somewhat larger than for $\eta\pi\pi$ in $J/\psi$ decay (Bolton, 1992; 
Augustin, 1990)  and for $\KKbar\pi$ in $\pbarp$ annihilation at rest (Bertin, 1995).
 
The  branching ratio to $\KKbar\pi$ has been measured earlier  
(Baillon, 1967): 
\begin{equation}
B(\pbarp\rightarrow E\pi\pi, E\rightarrow\KKbar\pi) = (2.0 \pm 0.2) 
\times 10^{-
3},
\end{equation}
while Crystal Barrel finds for the $\eta\pi\pi$ mode
\begin{equation}
B(\pbarp\rightarrow \eta(1410)\pi\pi, \eta(1410)\rightarrow\eta\pi\pi) = (3.3 \pm 1.0) 
\times 10^{-
3}.
\end{equation}
The fit yields $\eta\sigma$ and $a_0(980)(\rightarrow\eta\pi)\pi$ decay contributions 
with a relative rate of 0.78 $\pm$ 0.16. 
Assuming that 50\% of the $\KKbar\pi$ mode proceeds through  $\eta(1410)$ decaying to 
$a_0(980)\pi$ (Baillon, 1967) one can estimate from these branching ratios
\begin{equation}
\frac{B(a_0(980)\rightarrow\KKbar)}{B(a_0(980)\rightarrow\eta\pi)} \sim  0.54 \pm 
0.18,
\end{equation}
somewhat larger but not in violent disagreement with the result (\ref{77}).  

The observation of the $\eta\pi^0\pi^0$ decay mode also lifts the earlier isospin 
ambiguity for the $E$ meson  and clearly establishes that this state is isoscalar ($C=+1$). 
Note that the Crystal Barrel data do not exclude the presence of the other $I=0$ 
pseudoscalar, $\eta(1460)$,  since the latter was observed in $K^*\overline{K}$ and not in 
$\eta\pi\pi$. 

The first radial excitation of the $\eta$ could be $\eta(1295)$ decaying to $\eta\pi\pi$ 
(Barnett, 1996). Hence one of two pseudoscalars in the $\iota$ structure could 
be the radial excitation of the $\etap$. The near equality of the $\eta(1295)$ and 
$\pi(1300)$ masses suggests an ideally mixed nonet of $0^{-+}$ radials. This implies 
that the second isoscalar in the nonet should be mainly $\ssbar$ and hence decays   
to $K^*\overline{K}$, in accord with observations for $\eta(1460)$. This scheme then 
favours an exotic interpretation for $\eta(1410)$, perhaps gluonium mixed with $\qqbar$ 
(Close, 1997) or a bound state of gluinos (Farrar, 1996). The gluonium interpretation is, 
however, not favoured by lattice gauge theories, which predict the $0^{-+}$ state above 
2 GeV (see Szczpaniak (1996)).

\section{Summary and Outlook}
Crystal Barrel has collected  $10^8$ $\pbarp$ annihilation at rest in liquid hydrogen, three 
orders of magnitudes more than previous bubble chamber experiments. The results 
reviewed in this report  were achieved thanks to the availability of pure, cooled and 
intensive low energy antiproton beams which allow a good spatial definition of the 
annihilation source and thanks to the refinement of the analysis tools warranted by the 
huge statistical samples. The data processed so far concentrate on annihilations at rest into 
0-prong that had not been investigated before. The data collected with the 0-prong trigger 
correspond to $6.3 \times 10^8$ annihilations.

The measurement of the branching ratio for annihilation into $\pi^0\pi^0$ leads, together 
with a cascade calculation of the antiprotonic atom, to a fraction of (13 $\pm$ 4) \% P-
wave in liquid hydrogen. Therefore S-wave dominance has been, in general, assumed to 
analyze the data. 

The branching ratios for annihilation into two neutral light mesons ($\pi^0\eta$, 
$\pi^0\etap$,  $\eta\eta$, $\eta\etap$, $\omega\eta$, $\omega\etap$,  $\eta\rho^0$, 
$\etap\rho^0$) reveal the interplay of constituent quarks in hadrons. The non-planar 
quark rearrangement graph must play an important role in the annihilation process. 
Using the OZI rule the pseudoscalar mixing angle was determined to be $(-17.3 \pm 
1.8)^\circ$.  

However, the production of $\phi$ mesons is enhanced in nearly all channels compared to 
predictions by the OZI rule. The most significant deviation is found in the annihilation 
channel $\pi^0\phi$. After phase space correction, the $\pi^0\phi/\pi^0\omega$ ratio is 
(10.6 $\pm$ 1.2) \% while OZI predicts 0.42 \%.  Whether this enhancement can be 
explained by final state corrections or by $\ssbar$ pairs in the nucleon is not clear yet. 
The analysis of data in P-state annihilations,  in gas or at higher momenta,  will be helpful 
in settling the nature of this phenomenon.

In electromagnetic processes, the radiative annihilations $\pi^0\gamma$, $\eta\gamma$ 
and $\omega\gamma$ have been observed with rates consistent with predictions from 
VDM, but $\phi\gamma$ may be enhanced. The branching ratio for 
$\omega\rightarrow\eta\gamma$, $(6.6 \pm 1.7)\times 10^{-4}$, was measured 
independently of $\ro$ interference. This results solves the ambiguity in $e^+e^-$ 
formation experiments, selecting the constructive $\ro$ interference solution. The 
$\eta\rightarrow 3\pi^0$ Dalitz plot is not homogeneous but shows a negative slope of 
$\alpha$ =  0.052 $\pm$ 0.020. Crystal Barrel data also confirms the evidence for the direct 
decay $\etap\rightarrow\pi^+\pi^-\gamma$, in addition to 
$\etap\rightarrow\rho\gamma$.

A decisive progress has been achieved in understanding scalar mesons by studying 
annihilation into three pseudoscalars. An isovector state, $a_0(1450)$,  with mass and 
width $(m, \Gamma)$ = (1474 $\pm$ 19, 265 $\pm$ 13) MeV has been observed to decay 
into $\eta\pi$, $\etap\pi$ and $\KKbar$ with rates compatible with SU(3) flavor. The 
existence of $a_0(1450)$ adds evidence for $a_0(980)$ not being $\qqbar$, but perhaps a 
$\KKbar$ molecule. The ratio of $a_0(980)$ decay rates to $\KKbar$ and $\eta\pi$ was 
measured to be 0.24 $\pm$ 0.06.

An isoscalar state, $f_0(1370)$, with mass and width $(m, \Gamma)$ = (1360 $\pm$ 23, 351   
$\pm$ 41) MeV has been observed to decay into $\pi\pi$, $\eta\eta$, $\KKbar$ and $4\pi$. 
Obtaining accurate branching ratios for $f_0(1370)$ is difficult due to  interferences with 
the broad structure $f_0(400-1200)$. The nature of this structure (meson or slowly moving 
background phase) is unclear. Whether $f_0(400-1200)$ is really distinct from $f_0(1370)$ 
is not entirely clear. Both questions will probably remain with us for some time. The states 
$a_0(1450)$, $f_0(1370)$ and $K_0^*(1430)$ are broad, consistent with expectations for 
$\qqbar$ scalar mesons. The small coupling of $f_0(1370)$ to $\KKbar$ makes it an 
unlikely candidate for the $\ssbar$ meson, which is therefore still missing. More 
information on this state will hopefully emerge in the mass range above 1600 MeV from 
Crystal Barrel 
data in flight.

An additional isoscalar state, $f_0(1500)$,  with mass and width $(m, \Gamma)$ = (1505 
$\pm$ 9, 111  $\pm$ 12) MeV has been observed to decay into $\pi\pi$, $\eta\eta$, 
$\eta\etap$, $\KKbar$ and $4\pi$. The decay branching ratios are 29, 5, 1 , 3 and 62  \%, 
respectively. These rates exclude this state to be the missing $\ssbar$. Hence $f_0(1500)$ is 
supernumerary and anyway too narrow to be easily accommodated in the scalar nonet. The 
likely explanation is that $f_0(1500)$ is the ground state glueball predicted by QCD, mixed 
with the two nearby $\qqbar$  isoscalars, $f_0(1370)$ and the higher lying $\ssbar$ state.

The tensor $f_2(1565)$ is dominantly produced from P-states. It is, however, still required 
to fit the data when assuming pure S-wave annihilation: Its mass and width are $\simeq$ 
1552 and $\simeq$ 142 MeV. The systematic inclusion of P-wave annihilation at rest in all 
analyses is, however,  prevented by the large number of fit parameters. Data from Crystal 
Barrel and Obelix in liquid and  gaseous hydrogen might alleviate this problem, perhaps 
also modifying slightly some of the branching ratios obtained from liquid only.

A $0^{-+}$ state, $\eta(1410)$, with mass and width $(m, \Gamma)$ = (1409 $\pm$ 3, 80   
$\pm$ 10) MeV has been observed to decay into $\eta\pi^0\pi^0$ and $\eta\pi^+\pi^-$ 
with approximately equal rates 
through $a_0(980)\pi$ and $(\pi\pi)_S\eta$ . The neutral 
decay mode establishes this state as an isoscalar and, together with other experiments, 
strengthens the evidence for two $I=0$ pseudoscalars in the 1400 - 1500 MeV region.  

Crystal Barrel data also show evidence for the two excitations of the $\rho$ meson, 
$\rho(1450)$ and $\rho(1700)$ and for the radial excitation $a_2'(1650)$.  The analysis of 
in-flight data will hopefully reveal further radial excitations, hybrid mesons and higher 
mass glueballs.  For glueballs, a  more definitive progress will probably be achieved in 
radiative $J/\psi$ decay at a high luminosity $e^+e^-$ factory or in central collisions at the 
forthcoming Compass experiment at CERN. 

\section{Acknowledgements}
The results presented in this review have been obtained through the efforts of many 
physicists and technicians from the Crystal Barrel collaboration\footnote{Universities of 
California (Berkeley and Los Angeles), Bochum, Bonn, Hamburg, Karlsruhe, Mainz, 
Munich, Paris VI, Zurich, Carnegie Mellon University,  Academy of Science (Budapest), 
Rutherford Appleton Laboratory,  CERN, Queen Mary and Westfield College, Centre de 
Recherches Nucl\'eaires (Strasbourg).}  and from the supporting 
institutions. The author also wishes to thank C. Batty, F. Close, H. Genz, P. Giarritta, F. 
Heinsius, M. Heinzelmann, M. Locher, V. Markushin, L. Montanet, R. Ouared, S. Spanier, U. 
Wiedner and \v{C}. Zupan\v{c}i\v{c} for reading the manuscript and for helpful 
discussions.

\section*{References}
\begin{flushleft}
Abele, A., {\it et al.}, 1996a, Nucl. Phys. {\bf A 609}, 562.\\
Abele, A., {\it et al.}, 1996b, {\it  Observation of} $f_0(1500)$ {\it decay into} $K_LK_L$, 
Phys. Lett. {\bf B 385}, 425.\\
Abele, A., {\it et al.}, 1996c, {\it A study  of} $f_0(1500)$ {\it decays into} $4\pi^0$ {\it in} 
$\pbarp\rightarrow 5\pi^0$ {\it at rest}, Phys. Lett. {\bf B 380}, 453.\\
Abele, A., {\it et al.}, 1997a, $\ro$ {\it interference in} $\pbarp$ {\it annihilation at rest 
into} $\pi^+\pi^-\eta$, submitted to Phys. Lett. {\bf B}.\\
Abele, A., {\it et al.}, 1997b, {\it Momentum Dependence of the Decay} 
$\eta\rightarrow\pi^+\pi^-\pi^0$, submitted to Phys. Lett. {\bf B}.\\
Abele, A., {\it et al.}, 1997c, {\it Decay dynamics of the process} $\eta\rightarrow 3\pi^0$, 
submitted to Phys. Lett. {\bf B}.\\
Abele, A., {\it et al.}, 1997d, {\it Measurement of the} $\oeg$ {\it decay branching ratio}, 
submitted to Phys. Lett. {\bf B}.\\
Abele, A., {\it et al.}, 1997e, $\pbarp$ {\it annihilation at rest into} 
$K_LK^{\pm}\pi^{\mp}$, submitted to Phys. Lett. {\bf B}.\\
Abele, A., {\it et al.}, 1997f, {\it Antiproton-proton annihilation at rest into} 
$K_LK_S\pi^0\pi^0$, submitted to Phys. Lett. {\bf B}.\\
Abele, A., {\it et al.}, 1997g, {\it Study of the} $ \pi^0\pi^0\etap$ {\it final state in} 
$\pbarp$ {\it annihilation at rest}, Phys. Lett. {\bf B 404}, 179.\\
Abele, A., {\it et al.}, 1997h, {\it High-mass} $ \rho$ {\it meson states from} 
$\pbar d$ {\it annihilation at rest into} $\pi^-\pi^0\pi^0 p_{spectator}$, Phys. Lett. 
{\bf B 391}, 191.\\
Abele, A., {\it et al.}, 1997i, {\it Measurement of the decay distribution of} $ 
\etap\rightarrow\pi^+\pi^-\gamma$ {\it and evidence for the box anomaly}, Phys. Lett. 
{\bf B 402}, 195. \\
Ableev, V.G., {\it et al.}, 1994, Phys. Lett. {\bf B 334}, 237.\\
Ableev, V.G., {\it et al.}, 1995, Nucl. Phys. {\bf A 585}, 577.\\
Achasov, N.N., and G.N. Shestakov, 1978, Sov. J. Part. Nucl. {\bf 9}, 19.\\
Adamo, A., {\it et al.}, 1992, Phys. Lett. {\bf B 287}, 368.\\
Adiels, L., {\it et al.}, 1987, Z. Phys. {\bf C 35}, 15.\\
Adiels, L., {\it et al.}, 1989, Z. Phys. {\bf C 42}, 49.\\
Adomeit, J., {\it et al.}, 1996, {\it Evidence for two isospin zero} $J^{PC}=2^{-+}$ {\it mesons 
at 1645 and 1875 MeV}, Z. Phys. {\bf C 71}, 227.\\
Ahmad, S., {\it et al.}, 1985, Phys. Lett. {\bf 157 B}, 333.\\
Aitchison, I.J.R., 1972, Nucl. Phys. {\bf A 189}, 417.\\
Aker, E., {\it et al.}, 1991, {\it Observation of a} $2^{++}$ {\it resonance at 1515 MeV in 
proton-antiproton annihilations into} $3\pi^0$, Phys. Lett. {\bf B 260}, 249.\\
Aker, E., {\it et al.}, 1992, {\it The Crystal Barrel spectrometer at LEAR}, Nucl. Instrum. 
Methods {\bf A 321}, 69.\\
Alde, D., {\it et al.}, 1984, Z. Phys. {\bf C 25}, 225.\\
Alde, D., {\it et al.}, 1987, Phys. Lett. {\bf  B 198}, 286.\\
Alde, D., {\it et al.}, 1988a, Phys. Lett. {\bf B 205}, 397.\\
Alde, D., {\it et al.}, 1988b, Phys. Lett. {\bf B 201}, 160.\\
Alde, D., {\it et al.}, 1990, Phys. Lett. {\bf B 241}, 600.\\  
Alde, D., {\it et al.}, 1994, Z. Phys. {\bf C 61}, 35.\\
Amelin, D.V., {\it et al.}, 1996, Phys. At. Nucl. {\bf 59}, 976.\\  
Amsler, C., and J.C. Bizot, 1983, Comp. Phys. Comm. {\bf 30}, 21.\\
Amsler, C., 1987, in {\it Advances in Nuclear Physics}, edited by J.W. Negele and E. Vogt 
(Plenum, New York), Vol. 18, p. 183.\\ 
Amsler, C., and F. Myhrer, 1991, Ann. Rev. Nucl. Part. Sci. {\bf 41}, 219.\\
Amsler, C., {\it et al.}, 1992a, {\it P- versus S-wave} $\pbarp$ {\it annihilation at rest in}  
$LH_2$, Phys. Lett. {\bf B 297}, 214.\\
Amsler, C., {\it et al.}, 1992b, {\it The pseudoscalar mixing angle} $\Theta_{PS}$ {\it from}  
$\eta$ {\it and} $\etap$ {\it production in} $\pbarp$ {\it annihilation at rest}, Phys. Lett. 
{\bf B 294}, 451.\\
Amsler, C., {\it et al.}, 1992c, {\it Proton-antiproton annihilation into} $\eta\eta\pi$ {\it 
- observation of a scalar resonance decaying into}  $\eta\eta$, Phys. Lett. {\bf B 291}, 
347.\\
Amsler, C., {\it et al.}, 1993a, {\it Antiproton-proton annihilation at rest into} 
$\omega\pi^0\pi^0$, Phys. Lett. {\bf B 311}, 362.\\
Amsler, C., {\it et al.}, 1993b, {\it Antiproton-proton annihilation at rest into two-body 
final states}, Z. Phys. {\bf C 58}, 175.\\
Amsler, C., {\it et al.}, 1993c, {\it Radiative protonium annihilation into} $\gamma\gamma, 
\gamma\pi^0, \gamma\eta, \gamma\omega$ {\it and} $\gamma\etap$, Phys. Lett.  {\bf B 
311}, 371.\\
Amsler, C., {\it et al.}, 1993d, {\it Protonium annihilation into} $K^0_LK^0_S\pi^0$ {\it and}    
$K^0_LK^0_S\eta$, Phys. Lett.  {\bf B 319}, 373.\\
Amsler, C., {\it et al.}, 1994a, {\it Search for a new light gauge boson in decays of} $\pi^0$ 
{\it and} $\eta$, Phys. Lett.  {\bf B 333}, 271.\\
Amsler, C., {\it et al.}, 1994b, {\it Observation of a new} $I^G(J^{PC})=1^-(0^{++})$ 
{\it resonance at 1450 MeV}, Phys. Lett.  {\bf B 333}, 277.\\
Amsler, C., {\it et al.}, 1994c, {\it Study of} $\pbarp$  {\it annihilation at rest into} 
$\omega\eta\pi^0$, Phys. Lett.  {\bf B 327}, 425.\\
Amsler, C., {\it et al.}, 1994d, {\it Observation of a scalar resonance decaying to} $\pi^+\pi^-
\pi^0\pi^0$  {\it in} $\pbarp$ {\it annihilation at rest}, Phys. Lett.  {\bf B 322}, 431.\\
Amsler, C., {\it et al.,}, 1994e, in {\it Hadron '93}, Proceedings of the Hadron 93 Conf., edited 
by T. Bressani {\it et al.}, Nuovo Cimento {\bf 107 A}, 1815.\\
Amsler, C., {\it et al.}, 1994f, $\eta\etap$ {\it threshold enhancement in} $\pbarp$  {\it  
annihilation into} $\pi^0\eta\etap$ {\it at rest}, Phys. Lett.  {\bf B 340}, 259.\\
Amsler, C., {\it et al.}, 1995a, {\it First observations of Pontecorvo reactions with a 
recoiling neutron}, Z. Phys. {\bf A 351}, 325.\\
Amsler, C., {\it et al.}, 1995b, {\it First observation of the production of nucleon resonances 
in antiproton annihilation in liquid deuterium}, Phys. Lett. {\bf B 352}, 187.\\
Amsler, C., {\it et al.}, 1995c, {\it Observation of radiative} $\pbarp$ {\it annihilation into 
a} $\phi$ {\it meson}, Phys. Lett. {\bf B 346}, 363.\\
Amsler, C., {\it et al.}, 1995d, $\eta$ {\it decays into three pions}, Phys. Lett. {\bf B 346}, 
203.\\
Amsler, C., {\it et al.}, 1995e, {\it High statistics study of} $f_0(1500)$ {\it decay into}  
$\eta\eta$, Phys. Lett. {\bf B 353}, 571.\\
Amsler, C., {\it et al.}, 1995f, {\it High-statistics study of} $f_0(1500)$ {\it decay into}  
$\pi^0\pi^0$, Phys. Lett. {\bf B 342}, 433.\\
Amsler, C., {\it et al.}, 1995g, {\it Coupled channel analysis of} $\pbarp$ {\it annihilation 
into}  $\pi^0\pi^0\pi^0$, $\pi^0\eta\eta$ {\it and}  $\pi^0\pi^0\eta$, Phys. Lett. {\bf B 355}, 
425.\\
Amsler, C., {\it et al.}, 1995h, {\it E decays to} $\eta\pi\pi$ {\it in} $\pbarp$ {\it 
annihilation at rest}, Phys. Lett. {\bf B 358}, 389.\\
Amsler, C., {\it et al.}, 1996a, {\it Search for a new light gauge boson in} $\pi^0, \eta$ {\it 
and} $\etap$ {\it decays}, Z. Phys.  {\bf C 70}, 219.\\
Amsler C., and F.E. Close, 1996, Phys. Rev. {\bf D 53}, 295; see also Amsler C., and F.E. Close, 
1995, Phys. Lett. {\bf B 353}, 385. \\
Andrews, D.E., {\it et al.}, 1977, Phys. Rev. Lett. {\bf 38}, 198.\\
Anisovich, V.V., {\it et al.}, 1994, {\it Observation of two} $J^{PC} = 0^{++}$ {\it isoscalar 
resonances at 1365 and 1520 MeV}, Phys. Lett. {\bf B 323}, 233.\\
Antinori, F., {\it et al.}, 1995, Phys. Lett. {\bf B 353}, 589.\\
Armenteros, R., {\it et al.}, 1965, Phys. Lett. {\bf 17}, 170.\\
Armenteros R., and B. French, 1969, in $\NNbar$ {\it Interactions in High Energy 
Physics}, 
edited by E.H.S. Burhop (Academic, New York), Vol. 4, p. 237.\\
Armstrong, T., {\it et al.}, 1993, Phys. Lett. {\bf B 307}, 394, 399.\\
Aston, D., {\it el al.}, 1988a, Phys. Lett. {\bf B 201}, 573.\\
Aston, D., {\it el al.}, 1988b, Nucl. Phys. {\bf B 296}, 493.\\
Atiya, M.S., {\it et al.}, 1992, Phys. Rev. Lett. {\bf 69}, 733.\\
Au, K.L., D. Morgan, and M.R. Pennington, 1987, Phys. Rev. {\bf D 35}, 1633.\\
Augustin, J.-E., {\it et al.}, 1990, Phys. Rev. {\bf D 42}, 10.\\
Badalayan, A.M., L.P. Kok, M.I. Policarpov, and Yu. A. Simonov, 1982, Phys. Rep. {\bf 82},  
31.\\
Bai, Z., {\it et al.}, 1990, Phys. Rev. Lett. {\bf 65}, 2507.\\
Baillon, P., {\it et al.}, 1967, Nuovo Cimento {\bf 50 A}, 393.\\ 
Baker, C.A., {\it et al.}, 1988, Nucl. Phys. {\bf A 483}, 631.\\
Bali, G.S., {\it et al.}, 1993, Phys. Lett. {\bf B 309}, 378.\\
Baltay, C., {\it et al.}, 1966, Phys. Rev. {\bf B 145}, 1103.\
Baltrusaitis, R.M., {\it et al.}, 1987, Phys. Rev. {\bf D 35}, 2077.\\
Barash, N., {\it et al.}, 1965, Phys. Rev. {\bf B 139}, 1659.\\
Barnes, T., 1985, Phys. Lett. {\bf 165 B}, 434.\\
Barnes, T., F.E. Close, P.R. Page, and E.S. Swanson, 1997, Phys. Rev. {\bf D 55}, 4157.\\
Barnett, R.M. {\it et al.} (Particle Data Group), 1996, Phys. Rev. {\bf D 54}, 1.\\
Bassompierre, G., {\it et al.}, 1976, Phys. Lett. {\bf B 64}, 475.\\
Batty, C.J., 1989, Rep. Prog. Phys. {\bf 52}, 1165.\\
Batty, C.J., 1996, Nucl. Phys. {\bf A 601}, 425.\\
Beladidze, G.M., {\it et al.}, 1992, Sov. J. Nucl. Phys. {\bf 55}, 1535.\\
Benayoun, M., {\it et al.}, 1993, Z. Phys. {\bf 58}, 31. \\
Bertin, A., {\it et al.}, 1995, Phys. Lett. {\bf B 361}, 187.\\ 
Bertin, A., {\it et al.}, 1996, Phys. Lett.  {\bf B 388}, 450.\\
Bertin, A., {\it et al.}, 1997, Phys. Lett.  {\bf B 400}, 226.\\
Binon, F. , {\it et al.}, 1983, Nuovo Cimento {\bf 78 A}, 313.\\
Binon, F. , {\it et al.}, 1984, Nuovo Cimento {\bf 80 A}, 363.\\
Bityukov, S.I., {\it et al.}, 1987, Phys. Lett. {\bf B 188}, 383.\\
Bityukov, S.I., {\it et al.}, 1991, Z. Phys. {\bf C 50}, 451.\\
Bizzarri, R., {\it et al.}, 1969, Nucl. Phys. {\bf B 14}, 169. \\
Bizzarri, R., {\it et al.}, 1971, Nucl. Phys. {\bf B 27}, 140. \\
Bl\"um, P., 1996, Int. J. Mod. Phys. {\bf A 11}, 3003.\\
Bolton, T., {\it et al.}, 1992, Phys. Rev. Lett. {\bf 69}, 1328.\\
Bridges, D., I. Daftari, and T.E. Kalogeropoulos, 1987, Phys. Rev. Lett. {\bf 57}, 1534.\\
Bugg, D.V., V.V. Anisovich, A. Sarantsev, and B.S. Zou, 1994, Phys. Rev. {\bf D 50}, 4412.\\
Bugg, D.V., {\it et al.}, 1995, Phys. Lett. {\bf B 353}, 378.\\
Bugg, D.V., A.V. Sarantsev, and B.S. Zou, 1996, Nucl. Phys. {\bf B 471}, 59.\\
Carbonell, J., G. Ihle, and J.M. Richard, 1989, Z. Phys. {\bf A 334}, 329.\\
Chamberlain, O., E. Segr\`e, C.E. Wiegand, and T. Ypsilantis, 1955, Phys. Rev. {\bf 100}, 947.\\
Chew, F. G., and  S. Mandelstam, 1960, Phys. Rev. {\bf 119}, 467.\\
Chiba, M., {\it et al.}, 1987, in {\it Physics at LEAR with Low Energy Antiprotons}, Nuclear 
Science Research Conf. Series, Vol. 14, edited by C. Amsler {\it et al.}, (Harwood, Chur), p. 
401.\\
Chiba, M., {\it et al.}, 1988, Phys. Rev. {\bf D 38}, 2021, see also Chiba, M., {\it et al.}, 1989, 
Phys. Rev. {\bf D 39}, 3227. \\
Chung, S.U., {\it et al.}, 1995, Ann. Phys. {\bf 4}, 404.\\
Close, F.E., G.R. Farrar, and Z. Li, 1997,  Phys Rev {\bf D 55}, 5749.\\
Close, F.E., N. Isgur, and S. Kumano, 1993, Nucl. Phys. {\bf B 389}, 513.\\
Close, F.E., and P.R. Page, 1995, Nucl. Phys. {\bf B 443}, 233.\\
Conforto, B., {\it et al.}, 1967, Nucl. Phys. {\bf B 3}, 469.\\
Cooper, A. M. , {\it et al.}, 1978, Nucl. Phys. {\bf B 146}, 1.\\
Day, T.B., G.A. Snow, and T. Sucher, 1960, Phys. Rev. {\bf 118}, 864.\\
Delcourt, B., J. Layssac, and E. Pelaquier, 1984, in {\it Physics at LEAR with Low Energy 
Antiprotons}, edited by U. Gastaldi and R. Klapisch, (Plenum, New York), p. 305.\\
Devons, S., {\it et al,}, 1971, Phys. Rev. Lett. {\bf 27}, 1614.\\
Devons, S., {\it et al,}, 1973, Phys. Lett. {\bf 47 B}, 271.\\
Dobroliubov, M.I., and A. Yu. Ignatiev, 1988, Nucl. Phys. {\bf B 309}, 655.\\
Dobroliubov, M.I., 1990, Sov. J. Nucl. Phys. {52}, 352.\\
Dolinsky, S.I. {\it et al.}, 1989, Z. Phys. {\bf C 42}, 511.\\
Dombrowski, S. v., 1996, Ph.D. thesis (University of Zurich).\\
Donoghue, J.F., B.R. Holstein, and D. Wyler, 1992, Phys. Rev. Lett. {\bf 69}, 3444.\\
Doser, M., {\it et al.}, 1988, Nucl. Phys. {\bf A 486}, 493.\\
Dover, C.B., 1986, Phys. Rev. Lett. {\bf 57}, 1207.\\
Dover, C.B., and P. M. Fishbane, 1989, Phys. Rev. Lett. {\bf 62}, 
2917.\\
Dover, C.B., T. Gutsche, and A. Faessler, 1991, Phys. Rev. {\bf C 43}, 379.\\
Dover, C.B., and J.M. Richard, 1980, Phys. Rev. {\bf C 21}, 1466.\\
Ellis, J. , M. Karliner, D.E. Kharzeev, and M.G. Sapozhnikov, 1995, 
Phys. Lett. {\bf B 353}, 319.\\
Espigat, P., C. Guesqui\`ere, E. Lillestol, and L. Montanet, 1972, Nucl. Phys. {\bf B 36}, 93.\\
Farrar, G.R., 1996, Phys. Rev. Lett. {\bf 76},  4111.\\ 
Flatt\'e, S.M., 1976, Phys. Lett. {\bf 63 B}, 224.\\
Foster, M., {\it et al.}, 1968a, Nucl. Phys. {\bf B 8}, 174.\\
Foster, M., {\it et al.}, 1968b, Nucl. Phys. {\bf B 6}, 107.\\
Gaspero, M., 1993, Nucl. Phys. {\bf A 562}, 407.\\
Gasser, J., and H. Leutwyler, 1985, Nucl. Phys. {\bf B 250}, 539.\\
Genz, H., 1983, Phys. Rev. {\bf D 28}, 1094.\\
Genz, H., 1985, Phys. Rev. {\bf D 31}, 1136.\\
Genz., H., M. Martinis, and S. Tatur, 1990, Z. Phys. {\bf A 335}, 87.\\
Gilman, F.G., and R. Kauffman, 1987, Phys. Rev. {\bf D 36}, 2761.\\
Godfrey, S., and N. Isgur, 1985, Phys. Rev. {\bf D 32}, 189.\\
Goldhaber, A.S., G.C. Fox, and C. Quigg, 1969, Phys. Lett. {\bf 30 B}, 249.\\
Gortchakov, 0., M.P. Locher, V.E. Markushin, and S. von Rotz, 1996, Z. Phys. {\bf A 353}, 
447.\\
Gray, L., {\it et al.}, 1983, Phys. Rev. {\bf D 27}, 307.\\
Grayer, G., {\it et al.}, 1974, Nucl. Phys. {\bf B 75}, 189.\\
Guesqui\`ere, C., 1974, in {\it Antinucleon-Nucleon Interactions}, Proceedings of the 
Liblice Symposium, CERN Yellow Report 74-18, p. 436.\\
Hartmann, U., E. Klempt, and J. K\"orner, 1988, Z. Phys. {\bf A 331}, 217.\\
Heinzelmann, M., 1996, Diploma thesis (University of Zurich).\\
Hippel, F.v., and C. Quigg, 1972, Phys. Rev. {\bf D 5}, 624; see also Blatt, J.M.,  and V.F. 
Weisskopf, 1952, in {\it Theoretical Nuclear Physics}, (Wiley, New York), p. 361.\\
Isgur, N., and J. Paton, 1985, Phys. Rev. {\bf D 31}, 2910.\\
Jacob, M., and G.C. Wick, 1959, Ann. Phys. (NY)  {\bf 7}, 404.\\
Jaenicke, J., B. Kerbikov, and H.-J. Pirner, 1991, Z. Phys. {\bf A 339}, 297.\\
Kambor, J., C. Wiesendanger and D. Wyler, 1996, Nucl. Phys. {\bf B 465}, 215.\\
Klempt, E., 1990, Phys. Lett. {\bf B 244}, 122.\\
Kolybashov, V.M., I.S. Shapiro, and Yu.N. Sokolskikh, 1989, Phys. Lett. {\bf B 222}, 135.\\
Lacock, P., C. Michael, P. Boyle, and P. Rowland, 1997, Phys. Lett. {\bf B 401}, 308.\\
Landua, R., 1996, Ann. Rev. Nucl. Part. Sci. {\bf 46}, 351.\\
Layter, J.G., {\it et al.}, 1973, Phys. Rev. {\bf D 7}, 2565. \\
Locher, M.P., Y. Lu, and B.S. Zou, 1994, Z. Phys. {\bf A 347}, 281.\\
Markushin, V.E., 1997, Nucl. Phys. B (Proc. Suppl.) {\bf 56 A}, 303. \\
May, B., {\it et al.}, 1989, Phys. Lett. {\bf 225 B}, 450.\\
May, B., {\it et al.}, 1990, Z. Phys. {\bf C 46}, 191, 203.\\
Meijer Drees, R., {\it et al.}, 1992, Phys. Rev. Lett. {\bf 68}, 3845.\\
Morgan, D., and M.R. Pennington, 1993, Phys. Rev. {\bf D 48}, 1185.\\
O$'$Connell, H.B., {\it et  al.}, 1995,  Phys. Lett. {\bf B 354}, 14.\\
O$'$Donnell, P.J., 1981, Rev. Mod. Phys. {\bf  53}, 673.\\
Okubo, S., 1963, Phys. Lett. {\bf 5}, 165.\\
Ore, A., and J.L. Powell, 1949, Phys. Rev. {\bf 75}, 1696.\\
Orfanidis, S.J., and V. Rittenberg, 1973, Nucl. Phys. {\bf B59}, 570.\\
Pais, A., 1960, Ann. Phys. (NY), {\bf 9}, 548.\\
Peters, K., and E. Klempt, 1995, Phys. Lett. {\bf 352}, 467.\\
Pietra, C., 1996, Diploma thesis (University of Zurich).\\
Prokoshkin, Yu. D., and V.D. Samoilenko, 1995, Sov. Phys. Dokl. {\bf 342}, 273.\\
Regenfus, C., 1997, Nucl. Instrum. Methods {\bf A 386}, 60.\\
Reifenr\"other, J., {\it et al.}, 1991,  Phys. Lett. {\bf B 267}, 299.\\
Rosselet, L., {\it et al.}, 1977, Phys. Rev. {\bf D 15}, 574.\\
Scharre, D.L., {\it et al.}, 1980, Phys. Lett. {\bf B 97}, 329.\\
Schmid, B., 1991, Ph.D. thesis (University of Zurich).\\
Sedl\'ak, J., and V. \v{S}im\'ak, 1988, Sov. J. Part. Nucl. {\bf 19}, 191.\\
Sexton, J., A. Vaccarino, and D. Weingarten, 1995, Phys. Rev. Lett. {\bf 75}, 4563. \\
Spanier, S., 1994, Ph.D. thesis (University of Mainz).\\
Spanier, S., 1996, Yad. Fiz. {\bf 59}, 1352.\\
Spanier, S., 1997, Workshop on {\it The Strange Structure of the Nucleon}, CERN.\\
Szczepaniak, A., E.S. Swanson, C.R. Ji, and S.R. Cotanch, 1996, Phys. Rev. Lett. {\bf 76}, 
2011.\\
Thompson, D.R., {\it et al.}, 1997, Phys. Rev. Lett. (in print) \\
T\"ornqvist, N.A., 1991, Phys. Rev. Lett. {\bf 67}, 556.\\ 
T\"ornqvist, N.A., 1995, Z. Phys. {\bf C 68}, 647.\\
T\"ornqvist, N.A., and M. Roos, 1996, Phys. Rev. Lett. {\bf 76}, 1575.\\
Urner, D., 1995, Ph.D. thesis (University of Zurich).\\
Vandermeulen, J., 1988, Z. Phys. {\bf C 37}, 563.\\
Weidenauer, P., {\it et al.}, 1990, Z. Phys. {\bf C 47}, 353.\\
Weidenauer, P., {\it et al.}, 1993, Z. Phys. {\bf C 59}, 387.\\
Weinstein, J., and N. Isgur, 1990, Phys. Rev. {\bf D 41}, 2236.\\
Zemach, Ch., 1964, Phys. Rev. {\bf B 133}, 1201.\\ 
Zemach, Ch., 1965, Phys. Rev. {\bf B 140}, 97, 109.\\ 
\end{flushleft}

\clearpage

\begin{table}[p]
\caption[]{Summary of data (in millions of events) 
with a minimum bias trigger (MB), for 0-, 2-, 4-prong and with more specialized triggers at 
rest (first three rows) and in liquid hydrogen at high $\pbar$ momenta (in MeV/c).\\ 
LH$_2$: liquid hydrogen; LD$_2$: liquid deuterium; GH$_2$: gaseous hydrogen.\\
$^a$ $K_S(\rightarrow\pi^+\pi^-)X$, $^b$ $\pi^+\pi^-\pi^0\pi^0\eta$, $^c$ 
$\pi^0\eta\eta$, 
$^d$ 1-prong, $^e$ 3-prong, $^f$ $\pi^+\pi^-\pi^0\eta$,  $^g$ $\pi^+\pi^-\eta$}
\label{data}
\begin{center}
\begin{tabular}{r r r r r r r r r}
 & MB& 0  & 2 & 4 & \multicolumn{3}{c}{Triggers}  \\
\hline
LH$_2$ & 16.8 & 24.7 & 19.3  &  10.4 &  16.9$^a$ & 8.5$^b$ & 4.0$^c$ & 0.4$^g$ \\
LD$_2$ &  3.2 &  6.0  & 0.5  &   0.5 & 11.9$^a$ & 8.1$^d$ & 11.7$^e$ &\\
GH$_2$  &  8.6 & 18.0 & 14.3 &  8.2 & 6.4$^a$  && &\\
\hline
600 & 1.3 &  5.9 &  2.2 &  &  1.2$^a$  && &\\
900 &  & 20.0 & 19.4 & &  && &\\
1050 & 0.3 & 7.2 & & &   &&&\\
1200 & 1.6 & 10.4 & 6.0 & & 2.0$^f$  &&& \\
1350 & 1.0   &  11.5 &  &  &  &&&\\
1525 & 20.6 & 10.2 &  4.5 &  && \\
1642 &  0.1 & 11.1 & 12.5 &  && \\
1800 & & 6.8 & 3.6 &  &&  \\
1900 & 13.6 & 14.5  & 15.7 &  && \\
\end{tabular}
\end{center}
\end{table}

\begin{table}[p]
\caption[]{Branching ratios $B$ for $\pbarp$ annihilation at 
rest in liquid. See Amsler and Myhrer 
(1991) for annihilation in gaseous hydrogen. Further branching ratios from Dalitz plot 
analyses are listed in Table \ref{BRTB} below.\\
$^a$ From $\omega\rightarrow\pi^0\gamma$\\
$^b$ From $\omega\rightarrow\pi^+\pi^-\pi^0$\\
$^c$ average between Baltay (1966), Espigat (1972) and Foster (1968a) \\
$\ddagger$ Crystal Barrel experiment}
\begin{center}
\begin{tabular}{l r c r r l }
Channel & \multicolumn{4}{c}{$B$} & Reference \\
\hline
$e^+e^-$ & 3.2 &  $\pm$ & 0.9 & 10$^{-7}$ & Bassompierre 
(1976) \\
$\pi^0\pi^0$ & 6.93 & $\pm$ & 0.43 & 10$^{-4}$ & Amsler 
(1992a)$\ddagger$ \\
& 4.8 &$\pm$& 1.0 & 10$^{-4}$ & Devons (1971)\\
$\pi^+\pi^-$ & 3.33 &$\pm$& 0.17 & 10$^{-3}$ & Armenteros and French 
(1969) \\
$\pi^+\pi^-$ & 3.07 &$\pm$& 0.13 & 10$^{-3}$ & Amsler 
(1993b)$\ddagger$\\
$\pi^0\eta$ & 2.12 & $\pm$ & 0.12 & 10$^{-4}$ & Amsler 
(1993b)$\ddagger$\\
$\pi^0\etap$ & 1.23 & $\pm$ & 0.13& 10$^{-4}$ & Amsler 
(1993b)$\ddagger$\\
$\pi^0\rho^0$& 1.72 &$\pm$& 0.27 & 10$^{-2}$ & 
Armenteros and French (1969) \\
$\pi^{\pm}\rho^{\mp}$ & 3.44 &$\pm$& 0.54 & 10$^{-2}$ &
Armenteros and French (1969) \\
$\eta\eta$ & 1.64 & $\pm$ & 0.10 & 10$^{-4}$ & Amsler 
(1993b)$\ddagger$\\
$\eta\etap$ & 2.16 & $\pm$ & 0.25 & 10$^{-4}$ & Amsler 
(1993b)$\ddagger$\\
$\omega\pi^0$ & 5.73 & $\pm$ & 0.47 & 10$^{-3}$ & Amsler 
(1993b)$^a$ $\ddagger$\\
 & 6.16 & $\pm$ & 0.44 & 10$^{-3}$ & Schmid  (1991)$^b$ 
$\ddagger$\\
$\omega\eta$ & 1.51 & $\pm$ & 0.12 & 10$^{-2}$ & Amsler 
(1993b)$^a$ $\ddagger$\\
& 1.63 & $\pm$ & 0.12 & 10$^{-2}$ & Schmid  (1991)$^b$ 
$\ddagger$\\
$\omega\etap$ & 0.78 & $\pm$ & 0.08 & 10$^{-2}$ & Amsler 
(1993b)$\ddagger$\\
$\omega\omega$ & 3.32 & $\pm$ & 0.34 & 10$^{-2}$ & Amsler 
(1993b)$\ddagger$\\
$\eta\rho^0$ & 4.81 &$\pm$& 0.85 & 10$^{-3}$ & $^c$ \\
& 3.87 &$\pm$& 0.29 & 10$^{-3}$ & Abele (1997a)$\ddagger$ \\  
$\etap\rho^0$ &1.29 &$\pm$& 0.81 & 10$^{-3}$ & Foster (1968a) 
\\
&1.46 &$\pm$& 0.42 & 10$^{-3}$ & Urner (1995)$\ddagger$  
\\
$\rho^0\rho^0$ & 1.2 &$\pm$& 1.2 & 10$^{-3}$ & Armenteros and French 
(1969)\\
$\rho^0\omega$ & 2.26 &$\pm$& 0.23 & 10$^{-2}$ & Bizzarri 
(1969) \\
$K^+K^-$ &1.01 &$\pm$& 0.05 & 10$^{-3}$ & Armenteros and French (1969) \\
$K^+K^-$ &0.99 &$\pm$& 0.05 & 10$^{-3}$ & Amsler  
(1993b)$\ddagger$ \\
$K_SK_L$& 7.6 &$\pm$& 0.4 & 10$^{-4}$ & Armenteros and French (1969) \\
$K_SK_L$& 9.0 &$\pm$& 0.6 & 10$^{-4}$ & Amsler 
(1995c)$\ddagger$\\
\label{BR}
\end{tabular}
\end{center}
\end{table}

\begin{table}[p]
\caption[]{Pseudoscalar mixing angle $\theta_p$ derived from the 
measured ratios of two-body branching ratios ($\theta_i = 
35.3^\circ$). The first four rows   
assume only the QLR in the annihilation process. The last six rows 
assume in addition the dominance of the annihilation graph A.}
\label{QLR}
\begin{center}
\begin{tabular}{l  l r c r }
Ratio  & Prediction & \multicolumn{3}{c}{$\theta_p$ [$^{\circ}$]}\\
\hline
$\frac{\tilde{B}(\pi^0\eta)}{\tilde{B}(\pi^0\etap)}$   & 
$\tan^2(\theta_i - \theta_p)$ & 
-18.1 &$\pm$& 1.6\\
$\frac{\tilde{B}(\eta\eta)}{\tilde{B}(\eta\etap)}$  & 
$\frac{1}{2}\tan^2(\theta_i - 
\theta_p)$ & -17.7 &$\pm$& 1.9\\
$\frac{\tilde{B}(\omega\eta)}{\tilde{B}(\omega\etap)}$  & 
$\tan^2(\theta_i - \theta_p)$ 
& -21.1 &$\pm$&1.5\\
$\frac{\tilde{B}(\eta\rho^0)}{\tilde{B}(\etap\rho^0)}$  & 
$\tan^2(\theta_i - \theta_p)$ & 
-25.4 &$^{+}_{-}$& $^{5.0}_{2.9}$\\
\hline
$\frac{\tilde{B}(\eta\rho^0)}{\tilde{B}(\omega\pi^0)}$  & 
$\sin^2(\theta_i - \theta_p)$ & 
-11.9 &$\pm$& 3.2\\
$\frac{\tilde{B}(\etap\rho^0)}{\tilde{B}(\omega\pi^0)}$  &  
$\cos^2(\theta_i - \theta_p)$ & 
-30.5 &$\pm$& 3.5\\
$\frac{\tilde{B}(\eta\eta)}{\tilde{B}(\pi^0\pi^0)}$  & 
$\sin^4(\theta_i - \theta_p)$ & -6.2 & $^{+}_{-}$& $^{0.6}_{1.1}$\\
$\frac{\tilde{B}(\eta\etap)}{\tilde{B}(\pi^0\pi^0)}$  & 
$2\sin^2(\theta_i - \theta_p)$ & 14.6 &$\pm$& 1.8\\
&  $\times\cos^2(\theta_i - \theta_p)$ & or -34.0 &$\pm$& 1.8\\
$\frac{\tilde{B}(\omega\eta)}{\tilde{B}(\pi^0\rho^0)}$  & 
$\sin^2(\theta_i - \theta_p)$ & 
-23.7 &$^{+}_{-}$& $^{7.6}_{8.9}$\\
$\frac{\tilde{B}(\omega\etap)}{\tilde{B}(\pi^0\rho^0)}$  & 
$\cos^2(\theta_i - \theta_p)$ 
& -20.1 &$\pm$& 3.7\\
\end{tabular}
\end{center}
\end{table}

\begin{table}[p] 
\caption[]{Damping factors $F_L(p)$ where $z$ stands for $(p/p_R)^2$ and $p_R$ is 
usually taken as  197 MeV/c (after Hippel and Quigg (1972)).}
\begin{center}
\begin{tabular}{ l l }
$L$ & $F_L(p)$ \\
\hline
0 & 1 \\
1 & $\sqrt{\frac{2z}{z+1}}$ \\
2 & $\sqrt{\frac{13z^2}{(z-3)^2+9z}}$ \\
3 & $\sqrt{\frac{277z^3}{z(z-15)^2+9(2z-5)^2}}$ \\
4 & $\sqrt{\frac{12,746z^4}{(z^2 - 45z + 105)^2 + 25z(2z - 21)^2}}$ 
\\
\end{tabular}
\end{center}
\label{Blatt}
\end{table}

\begin{table}[p]
\caption[]{Branching ratios $B$ for radiative $\pbarp$ annihilation 
at rest in liquid from Crystal Barrel (Amsler, 1993c, 1995c). The lower and upper
limits $L$ and $U$, calculated from VDM, are given in the third and fourth column, 
respectively.\\
$\dagger$ 95\% confidence upper limit.}
\begin{center}
\begin{tabular}{l r c r l c r }
Channel & \multicolumn{4}{c}{$B$} & $L$ & $U$ \\
\hline
$\pi^0\gamma$ & 4.4 & $\pm$ & 0.4 & $\times 10^{-5}$ & 
$3.1\times 10^{-5}$ &  $6.8 \times 10^{-5}$ \\
$\eta\gamma$ & 9.3 & $\pm$ & 1.4 & $\times 10^{-6}$ & $1.0\times 10^{-6}$ & $2.5\times 
10^{-5}$  \\
$\omega\gamma$ & 6.8 & $\pm$ & 1.8 & $\times 10^{-5}$ & $8.5\times 10^{-6}$ & $1.1\times 
10^{-4}$ \\
$\etap\gamma$ & $<$ 1.2 &   &   & $\times 10^{-5}\dagger$ & $2.7\times 10^{-7}$ 
& $10^{-5}$ \\
$\gamma\gamma$ & $<$ 6.3 &   &  & $\times 10^{-7}\dagger$ &  &   \\
$\phi\gamma$& 2.0 & $\pm$ & 0.4 & $\times 10^{-5}$ & $2.1\times 
10^{-7}$ & $1.5\times 10^{-6}$ \\
\end{tabular}
\label{BRG}
\end{center}
\end{table}

\begin{table}[p]
\caption[]{Branching ratios for $\phi$ production at 
rest in liquid.\\
$^a$ updates Amsler (1995c)\\
$^b$ annihilation in gas extrapolated to pure S-wave annihilation\\
$^d$ using Chiba (1988)  in liquid\\
$^d$ using Bizzarri (1971) in liquid\\
$\ddagger$ Crystal Barrel experiment}
\begin{center}
\begin{tabular}{l r c r r l }
Channel & \multicolumn{4}{c}{$B$} & Reference \\
\hline
$\pi^0\phi$& 6.5 &$\pm$& 0.6 & 10$^{-4}$ & $^a\ddagger$\\
$\pi^0\phi$& 3.0 &$\pm$& 1.5 & 10$^{-4}$ & Chiba (1988) \\
$\pi^0\phi$& 4.0 &$\pm$& 0.8 & 10$^{-4}$ & Reifenr\"other  (1991)$^{bc}$\\
$\eta\phi$& 7.8 &$\pm$& 2.1 & 10$^{-5}$ & Amsler (1995c)$\ddagger$\\
$\eta\phi$& 3.0 &$\pm$& 3.9 & 10$^{-5}$ & Reifenr\"other  (1991)$^b$\\
$\omega\phi$& 6.3 &$\pm$& 2.3 & 10$^{-4}$ & Bizzarri (1971)\\
$\omega\phi$& 5.3 &$\pm$& 2.2 & 10$^{-4}$ & Reifenr\"other  (1991)$^{bd}$\\
$\rho^0\phi$& 3.4 &$\pm$& 1.0 & 10$^{-4}$ & Reifenr\"other  (1991)$^b$\\
$\gamma\phi$& 2.0&$\pm$& 0.4 & 10$^{-5}$ &  $^a\ddagger$\\
\end{tabular}
\label{BRR}
\end{center}
\end{table}

\begin{table}[p]
\caption[]{Ratio of $\phi$ to $\omega$ production in low energy 
annihilation in liquid.}
\begin{center}
\begin{tabular}{l c c l}
$X$ & \multicolumn{3}{c}{$\tilde{R}_X$ [$10^{-2}$]} \\
\hline
$\gamma$ & 29.4 & $\pm$ & 9.7\\
$\pi^0$ & 10.6 & $\pm$ & 1.2 \\
$\eta$ & 0.46 & $\pm$ & 0.13 \\
$\omega$ & 1.02 & $\pm$ & 0.39 \\
$\rho^0$ & 1.57 & $\pm$ & 0.49 \\
$\pi^-$ & 13.0 & $\pm$ & 2.5 \\
$\pi^+$ & 10.8 & $\pm$ & 1.5 \\
$\sigma$ & 1.75 & $\pm$ & 0.25 \\
$\pi^+\pi^-$ & 1.65 & $\pm$ & 0.35 \\
\end{tabular}
\label{R_X}
\end{center}
\end{table}

\begin{table}[p]
\caption[]{Relative sign of $\alpha_1$ and $\alpha_2$ for $\pbarp$ 
annihilation into $\pi\KKbar$  (see text).}
\begin{center}
\begin{tabular}{l l l l l}
Channel & \multicolumn{2}{c}{$^1S_0(\pbarp)$} & \multicolumn{2}{c}{$^3S_1(\pbarp)$} 
\\
& $I=0$ & $I=1$ & $I=0$ & $I=1$ \\
\hline
$\pi^{\pm} K^{\mp} K^0$ & + & -- & -- & + \\
$\pi^0K^+K^-$ & + & + & -- & -- \\
$\pi^0K^0\overline{K^0}$ & + & + & -- & -- \\
\end{tabular}
\label{symm}
\end{center}
\end{table}

\begin{table}[p] 
\caption[]{Branching ratios for $\pbarp$ annihilation at rest into three narrow mesons. 
Mesons in parentheses were not detected.\\
$^a$ using $B(\pi^0\phi)$ from Table \ref{BRR} and Eq. (\ref{40})\\
$^b$ average between Armenteros (1965) and Barash (1965)\\
$\ddagger$ Crystal Barrel experiment}
\begin{center}
\begin{tabular}{ l l l l l l l}
Channel & Final state & \multicolumn{4}{c}{$B$} & Reference \\
\hline
$\pi^0\pi^0\pi^0$ & $6\gamma$ & 6.2  & $\pm$ & 1.0 & $10^{-3}$ & Amsler 
(1995f)$\ddagger$ 
\\
$\pi^+\pi^-\pi^0$ & $\pi^+\pi^-(\pi^0) $ & 6.9 & $\pm$ & 0.4 & $10^{-2}$ & Foster (1968b)\\
$\pi^0\eta\eta$ & $6\gamma$ & 2.0  & $\pm$ & 0.4 & $10^{-3}$ & Amsler (1995e)$\ddagger$ 
\\
$\pi^0\pi^0\omega$ & $7\gamma$ &2.00  & $\pm$ & 0.21 & $10^{-2}$ & Amsler 
(1993a)$\ddagger$ \\
& $\pi^+\pi^- 6\gamma$ & 2.57  & $\pm$ & 0.17 & $10^{-2}$ & Amsler (1994d)$\ddagger$ \\
$\pi^+\pi^-\omega$ & $2\pi^+2\pi^-(\pi^0) $ & 6.6 & $\pm$ & 0.6 & $10^{-2}$ & Bizzarri 
(1969)\\
$\omega\eta\pi^0 $ & $7\gamma$ & 6.8  & $\pm$ & 0.5 & $10^{-3}$ & Amsler 
(1994c)$\ddagger$ \\
$\pi^0\pi^0\eta$ & $6\gamma$ & 6.7  & $\pm$ & 1.2 & $10^{-3}$ & Amsler 
(1994b)$\ddagger$ \\
& $\pi^+\pi^- 6\gamma$ & 6.50  & $\pm$ & 0.72 & $10^{-3}$ & Amsler (1994d)$\ddagger$ \\
$\pi^+\pi^-\eta$ & $\pi^+\pi^- 2\gamma$ & 1.63  & $\pm$ & 0.12 & $10^{-2}$ & Abele  
(1997a)$\ddagger$ \\
& $\pi^+\pi^- 6\gamma$ & 1.33  & $\pm$ & 0.16 & $10^{-2}$ & Amsler (1994d)$\ddagger$ \\
& $2\pi^+2\pi^-(\pi^0) $ & 1.38 & $\pm$ & 0.17 & $10^{-2}$ & Espigat (1972)\\
& $2\pi^+2\pi^-(\pi^0) $ & 1.51 & $^+_-$ & $^{0.17}_{0.21}$ & $10^{-2}$ & Foster (1968a)\\
$\pi^0\pi^0\etap$ & $10\gamma$ & 3.2  & $\pm$ & 0.5 & $10^{-3}$ & Abele 
(1997g)$\ddagger$ \\
& $6\gamma$ & 3.7  & $\pm$ & 0.8 & $10^{-3}$ & Abele 
(1997g)$\ddagger$\\
$\pi^+\pi^-\etap$ & $\pi^+\pi^- 6\gamma$ & 7.5  & $\pm$ & 2.0 & $10^{-3}$ & Urner 
(1995)$\ddagger$ \\
& $3\pi^+3\pi^-(\pi^0) $ & 2.8 & $\pm$ & 0.9 & $10^{-3}$ & Foster 
(1968a)\\
$\pi^0\eta\etap$ & $6\gamma$ & 2.3  & $\pm$ & 0.5 & $10^{-4}$ & Amsler 
(1994f)$\ddagger$ 
\\
$\pi^0\pi^0\phi$ & $ 8\gamma (K_L) $ & 9.7   & $\pm$ & 2.6  & $10^{-5}$ & Abele 
(1997f)$\ddagger$ \\
$\pi^+\pi^-\phi$ & $2\pi^+2\pi^-(K_L)$ & 4.6   & $\pm$ & 0.9  & $10^{-4}$ & Bizzarri (1969) 
\\
$\pi^0K_SK_L$ & $3\pi^0(K_L)$ & 6.7 & $\pm$ & 0.7 & $10^{-4}$ & Amsler 
(1993d)$^a\ddagger$\\
$\pi^0K_SK_S$ & $2\pi^+2\pi^-(\pi^0)$ &7.5 & $\pm$ & 0.3 & $10^{-4}$ & $^b$\\
$\pi^{\pm}K^{\mp}K_S $ & $\pi^+\pi^-\pi^{\pm}K^{\mp}$ & 2.73 & $\pm$ & 0.10 & 
$10^{-3}$ & $^b$\\
$\pi^{\pm}K^{\mp}K_L $ & $\pi^{\pm}K^{\mp} (K_L)$ & 2.91 & $\pm$ & 0.34 & 
$10^{-3}$ &  Abele (1997e)$\ddagger$\\
$\omega K_SK_S$ & $3\pi^+3\pi^-(\pi^0)$ & 1.17 & $\pm$ & 0.07 & $10^{-3}$ & Bizzarri 
(1971)\\
$\omega K^+K^-$ & $K^+K^-\pi^+\pi^-(\pi^0)$ & 2.30 & $\pm$ & 0.13 & $10^{-3}$ & Bizzarri 
(1971)\\ 
\end{tabular}
\end{center}
\label{TBBR}
\end{table}

\begin{table}[p]
\caption[]{Branching ratios for $\pbarp$ annihilation at rest in liquid 
determined from Dalitz plot analyses. The branching ratios include all decay 
modes of the final state stable particles (e.g. $\pi^0$,  $\eta$, $\etap$) but only 
the decay mode of the intermediate resonance leading to the observed final 
state.}
\begin{center}
\begin{tabular}{ l l }
Channel &  Contributing resonances \\
Subchannel & Branching ratio \\
\hline
\hline
$\pi^0\eta\eta$ & $a_0(980), f_0(1370), f_0(1500), a_2(1320), 
X_2(1494)$  \\
\hline
$f_0(1370)\pi^0$ &  $\sim 3.5 \times 10^{-4}$ \\
$f_0(1500)\pi^0$ &  $(5.5 \pm 1.3)\times 10^{-4}$ \\
$a_2(1320)\eta$ & $\sim 5.6 \times 10^{-5}$ \\
$X_2(1494)\pi^0$ & $\sim 4.0 \times 10^{-4}$ (dominantly P-wave annihilation)\\
\hline
\hline
$\pi^0\pi^0\eta$ &  $a_0(980), a_0(1450), a_2(1320),
a_2'(1650), (\eta\pi)_P $\\ 
& $(\pi\pi)_S\equiv f_0(400-1200) + f_0(980) + f_0(1370)$\\
\hline
$a_0(980)\pi^0$ & $(8.7 \pm 1.6)\times 10^{-4}$ \\
$a_0(1450)\pi^0$ &  $(3.4 \pm 0.6)\times 10^{-4}$ \\
$(\pi\pi)_S\ \eta$ &  $(3.4 \pm 0.6)\times 10^{-3}$ \\
$(\eta\pi)_P\ \pi$ & $\sim 1.0 \times 10^{-4}$ \\
$a_2(1320)\pi^0$ &  $(1.9 \pm 0.3)\times 10^{-3}$ \\
$a_2'(1650)\pi^0$ &  $\sim 1.3 \times 10^{-4}$ \\
\hline
\hline
$\pi^0\pi^0\pi^0$ &  $(\pi\pi)_S + f_0(1500)$, $f_2(1270), 
f_2(1565)$\\
\hline
$(\pi\pi)_S \ \pi^0$ &  $\sim 2.6 \times 10^{-3}$ \\
$f_0(1500)\pi^0$ &  $(8.1 \pm 2.8)\times 10^{-4}$ \\
$f_2(1270)\pi^0$ &  $\sim 1.8 \times 10^{-3}$ \\
$f_2(1565)\pi^0$ &  $\sim 1.1 \times 10^{-3}$ \\
\hline
\hline
$\pi^0\eta\eta, \pi^0\pi^0\eta, 3\pi^0$ &  Coupled channels (S-wave annihilation only)\\
& $(\eta\eta)_S\equiv f_0(400-1200) + f_0(1370)$\\
\hline
$(\pi\pi)_S\ \pi^0$ &  $(3.48 \pm 0.89)\times 10^{-3}$ \\
$(\pi\pi)_S\ \eta$ &  $(3.33 \pm 0.65)\times 10^{-3}$ \\
$(\eta\eta)_S\ \pi^0$  &$ (1.03 \pm 0.29)\times 10^{-3}$ \\
$f_0(1500)(\rightarrow\pi^0\pi^0)\pi^0$ &  $(1.27 \pm 0.33)\times 10^{-3}$ 
\\
$f_0(1500)(\rightarrow\eta\eta)\pi^0$ &   $(0.60 \pm 0.17)\times 10^{-3}$ \\
$f_2(1270)(\rightarrow\pi^0\pi^0)\pi^0$   & $(0.86 \pm 0.30)\times 10^{-3}$ 
\\
$f_2(1565)(\rightarrow\pi^0\pi^0)\pi^0$   & $(0.60 \pm 0.20)\times 10^{-3}$ 
\\
$f_2(1565)(\rightarrow\eta\eta)\pi^0$  & $(8.60 \pm 3.60)\times 10^{-5}$ (may be more 
than one object)  \\
$a_0(980)(\rightarrow\pi^0\eta)\pi^0$ &  $(0.81 \pm 0.20)\times 
10^{-3}$ \\
$a_0(980)(\rightarrow\pi^0\eta)\eta$ &  $(0.19 \pm 0.06)\times 
10^{-3}$ \\
$a_0(1450)(\rightarrow\pi^0\eta)\pi^0$ &  $(0.29 \pm 
0.11)\times 10^{-3}$ \\
$a_2(1320)(\rightarrow\pi^0\eta)\pi^0$ & $(2.05 \pm 0.40)\times 10^{-3}$ (including 
$a_2'(1650)$) \\
\hline
\hline
$\pi^0\eta\etap$ &  $ f_0(1500)$ \\
\hline
$f_0(1500)\pi^0$ &  $(1.6 \pm 0.4)\times 10^{-4}$ \\
\hline
\hline
$\pi^0\pi^0\etap$ & $ (\pi\pi)_S, a_2(1320), a_0(1450)$ \\
\hline
$(\pi\pi)_S\ \etap$ &  $(3.1 \pm 0.4)\times 10^{-3}$ \\
$a_2(1320)\pi^0$ &  $(6.4 \pm 1.3)\times 10^{-5}$ \\
$a_0(1450)\pi^0$ &   $(1.16 \pm 0.47)\times 10^{-4}$ \\
\label{DPBR}
\end{tabular}
\end{center}
\end{table}

\begin{table}[p]
\caption[]{Branching ratios for $\pbarp$ annihilation at rest in liquid 
into kaonic channels. The branching ratios include only 
the decay mode of the intermediate resonance leading to the observed final 
state.\\
$^a$ from the corresponding $K_S$ channels (Armenteros, 1965; Barash, 
1965)\\
$^b$ includes low energy $K\pi$ scattering\\
$^c$ fixed by $\pi^{\pm} K^\mp K_L$ data}
\begin{center}
\begin{tabular}{l l l }
Channel & $\pbarp (I)$ & Contributing resonances \\
Subchannel & & Branching ratio \\
\hline
\hline
$\pi^0K_LK_L$ &  & $ K^*(892), K^*_0(1430), a_2(1320), f_2(1270), 
f_2'(1525)$\\
& & $f_0(1370), f_0(1500), a_0(1450)$\\
Total$^a$ & & $(7.5 \pm 0.3)\times 10^{-4}$ \\
\hline
$K^*(892)\overline{K}$ & $^1S_0(0,1)$ & $(8.71 \pm 0.68)\times 10^{-5}$ \\
$K^*_0(1430)\overline{K}$ & &  $(4.59 \pm 0.46)\times 10^{-5}$ $^b$ \\
$a_2(1320)\pi^0$ & $^1S_0(0)$ & $(6.35 \pm 0.74)\times 10^{-5}$ \\
$a_0(1450)\pi^0$ & & $(7.35 \pm 1.42)\times 10^{-5}$ $^c$ \\
$f_2(1270)\pi^0$ & $^1S_0(1)$ & $(4.25 \pm 0.59)\times 10^{-5}$ \\
$f_2'(1525)\pi^0$ & & $(1.67 \pm 0.26)\times 10^{-5}$ \\
$f_0(1370)\pi^0$ & & $(2.20 \pm 0.33)\times 10^{-4}$ \\
$f_0(1500)\pi^0$ & & $(1.13 \pm 0.09)\times 10^{-4}$ \\
\hline
\hline
$\pi^{\pm}K^\mp K_L$ & & $ K^*(892), K^*_0(1430), 
a_2(1320)$\\
&  & $ a_0(980), a_0(1450)$, $\rho(1450/1700)$\\
Total$^a$ & & $(2.73 \pm 0.10)\times 10^{-3}$ \\
\hline
$K^*(892)\overline{K}$ & $^1S_0(0)$ &  $(2.05 \pm 0.28)\times 10^{-4}$ \\
$K^*_0(1430)\overline{K}$ & & $(8.27 \pm 1.93)\times 10^{-4}$ $^b$  \\
$a_0(980)\pi$ & & (1.97 $  ^{+ \ 0.15}_{- \ 0.34})\times 10^{-4}$\\
$a_2(1320)\pi$ & & (3.99 $  ^{+ \ 0.31}_{- \ 0.83})\times 10^{-4}$\\
$a_0(1450)\pi$ & & $(2.95 \pm 0.56)\times 10^{-4}$\\
$K^*(892)\overline{K}$ & $^1S_0(1)$ & $(3.00 \pm 1.10)\times 10^{-5}$ \\
$K^*_0(1430)\overline{K}$ & &  $(1.28 \pm 0.55)\times 10^{-4}$ $^b$\\
$\rho(1450/1700)\pi$ & & (8.73 $  ^{+ \ 1.40}_{- \  2.75})\times 10^{-5}$\\ 
\hline
$K^*(892)\overline{K}$ & $^3S_1(0)$ & $(1.50 \pm 0.41)\times 10^{-4}$ \\
$\rho(1450/1700)\pi$ & & $(8.73 \pm 2.75)\times 10^{-5}$\\ 
$K^*(892)\overline{K}$ & $^3S_1(1)$ & $(5.52 \pm 0.84)\times 10^{-4}$ \\
$a_2(1320)\pi$ & & $(1.42 \pm 0.44)\times 10^{-4}$ \\
\end{tabular}
\label{DPBRK}
\end{center}
\end{table}

\begin{table}[p]
\caption[]{Weights of the channels contributing to 
$\pbarp\rightarrow\pi\KKbar$. $I$ refers to the $\pbarp$ isospin and $i$ to the 
$\KKbar$ isospin.}
\begin{center}
\begin{tabular}{l l l l l}
Channel & $i=1/2$ & $i=1$ & $i=0$ & $i=1$\\
& $I=0,1$ & $I=1$ & $I=1$ & $I=0$\\
\hline
$\pi^{\pm} K^\mp K^0$ & 4 & 2 & 0 & 4 \\
$\pi^0K^+K^-$ & 1 & 0 & 1 & 1 \\
$\pi^0K^0\overline{K^0}$ & 1 & 0 & 1 & 1 \\
\end{tabular}
\label{iweight}
\end{center}
\end{table}

\begin{table}[p]
\caption[]{Branching ratios $B$ for two-body $\pbarp$ annihilation at 
rest in liquid hydrogen (including all decay modes), calculated from 
the final states given in the last column.\\
$^a$ assumes 100\% $b_1$ decay to $\pi\omega$\\
$^b$ using $B(\pi^0\phi)$ from Table \ref{BRR} and Eq. (\ref{40})}
\begin{center}
\begin{tabular}{l r c r r l }
Channel & \multicolumn{4}{c}{$B$} & Final state or ref. \\
\hline
$f_2(1270)\pi^0$ & 3.1 &  $\pm$ & 1.1 & 10$^{-3}$ & $\pi^0\pi^0\pi^0$\\ 
$f_2(1270)\pi^0$ & 3.7 &  $\pm$ & 0.7 & 10$^{-3}$ & $\pi^0K_LK_L$ \\
 & 4.3 &  $\pm$ & 1.2 & 10$^{-3}$ & Foster (1968b) \\
$f_2(1270)\omega$ & 3.26 &  $\pm$ & 0.33 & 10$^{-2}$ & Bizzarri 
(1969) \\
$f_2(1270)\omega$ & 2.01 &  $\pm$ & 0.25 & 10$^{-2}$ & Amsler 
(1993a) \\
$f_0(1500)\pi^0$ & 1.29 &  $\pm$ & 0.11 & 10$^{-2}$ & $\pi^0K_LK_L$ \\
$f_2'(1525)\pi^0$ & 7.52 &  $\pm$ & 1.20 & 10$^{-5}$ & $\pi^0K_LK_L$ \\
\hline
$a_2(1320)\pi $ & 3.93 &  $\pm$ & 0.70 & 10$^{-2}$ & $\pi^0\pi^0\eta$\\
$(^1S_0)$ & 3.36 &  $\pm$ & 0.94 & 10$^{-2}$ & $\pi^0\pi^0\etap$ \\
& 1.55 &  $\pm$ & 0.31 & 10$^{-2}$ &  $\pi^0K_LK_L$ \\
& 2.44 &  $^+_-$ & $^{0.44}_{0.64}$ & 
10$^{-2}$ & $\pi^{\pm}K^{\mp}K_L$\\
& 1.32 &  $\pm$ & 0.37   & 10$^{-2}$ & Conforto (1967) \\
$a_2^\pm(1320)\pi^{{\mp}}$ & 5.79 &  $\pm$ & 2.02 & 10$^{-3}$ 
& $\pi^{\pm}K^{\mp}K_L$ \\
$(^3S_1)$ &  4.49 &  $\pm$ &  1.83 & 10$^{-3}$ & Conforto (1967) 
\\
\hline
$b_1^\pm(1235)\pi^{{\mp}}$ &  7.9 &  $\pm$ & 1.1  &10$^{-3}$ & Bizzarri  
(1969) $^a$ \\
$b_1^0(1235)\pi^0$ & 9.2 &  $\pm$ & 1.1 & 10$^{-3}$ & Amsler 
(1993a) $^a$ \\
$a_2^0(1320)\omega$ & 1.70 &  $\pm$ & 0.15 & 10$^{-2}$ & 
Amsler (1994c) \\
\hline
$K^*(892)\overline{K} $ &  7.05 &  $\pm$ &  0.90& 10$^{-4}$ 
& $\pi^{\pm}K^{\mp}K_L$ \\
$(^1S_0)$ &  1.05 &  $\pm$ &  0.08 & 10$^{-3}$& $\pi^0K_LK_L$ \\
&  1.5 &  $\pm$ &  0.3 & 10$^{-3}$ & Conforto (1967) \\
$K^*(892)\overline{K} $ &  2.11 &  $\pm$ &  0.28 & 10$^{-3}$ & 
$\pi^{\pm}K^{\mp}K_L$ \\
$(^3S_1)$ &  2.70 &  $\pm$ &  0.37 & 10$^{-3}$ & $\pi^0K_SK_L$ $^b$ \\
&  $\geq$ 2.51 &  $\pm$ &  0.22 & 10$^{-3}$ & Conforto (1967) \\
\label{BRTB}
\end{tabular}
\end{center}
\end{table}

\clearpage

\begin{figure}
\vspace*{1mm}
\caption[]{Pion multiplicity distribution for $\pbarp$ annihilation 
at rest in liquid hydrogen. Open squares: statistical distribution;  full circles: data; open 
circles: estimates from Guesqui\`ere (1974). The curve is a Gauss fit assuming $\langle 
N\rangle = 5$.}
\label{mult}
\end{figure}

\begin{figure}
\vspace*{1mm}
\caption[]{Fraction $f_P$ of P-wave  annihilation as a 
function of hydrogen density (curve). The dots with error bars give the 
results from one particular optical model (Dover and Richard, 1980) using two-body 
branching ratios (adapted from Batty (1996)).}
\label{fp}
\end{figure}

\begin{figure}
\vspace*{1mm}
\caption[]{The Crystal Barrel detector. 1,2 - yoke, 3 - coil, 4 - CsI(Tl) 
barrel, 5 - JDC, 6 - PWC's, 7 - LH$_2$ target.}
\label{Barrel}
\end{figure}

\begin{figure}
\vspace*{1mm}
\caption[]{The silicon vertex detector. 1 - microstrip detectors, 2 -
hybrids, 3 - readout electronics, 4 - cooling ring (from Regenfus 
(1997)).}
\label{SVX}
\end{figure}

\clearpage

\begin{figure}
\vspace{1mm}
\caption[]{$2\gamma$ invariant mass distribution for a sample of  
$4\gamma$ events (6 entries/event).}
\label{gaga}
\end{figure}

\begin{figure}
\vspace{1mm}
\caption[]{$\pi^0\gamma$ momentum distribution in $\pbarp\rightarrow 
4\pi^0\gamma$ (4 entries/event). The peak is due to $\pbarp\rightarrow\omega\eta $. 
The inset shows the $\omega$-region and a fit (Gaussian and polynomial background).}
\label{ninega}
\end{figure}

\begin{figure}
\vspace{1mm}
\caption[]{$\pi^+\pi^-\pi^0$ invariant mass distribution  
for $\pi^+\pi^-\pi^0\eta$ events. The peak is due to 
$\pbarp\rightarrow\omega\eta $. The inset shows the background subtracted angular 
distribution in the $\omega$-rest frame (see text).}
\label{Schmid}
\end{figure}

\begin{figure}
\vspace{1mm}
\caption[]{$\gamma$ angular distribution in the $\omega$ rest 
frame for $\omega\eta (\omega\rightarrow\pi^0\gamma)$ (see text).}
\label{Strassb}
\end{figure}

\clearpage 

\begin{figure}
\vspace{1mm}
\caption[]{Annihilation graph A   and rearrangement graph R  for 
$\pbarp$ annihilation into two mesons.}
\label{AR}
\end{figure}

\begin{figure}
\vspace{1mm}
\caption[]{Following VDM, radiative annihilation can be 
described by a superposition of two isospin amplitudes with unknown relative phase 
$\beta$. $X$ stands for any neutral meson.}
\label{Isospin}
\end{figure}

\begin{figure}
\vspace{1mm}
\caption[]{Energy deposits in the barrel 
versus polar angle $\Theta$ and azimuthal angle $\Phi$ for a $\pi^0\gamma$ event. The 
two $\gamma$'s from $\pi^0$ decay cluster near the minimum 
opening angle ($16.5^\circ$).}
\label{pi0gam}
\end{figure}

\begin{figure}
\vspace{1mm}
\caption[]{Energy distribution (24,503 events) of the single 
$\gamma$ in the missing $\pi^0$ rest frame for events satisfying 
the kinematics $\pbarp\rightarrow 2\pi^0$ and a missing $\pi^0$.
The full line is a fit. The dotted curve shows the expected signal for a 
branching ratio of $5\times 10^{-4}$.}
\label{pi0X}
\end{figure}

\begin{figure}
\vspace{1mm}
\caption[]{90 \% confidence level upper limits for radiative pseudoscalar decays 
as a function of missing mass.}
\label{upplim}
\end{figure} 

\clearpage

\begin{figure}
\vspace{1mm}
\caption[]{Squared matrix element for $\eta\rightarrow 3\pi^0$. The straight line shows 
the fit according to Eq. (\ref{72}).}
\label{zplot}
\end{figure}

\begin{figure}
\vspace{1mm}
\caption{(a) $\omega \rightarrow 3 \gamma$ Dalitz plot for $\pi^0
 \omega$ events (62,853 events, 6 entries/event); (b) $\omega \rightarrow 3 \gamma$ 
Dalitz plot for $\eta\omega$ events (54,865 events, 6 entries/event).}
\label{Omega}
\end{figure}

\begin{figure}
\vspace{1mm}
\caption[]{Dalitz plot of the final state $K_SK_L\pi^0$ 
(2,834 events).}
\label{kskl}
\end{figure}

\begin{figure}
\vspace{1mm}
\caption[]{Ratio of $\phi$ to $\omega$ production in low energy 
annihilation. The measured branching ratios have been divided by the factor $W$ (Eq. 
(\ref{10})). The expectation 
from the OZI  rule using the quadratic mass formula ($4.2\times 10^{-3}$) is 
shown by the horizontal line.}
\label{figR_X}
\end{figure}

\begin{figure}
\vspace{1mm}
\caption[]{(a,b): OZI allowed $\phi$ production with $\ssbar$ pairs in the 
nucleon. In (c) an intermediate four-quark state is excited below 
threshold. The production of $\phi$ mesons can also be enhanced 
by final state rescattering, $K^*\overline{K}\rightarrow\pi\phi$  or 
$\rho\rho\rightarrow\pi\phi$ (d).}
\label{OZIphi}
\end{figure}

\clearpage

\begin{figure}
\vspace{1mm}
\caption[]{$\eta\pi$ and $\KKbar$ mass distributions for 
the $a_0(980)$ resonance in $\pbarp\rightarrow\eta\pi X$ and 
$\KKbar X$ (in arbitrary units and assuming that no other resonance is produced in these 
channels). The dashed line shows the $\eta\pi$ mass distribution for the same width 
$\Gamma_0'$ in the absence of $\KKbar$ coupling ($g_2=0$).}
\label{Flatte}
\end{figure}

\begin{figure}
\vspace{1mm}
\caption[]{Invariant mass distributions for $\pi^0\eta\eta $; (a) $\eta\eta$ 
mass distribution for $6\gamma$ events showing the two new scalar mesons; 
(b) $\eta\eta$ mass distribution for one $\eta$ decaying to $3\pi^0$ ($10\gamma$ final 
state, not corrected for acceptance); (c) $\pi^0\eta$ mass distribution for $6\gamma$ 
events (2 entries/event) showing the $a_0(980)$. The solid lines in (a) and (c)  represent 
the best fit described in Amsler (1992c).}
\label{etaetapi}
\end{figure}

\begin{figure}
\vspace{1mm}
\caption[]{Dalitz plots of 3-pseudoscalar channels. Red and blue  regions correspond to 
high, respectively low, event densities;  (a) $\pi^0\eta\eta$ (198,000 events). The Dalitz plot  
is symmetrized across the main diagonal; (b) $\pi^0\pi^0\eta$ 
(symmetrized, 280'000 events); (c): $3\pi^0$  (712,000 events). Each event 
is entered six times for symmetry reasons; (d) $\pi^0K_LK_L$ (37,358 events). }
\label{Dalitz}
\end{figure}

\begin{figure}
\vspace{1mm}
\caption[]{$\pi^0\pi^0$ mass projection in $\pbarp\rightarrow 3\pi^0$ (3 entries/ event) 
with the fit (solid line) described in the text.}
\label{pipipi}
\end{figure}

\clearpage

\begin{figure}
\vspace{1mm}
\caption[]{Argand diagram of the $\pi\pi$ scattering amplitude $T$ obtained
from a common fit to production and scattering data (from Spanier, 1994).}
\label{Argand}
\end{figure}

\begin{figure}
\vspace{1mm}
\caption[]{Isoscalar S-wave production  
intensities $|{\cal T}|^2$ in $3\pi^0$ (full curve), $2\pi^0\eta$ (dashed curve) and 
$2\eta\pi^0$ (dotted curve) before multiplying by the phase space factor $\rho$. The 
vertical scale is arbitrary (from Spanier, 1994).}
\label{Intense}
\end{figure}

\begin{figure}
\vspace{1mm}
\caption[]{$\eta\etap$ mass projection in $\pi^0\eta\etap$. The full curve is the fit to the 
6$\gamma$ data  with a scalar resonance close to threshold and the dashed curve shows 
the expected phase space distribution. The inset shows the  $\eta\etap$ mass distribution 
from $\pi^0\eta(\rightarrow 2\gamma)\etap(\rightarrow\eta\pi^+\pi^-)$.}
\label{peep}
\end{figure}

\begin{figure}
\vspace{1mm}
\caption[]{$2\pi^0\eta$ mass distribution recoiling against $\pi^0\pi^0$ for events with 10 
reconstructed photons (6 entries/event).}
\label{fourpe}
\end{figure}

\begin{figure}
\vspace{1mm}
\caption[]{$\etap\pi^0$ mass projection in 
$\pi^0\pi^0\etap\rightarrow 6\gamma$  for data 
(dots) and fits (histograms); (a) ($\pi\pi$) S-wave and  
$a_2(1320)$; (b) including $a_0(1450)$ (best fit).}
\label{ppetpfit}
\end{figure}

\clearpage

\begin{figure}
\vspace{1mm}
\caption[]{The phase shift $\delta$ in elastic $K\pi$ scattering from Aston  
(1988b). The curve shows the fit using prescription (\ref{62}).}
\label{LASS}
\end{figure}
 
\begin{figure}
\vspace{1mm}
\caption[]{$\KKbar$ mass projection in $\pi^0K_LK_L$. The dashed line shows the fit with 
one scalar resonance, the full line the fit with two scalar resonances. 
The peak on the left is due to $f_2(1270)$ and $a_2(1320)$ and the 
peak on the right to $K^*(892)$ reflections. The central peak is due to 
interference from various amplitudes.}
\label{KLKLp}
\end{figure}

\begin{figure}
\vspace{1mm}
\caption[]{Branching ratio for $f_0(1370)$ and $f_0(1500)$ decay into $\KKbar$ as a 
function of $a_0(1450)$ contribution to $\pi^0K_LK_L$  (from Dombrowski (1996)).}
\label{correl}
\end{figure}

\begin{figure}
\vspace{1mm}
\caption[]{$dE/dx$ distribution in the jet drift chamber for 2-prong events with a 
missing $K_L$. The curve shows the expected (Bethe-Bloch) dependence.}
\label{dedx}
\end{figure}

\begin{figure}
\vspace{1mm}
\caption[]{$\pi^{\pm}K^{\mp}K_L$ Dalitz plot (11,373 events).}
\label{KKp}
\end{figure}

\clearpage

\begin{figure}
\vspace{1mm}
\caption[]{$\chi^2$ dependence on the fractional contribution from $a_0(1450)$ 
to  $\pi^{\pm}K^{\mp}K_L$.}
\label{a0toKK}
\end{figure}

\begin{figure}
\vspace{1mm}
\caption[]{Tangent of the nonet mixing angle $\alpha$ as a function of $R_1$ (solid), 
$R_2$ (dotted) and $R_3$ (dashed curve). The shaded areas show the experimentally 
allowed regions for $f_0(1500)$, assuming that this state is $\qqbar$.}
\label{tanal}
\end{figure}

\begin{figure}
\vspace{1mm}
\caption[]{$\pi^0\pi^0\eta$ (a) and $\pi^+\pi^-\eta$ (b) mass distributions in $\pbarp$ 
annihilation at rest into $\pi^+\pi^-\pi^0\pi^0\eta$, showing the $\etap$ and $\eta(1410)$ 
signals. The dashed line shows the result of the fit.}
\label{Espec}
\end{figure}

\begin{figure}
\vspace{1mm}
\caption[]{Angular distribution of $a_0(980)^\pm$ in the $\eta(1410)$ rest frame. The data 
are shown with error bars. The full curve shows the fit for a $0^{-+}$ state and the dashed 
curve the prediction for a $1^{++}$ state produced with the same intensity.}
\label{Ange}
\end{figure}

\end{document}